\documentclass[reprint,amsmath,amssymb,rmp]{revtex4-2}
\usepackage{graphicx}
\usepackage[colorlinks]{hyperref}
\usepackage{wasysym}
\usepackage{ulem}
\usepackage[mathlines,columnwise,modulo]{lineno}

\newcommand{\add}[1]{{#1}}
\newcommand{\rem}[1]{{}}

\newcommand{\addr}[1]{{\color{blue}#1}}

\newcommand{\tavg}[1]{\langle#1\rangle}
\newcommand{\tavgb}[1]{\left\langle #1 \right\rangle}

\begin{document}

\title{The climate system and the second law of thermodynamics}


\author{Martin S. Singh}
\email{Martin.Singh@monash.edu}
\affiliation{School of Earth, Atmosphere, and Environment, Monash University, Victoria, Australia}

\author{Morgan E O'Neill}
\affiliation{Department of Earth System Science, Stanford University, California, USA}

\date{\today}

\begin{abstract}
The second law of thermodynamics implies a relationship between the net entropy export by the Earth and its internal irreversible entropy production. 
The application of this constraint for the purpose of understanding Earth's climate is reviewed. 
Both radiative processes and material processes are responsible for irreversible entropy production in the climate system. 
Focusing on material processes,
an entropy budget for the climate system is derived which accounts for the multi-phase nature of the hydrological cycle.
The entropy budget facilitates a heat-engine perspective of atmospheric circulations that has been used to propose theories for convective updraft velocities, tropical cyclone intensity, and the atmospheric meridional heat transport. Such theories can only be successful, however, if they properly account for the irreversible entropy production associated with water in all its phases in the atmosphere.
Irreversibility associated with such moist processes is particularly important in the context of global climate change, for which the concentration of water vapor in the atmosphere is expected to increase, and recent developments toward understanding the response of the atmospheric heat engine to climate change are discussed.
Finally, the application of variational approaches to the climate and geophysical flows is briefly reviewed, including the use of equilibrium statistical mechanics to predict behavior of long-lived coherent structures, and the controversial maximum entropy production principle.

\end{abstract}

\maketitle
\tableofcontents

\section{Introduction}


\subsection{Motivation}




The Earth is a highly irreversible thermodynamic system.
It receives energy and entropy from the sun and radiates energy and entropy to outer space. But while the incoming and outgoing fluxes of energy are roughly in balance, the Earth exports vastly more entropy than it receives \citep{Peixotoetal1991,StephensOBrien1993}. \add{For a climate whose statistics are stationary,}
the second law of thermodynamics requires that this net export of entropy be balanced by irreversible production of entropy within the climate system. The second law therefore provides a fundamental steady-state constraint on the climate system, relating a measure of its internal activity, the irreversible production of entropy, to fluxes of entropy at its boundaries. 

In fact, the Earth's climate is not steady; 
it has undergone vast changes over Earth's history, from the icy cold of Snowball Earth episodes \citep{Hoffman1998} to the extreme warmth of the Late Cretaceous that allowed crocodile-like reptiles to roam the Arctic \citep{Tarduno1998}. Such climate variability occurs on a range of timescales \citep{Ghil2020} and implies imbalances in the planetary energy and entropy budgets. In the context of the climate change \add{observed in recent decades and projected over the next century}, these energy imbalances are small relative to the total incoming and outgoing fluxes \citep{Trenberth2014}, and the steady-state assumption provides a useful framework for understanding the second law as applied to the climate system. 


A range of processes are involved in the irreversible production of entropy within the climate system including the absorption and emission of radiation, the frictional dissipation of winds and ocean currents, \add{molecular diffusion of heat and mass, and phase changes of water within the hydrological cycle}. Indeed, life itself is an irreversible process, although we will not discuss the biotic generation of entropy in this review. But despite their ubiquity,
irreversible processes are often treated in simplified ways in studies of the large-scale atmospheric circulation \citep{Emanuel2001}, \add{while numerical models of the climate system often treat irreversible processes in a physically inconsistent way \citep[e.g.,][]{Becker2007}, neglect certain irreversible processes altogether \citep[e.g.,][]{Pauluis2002,Pascaleetal2011}, or include spurious numerical sources of entropy \citep{WoollingsThuburn2006}}. Moreover, interaction between the communities of climate scientists and physicists developing tools for the understanding of irreversible processes remains limited. Fostering collaborations between these communities has the potential to reveal new methods for analyzing and understanding the climate system, particularly in the context of \add{a rapidly changing climate}
\citep{Lucarini2010}.

The purpose of this review is to provide an introduction to the application of the second law of thermodynamics to the climate system suitable both for scientists active within climate research and for a more general audience of physicists. The research frontier in climate physics is rich with fascinating, complex problems \add{in their own right.  Amidst a period of rapid anthropogenic climate change \cite{IPCCAR5WG1techsumm}, many of the most societally urgent problems, such as predicting future freshwater availability, crop viability, or storm frequency, are also the most difficult, requiring collaboration with public and private decision-makers}. A training in traditional physics is excellent preparation for climate research, and researchers with expertise in areas including statistical mechanics, fluid dynamics and physical chemistry have much to offer as part of a vibrant, interdisciplinary climate science community \cite{Marston2011,Wettlaufer2016}. 


\subsection{Applications of the second law in climate research}

\add{The second law has been applied in diverse ways within the broad field of climate science. An important strand of research focuses on the quantification of irreversibility within the climate system through analysis of its entropy budget.} The bulk of Earth's irreversible entropy production occurs as a result of radiative processes \citep{StephensOBrien1993}. 
\add{But applications of the second law often focus on a subset of the climate system that includes only matter and considers radiation as part of the system's surroundings. This perspective allows for the definition of the material entropy budget, in which radiation acts as an external (and reversible) heat source, and irreversible radiative processes play no role \citep{Goody2000}.}
The steady-state material entropy budget
requires a balance between \add{material sources of entropy, such as frictional dissipation, heat and mass diffusion, and irreversible phase changes,} and the net sink of entropy owing to radiative heating at high temperature and radiative cooling at low temperature \citep{Pauluis2002}.

The material entropy budget provides a framework for analyzing the climate system as a heat engine. A number of studies have used a heat-engine based perspective to \addr{derive} theoretical constraints on the behavior of atmospheric circulations of various scales, including convective clouds \citep{Emanuel1996,Renno1996}, tropical cyclones \add{\citep{Emanuel1986,WangLin2021}}, and the global circulation \citep{Barryetal2002}. Like a heat engine, the climate system ingests heat in a warm region, transports it to a cool region where it is expelled, and performs an amount of work in the process.  
But unlike a traditional heat engine, the work performed by the climate system must be dissipated within the system itself \citep{Johnson2000,Lucarini2009}. \add{This cycle of kinetic energy production and dissipation may also be described through the Lorenz energy cycle and the concept of available potential energy \citep[APE; see section \ref{sec:APE} and][]{Lorenz1955,Lorenz1967}. The APE, and the related concept of exergy \citep{Tailleux2013a}, provide measures of the climate system's ability to perform work. Such concepts allow the second law, usually formulated in terms of entropy, to be recast in terms of transformations between different energy reservoirs.}

A major challenge for heat-engine based theories applied to the atmosphere is that they must properly account for the influence of ``moist'' processes---processes associated with water in the atmosphere.  
Moist processes are responsible for the bulk of the irreversible material entropy production in Earth's atmosphere, and this limits the efficiency with which the climate system's heat engine may generate kinetic energy in winds and ocean currents \citep{Pauluis2000,Pauluis2002,Romps2008}. The effects of irreversible moist processes are particularly relevant in the context of global climate change \citep{Laliberteetal2015,SinghOGorman2016}, as the concentration of water vapor in the atmosphere is expected to increase with warming roughly following the Clausius-Clapeyron relation \citep{OGorman2010b}.

\add{The second law has also been applied in climate research in ways that go beyond classical thermodynamics. In its most general form, the second law governs the macroscale evolution of an isolated system with many degrees of freedom toward a more probable state. In the field of statistical geophysical fluid dynamics, the system is a two-dimensional ideal fluid, and its degrees of freedom are the set of possible flow fields. Tools from statistical mechanics may then be applied to find the equilibrium flow structures based on the maximization of an entropy variable, subject to appropriate constraints. This approach has provided a range of insights into  nonequilibrium, steady-state geophysical flows on Earth and other planets \citep[e.g.,][]{BouchetVenaille2012,MajdaWang2006} in situations where a heat engine analysis is not applicable.

An advantage of the statistical approach is that it fundamentally involves a maximization problem, and it is therefore amenable to the powerful techniques described by the calculus of variations. A disadvantage is that it is only formally valid for equilibrium systems and cannot be applied to the climate system as a whole. A generalization of the entropy maximum formalism to non-equilibrium systems would therefore be of considerable value to climate research. Such a generalization was proposed by \citet{Paltridge1975,Paltridge1978}, who suggested that the climate system evolves to a state that maximizes its entropy \emph{production rate}. We briefly review the maximum entropy production (MEP) principle, \add{but we emphasize that there are a range of theoretical and modeling issues that pre-empt its broad acceptance in the field} (section \ref{sec:MEP}).}

\subsection{Structure of the review}
   
 The bulk of this review is focused on Earth's atmosphere, where most irreversible entropy production within the (material) climate system occurs.
 While this review primarily adopts a view of the second law focused on entropy production, irreversibility in the climate system
 may also be framed in energetic terms through the  concepts of exergy \citep[e.g.,][]{Bannon2005}, and available potential energy \citep{Lorenz1955}. We briefly discuss these approaches in section \ref{sec:entropic_energies}; the reader is referred to \citet{Tailleux2013a} for a more complete treatment. 
 Finally, we emphasize that this review covers only a small fraction of the broader research field of climate dynamics; a thorough review of the physics of climate change has been recently published in this journal by \citet{Ghil2020}. The remainder of the review is structured as follows.
 
 Section \ref{sec:second_law} 
 introduces the basic thermodynamic properties of the climate system. 
 We discuss methods of defining the boundaries of the system, including the planetary and material definitions used most commonly in the literature.
 We also describe the climate system as a heat engine, and we show how classical engineering concepts such as the work performed and the efficiency may be meaningfully applied to the climate system.
 
 
 Section \ref{sec:irr_procs} sketches a derivation of the entropy budget of the climate system. We focus on the material entropy budget of the atmosphere, and we describe the physical and mathematical origins of the main irreversible processes.
 We also briefly discuss the oceanic entropy budget and recent work estimating irreversible processes in the ocean.
 
 Sections \ref{sec:conv} and \ref{sec:TCs}
 review applications of the second law of thermodynamics to atmospheric convection and tropical cyclones,
 respectively. 
 In particular, we highlight how the irreversibility of moist processes fundamentally changes the fluid dynamics of the atmosphere.  
 
 Section \ref{sec:global_atmosphere} considers the global atmospheric circulation from a thermodynamic perspective. We consider theories of the global atmospheric heat engine and we discuss how it may change under climate change. We also review research describing the heat engines of other planets and bodies in the Solar System and beyond.
 
 
 

 Section \ref{sec:models} discusses some of the challenges faced in developing numerical models of the climate system that accurately represent the second law of thermodynamics. We describe practical and theoretical limitations of present modeling frameworks, and we suggest strategies to aid future model development.
 
 
 Section \ref{sec:Hamiltonian} provides an introduction to variational approaches to understanding geophysical fluid dynamics and the climate generally. We discuss the application of such approaches to atmospheric energetics and turbulence in large-scale geophysical flows. Here, both the classical thermodynamic definition of entropy, as well as the Boltzmann entropy of statistical mechanics, are employed.
 We also discuss the controversial maximum entropy production (MEP) principle, which has motivated much research into the climate system's entropy budget.
 
 Section \ref{sec:conclusions} concludes this review with a summary and discussion of outstanding research questions. We particularly highlight those areas that are likely to benefit from engagement with a broader community of physicists.

\section{The second law of thermodynamics applied to the climate system}
\label{sec:second_law}



\add{The second law of thermodynamics 
is fundamentally concerned with irreversibility; certain physical processes or transformations proceed spontaneously in one direction, but not in the reverse direction.
Common everyday examples include the cooling of a cup of tea to room temperature when it is left out or the evaporation of water from wet clothes hung out on a dry day. We do not expect a cup of tea to extract heat from its surroundings and spontaneously boil, and neither do we expect liquid water to condense out of the air on already wet clothes. All real macroscopic physical processes involve some degree of irreversibility, and the second law provides a framework for understanding such irreversible processes.}

A modern expression of the second law
states that the entropy $S$ of an isolated system must \add{not decrease} with time \citep{deGroot1984}:
\begin{equation} 
\frac{dS}{dt} \ge 0. \hspace{20pt} \text{(isolated system)}
\label{eq:second_law_general}
\end{equation}
\add{Here, we refer to an isolated system as one that does not exchange mass or energy with its environment.} The entropy is a function of the state of the system. \add{If an isolated system's entropy $S$ does not change, it is said to be reversible, while irreversible processes cause an increase in $S$.} 

The entropy may be defined using statistical mechanics as a measure of the number of microstates corresponding to a given macrostate, 
or in classical thermodynamics by
the relationship
\begin{equation}
    \frac{dS}{dt} = \frac{\dot{Q}_{\text{rev}}}{T}, \hspace{20pt} \text{(closed, reversible system)} \label{eq:entropy_def}
\end{equation}
valid for a closed, reversible system.
\add{Here, a closed system may exchange energy but not mass with its environment,} $\dot{Q}_\text{rev}$ represents a reversible heat transport from the surroundings to the system, and $T$ is the temperature at which this heat is transported 
\citep[e.g.,][]{deGroot1984,Iribarne1981}. 
\add{While \eqref{eq:entropy_def} is valid only for a closed, reversible system, as a state function, the entropy $S$ remains well defined under both reversible and irreversible conditions.}
The climate system exchanges energy with space in the form of radiation, and it is therefore not isolated.
The second law for a non-isolated system may be written in the more general form,
\begin{equation} 
\frac{dS}{dt} = \dot{S}_e + \dot{S}_i, 
\label{eq:second_law}
\end{equation}
where $\dot{S}_e$ is the net import of entropy from the surroundings and $\dot{S}_i$ is the production of entropy within the system owing to irreversible processes \citep{deGroot1984}. The second law of thermodynamics requires that $\dot{S}_i \ge 0$.



A simplification to \eqref{eq:second_law} may be made 
for systems close to steady state, where $dS/dt \approx 0$. 
At a given instant, this assumption is likely to be poor for the climte system; \citet{HuangMcElroy2015} computed observational estimates of various measures of the atmosphere's thermodynamic disequilibrium and found substantial seasonal variation. On longer timescales, however, 
the magnitude of the entropy tendency due to 
internally generated and forced climate variability 
is likely to be a small fraction of the total irreversible production of entropy by the climate system. 
\citet{Peixotoetal1991} argued that
for time averages over periods longer than a year, 
$dS/dt \approx 0$, 
and the second law of thermodynamics as applied to the climate system may be written,
\begin{equation} 
\tavg{\dot{S}_i} = -\tavg{\dot{S}_e}, \label{eq:entropy_balance}
\end{equation}
\add{where the angle brackets refer to a time average over a suitably long period. According to \eqref{eq:entropy_balance}, the time-mean irreversible entropy production rate of the climate system is equal to the time-mean net rate of export of entropy to space.} For applications to the Earth, it will prove useful to measure entropy exchanges per unit area of the Earth's surface, giving $\dot{S}_i$ the units of W m$^{-2}$ K$^{-1}$.

The steady-state entropy budget \eqref{eq:entropy_balance} states that, in order to maintain entropy producing processes such as those associated with winds, ocean currents, and the hydrological cycle, 
the climate system must export a greater quantity of entropy than it receives. This is manifest in the relatively high entropy contained in the radiation emitted from Earth to space compared to the lower entropy of the solar beam. 
More generally, the entropy budget places a fundamental constraint on the climate system by relating a measure of its internal activity, the total irreversible entropy production, to fluxes at its boundaries. 
One of the main purposes of studies of the climate's entropy budget is to leverage this constraint to better understand aspects of the climate system's behavior.

In the remainder of this section, we describe different methods of evaluating the entropy fluxes into and out of the climate system depending on how the system's boundaries are defined (section \ref{sec:fluxes}).
We also introduce the concept of the climate system as a heat engine, and we define the work done by the climate system and its thermodynamic and mechanical efficiency (section \ref{sec:heat_engine}).

\subsection{The boundaries of the climate system}
\label{sec:fluxes}

\add{Applying the second law to the climate system} requires a proper definition of the climate system's boundaries;
where does the Earth's climate system end and ``the surroundings'' begin? \add{In most applications, the climate system is defined to include the atmosphere, oceans, and the uppermost few meters of the land surface. While this definition excludes the solid Earth, the smallness of the geothermal heat flux indicates that irreversible processes in Earth's interior are likely to be weak compared to those in the atmosphere and ocean. The irreversible entropy production in the climate system is therefore approximately equal to that of the entire Earth system. }


\add{Defining the boundaries of the climate system also requires consideration of the role of radiation within it. Like matter, radiation obeys the second law of thermodynamics, and the interaction between radiation and matter may be shown to be an irreversible source of entropy \citep[e.g.,][]{Callies1988,Callies1984}. The extent to which the irreversibility of radiative processes is included in \eqref{eq:second_law} depends on the extent to which radiation is included as part of the climate system or excluded as part of the surroundings.}

\begin{figure}
\centering
\includegraphics[width=8.6cm]{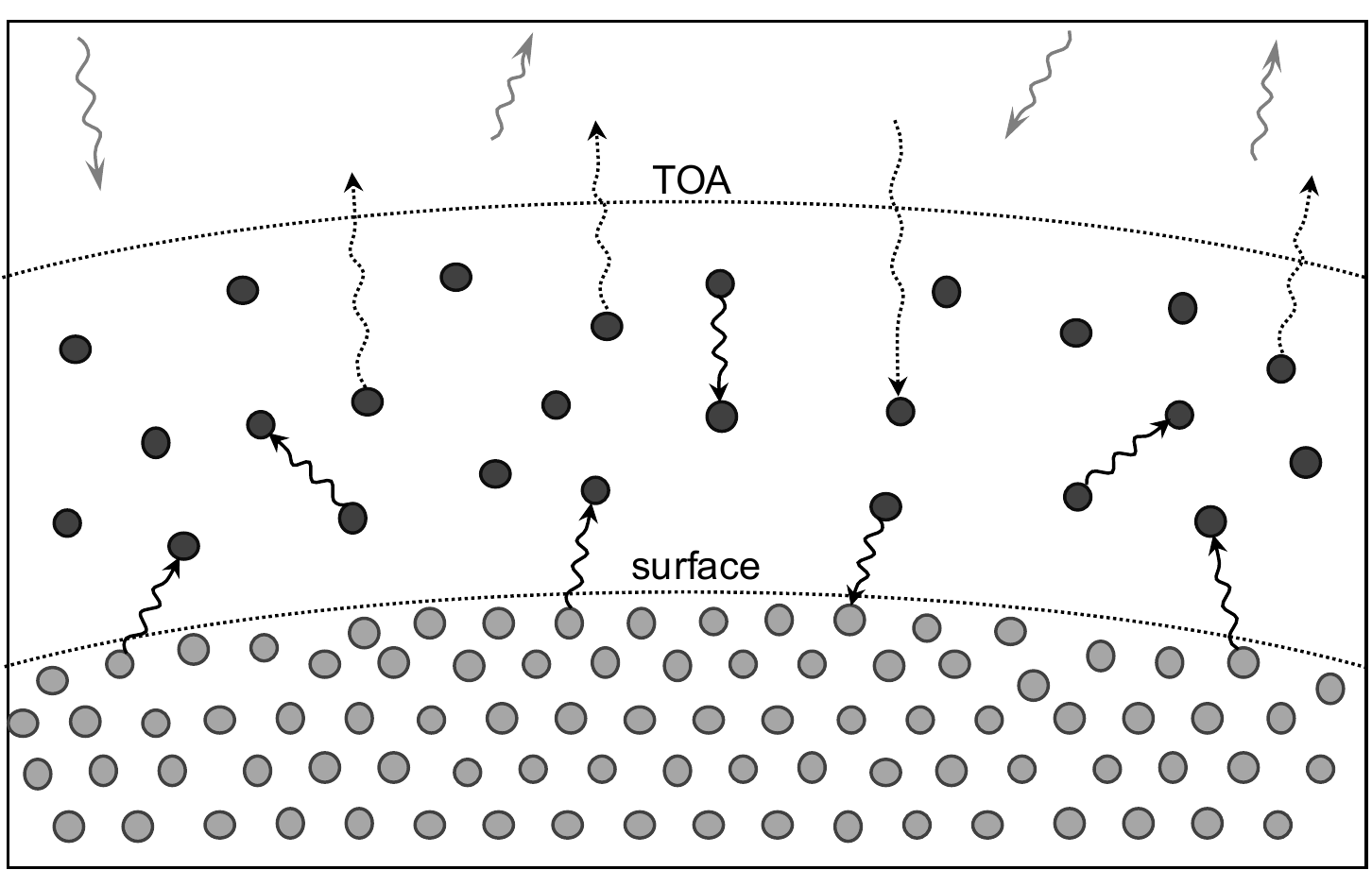}
\caption{Schematic of different definitions of the climate system and its boundary in the application of the second law of thermodynamics. The planetary system is defined as a control volume including all particles (matter and radiation) encompassed by a fictitious surface denoted ``top of the atmosphere'' (TOA). The material system includes only matter (circles) and excludes all radiation (arrows) as part of the surroundings. The transfer system \citep{Gibbins2020} includes internal radiative transfer (photons that are emitted and absorbed by matter within the system; solid black arrows), but excludes external radiative transfer (photons that are incident from the sun or emitted directly to outer space; dashed black arrows). 
}
		\label{fig:system_boundary}
\end{figure}

\citet{Bannon2015} summarizes a number of possible definitions of the climate system, but here we limit our discussion to three common definitions used in studies of the Earth's entropy budget (Fig. \ref{fig:system_boundary}): 
\begin{enumerate}
    \item \textit{The planetary climate system:} the Earth and its atmosphere is treated as a control volume, and the climate system is defined as all substances, both matter and radiation, within this volume \citep[e.g.,][]{Bannon2015}. \add{This is the most expansive definition, and it leads to the largest value of the irreversible entropy production $\dot{S}_i$}.
    \item \textit{The material climate system:} the climate system is defined to include only matter within the Earth and atmosphere, and all photons are considered part of the surroundings \citep[e.g.,][]{Goody2000}.
    \item \textit{The transfer climate system:} discussed in \citet{Bannon2015}, and recently advocated for by \citet{Gibbins2020}, the transfer climate system is defined to include matter plus internal radiative  transfer (photons  that  are  emitted  and  absorbed  by  matter  within  the  system) but to exclude external radiative transfer (photons that are incident from the sun or emitted directly to space). \add{Unlike the planetary climate system, the transfer climate system cannot be defined using a control volume approach because it excludes some photons present within the atmosphere (dashed black arrows on Fig. \ref{fig:system_boundary}).}
\end{enumerate}

Although each perspective provides a consistent description of the climate system, the magnitude of the entropy \add{export $-\dot{S}_e$} and the irreversible entropy production $\dot{S}_i$ differ greatly between the planetary, material, and transfer definitions, and previous authors have disagreed on which perspective is most relevant for \add{understanding climate system behavior}  \citep[e.g.,][]{Essex1984,Essex1987,Goody2000}.
\add{Below we briefly outline the calculation of the entropy \add{export $-\dot{S}_e$} according to each perspective.} Following \citet{Goody2000} and a number of other authors \citep[e.g.][]{Ozawa2003,Lucarini2010,Pascaleetal2011}, we will argue that the material climate system is most relevant to understanding the dynamics of the atmosphere and ocean, and the material entropy budget will be the focus of much of the later sections of this review.

\subsubsection{The planetary climate system}\label{sec:diffsystems}
    
The planetary climate system consists of a control volume bounded by a fictitious surface beyond the atmosphere which we will refer to as the ``top of the atmosphere'' (TOA; Fig. \ref{fig:system_boundary})\footnote{\add{Since the density of the atmosphere decreases exponentially with height, there is no precise dividing line between the atmosphere and space. Conceptually, it is useful to consider the TOA to be at roughly  80 km above sea level, where the gas density becomes so low that the approximation of local thermodynamic equilibrium breaks down (see section \ref{sec:single}).}}. \add{Within this control volume exists all matter within the climate system (and indeed the entire Earth system), as well as photons emitted by the sun (shortwave radiation) and those emitted by the Earth and atmosphere (longwave radiation)\footnote{\add{The shortwave/longwave nomenclature is motivated by the fact that the spectra of solar and terrestrial radiation have practically no overlap.
}}. A full account of the second law applied to the planetary climate system must consider the entropy embodied in both matter and radiation; the irreversible entropy production by the planetary climate system may then be divided into a component $\dot{S}_i^\text{mat}$ associated with material processes, and a component $\dot{S}_i^\text{rad}$ owing to the interaction of matter with radiation.}

\add{The radiative component $\dot{S}_i^\text{rad}$ is a result of the irreversibility of absorption, emission, and scattering processes that occur within the climate system. In particular, the transformation of a focused beam of shortwave radiation, with an effective emission temperature of $\sim$ 6000 K, into diffuse emission of longwave radiation, with an effective emission temperature of $\sim 250$ K, is highly irreversible. \citet{Callies1988} provide a derivation of the equations governing the entropy of the radiation field, showing how $\dot{S}_i^\text{rad}$ may be expressed in the classic form of the product of a generalized thermodynamic flux and a generalized thermodynamic force. The authors further demonstrate the irreversibility of radiative interactions by showing that $\dot{S}_i^\text{rad}$ is positive definite for the separate cases of absorption/emission and scattering. Here we do not provide a detailed account of the various irreversible radiative processes in the climate system \citep[see e.g.,][for more detailed treatments]{Li1994,Goody1996,Wu2010,Pelkowski1994,Pelkowski2012}. Instead, we characterize the planetary entropy budget through the \add{time-mean} net export of entropy out of the climate system $-\tavg{\dot{S}_e}$, given by the net flux of entropy across its boundary.}

\add{For the planetary climate system, the relevant boundary is the TOA, and the relevant fluxes are those carried by shortwave and longwave radiation.} 
Defining $\Omega$ as the volume of the climate system, we may write the net \add{export} of entropy \add{out of} the system as,
    \begin{linenomath}\begin{equation*}
        -\tavg{\dot{S}_e} = \frac{1}{A}\int_{\partial \Omega} \tavgb{J_{SW} + J_{LW} } \, \mathrm{d}A, 
    \end{equation*}\end{linenomath}
where $\partial \Omega$ represents the boundary of $\Omega$, in this case the TOA, \add{the angle brackets represent a time mean,} $A$ is the surface area  of Earth, $J$ is the radiant flux of entropy out of the climate system, and subscripts $SW$ and $LW$ refer to shortwave and longwave radiation, respectively. \add{Here we follow the convention in atmospheric science to refer to a flux as the transport of a quantity per unit area (also known as flux density), and we divide the integral on the right-hand side by $A$ to express $\tavg{\dot{S}_e}$ per unit area of the Earth's surface.}
For a system approximately in steady state, we must also have a \add{time-mean} balance between the shortwave and longwave radiant energy fluxes $F$,
    \begin{linenomath}\begin{equation*}
        \frac{1}{A}\int_{\partial \Omega} \tavgb{ F_{SW} + F_{LW} } \, \mathrm{d}A = 0. 
    \end{equation*}\end{linenomath}

Previous authors have estimated the entropy fluxes $J_{LW}$ and $J_{SW}$ from both observations \citep{StephensOBrien1993,KatoRose2020} and climate models \citep{Li1994,Pascaleetal2011}. Before we discuss these estimates, however, it is useful to consider the planetary entropy budget for a simplified model of the climate system in order to build some intuition of the magnitude and behavior of various components of $\tavg{\dot{S}_e}$.

\begin{figure*}
\centering
\includegraphics[width=14cm]{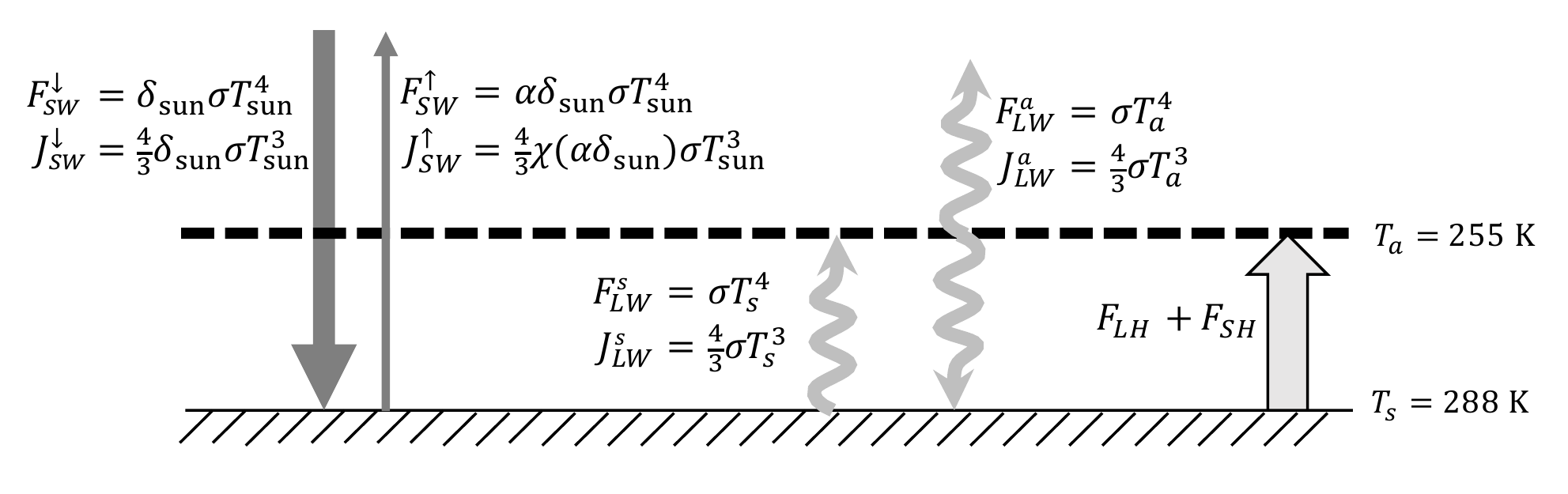}
\caption{A simple model for Earth's climate and the associated vertical energy and entropy fluxes. The climate system is assumed to be horizontally homogeneous, and it includes a surface, with temperature $T_s$, and a single-layer atmosphere, with  temperature $T_a$. 
Arrows show energy fluxes $F$ and their corresponding entropy fluxes $J$ expressed per unit area. The subscripts $LW$ and $SW$ denote shortwave and longwave radiant fluxes,  respectively, and the subscripts $s$ and $a$ refer to the surface and atmosphere, respectively. Shortwave radiation is divided into its upward and downward components. The turbulent flux of enthalpy from the surface to the atmosphere is made up of a sensible heat flux $F_{SH}$ and a latent heat flux $F_{LH}$. Other terms in the equations are described in the text.}
		\label{fig:simple_budget}
\end{figure*}

The simple model is described schematically in Fig. \ref{fig:simple_budget}; it is similar to models presented in \citet{Bannon2015} and \citet{KatoRose2020}, and our discussion of the different entropy production rates follows that of \citet{Gibbins2020}. The model is horizontally homogeneous, representing globally-averaged conditions, and it consists of a surface and a single-layer atmosphere. Both the surface and atmosphere are assumed to be completely opaque to longwave radiation and to behave as blackbodies for radiation in the longwave portion of the electromagnetic spectrum.
The atmosphere is assumed to be transparent to shortwave radiation, and the surface has a fixed shortwave albedo of $\alpha$, reflecting a fraction $\alpha$ of the incoming solar radiation to space and absorbing the rest. Energy and entropy transports in this model occur via radiative fluxes between the surface, atmosphere, and space, and via turbulent fluxes of latent\footnote{Latent heat refers to the energy embodied in water vapor that is released \add{upon} condensation.} and sensible heat between the surface and the atmosphere.

Assuming steady state, we may write energy balance equations for the TOA, atmosphere, and surface, respectively, given by \citet{KatoRose2020}:
\begin{subequations}
\label{eq:one_layer}
\begin{linenomath}\begin{align}
    (1-\alpha)F^\downarrow_{SW} &= F_{LW}^a, \\
    F_{LW}^s +F_{LH}+F_{SH} &= 2F_{LW}^a, \label{eq:one_layer_atm}\\
    (1-\alpha)F^\downarrow_{SW} \add{+ F_{LW}^a} &= F_{LW}^s +F_{LH}+F_{SH}, 
\end{align}\end{linenomath}
\end{subequations}
where \add{$F_{SW}^\downarrow$} is the downward solar energy flux input to the Earth, \add{$F^s_{LW}$} and \add{$F^a_{LW}$} are the longwave energy fluxes from the surface and atmosphere, respectively, and $F_{LH}$ and $F_{SH}$ are the latent and sensible heat fluxes from the surface to the atmosphere, respectively (Fig. \ref{fig:simple_budget}). \add{Since the model is time-invariant, we omit the time-averaging operator in equations \eqref{eq:one_layer}, but the fluxes should be interpreted as time means.}

The longwave energy fluxes from the surface and atmosphere are given by the well-known Stefan-Boltzmann law
\begin{equation}
    F = \sigma T^4,
    \label{eq:stefanboltzmann}
\end{equation}
where $F$ is the energy flux, $T$ is the temperature of the emitting body and $\sigma$ is the Stefan-Boltzmann constant. Approximating the sun as a blackbody, the downward solar energy flux at the TOA is given by,
    \begin{linenomath}\begin{equation*}
         F^\downarrow_{SW} = \delta_\text{sun}\sigma T_\text{sun}^4,
    \end{equation*}\end{linenomath}
where $\delta_\text{sun}= \frac{\Omega_0\cos\phi}{\pi}$, with $\Omega_0 = 6.77\times10^{-5}$ being the solid angle subtended by the sun's disk \citep{StephensOBrien1993}, and $\phi$ the zenith angle of the sun's rays. Here we take a global mean value of $\cos\phi=0.25$. 

To apply the one-layer model to the Earth, we set the solar temperature to be $T_\text{sun} = 5777$ K \citep{Peixotoetal1991}, and we set $\alpha = 0.3$ to roughly match the planetary albedo of Earth \citep{Stephens2015}. \add{Using the energy balance equations \eqref{eq:one_layer}, this constrains the atmospheric temperature, which for this model is equal to the effective emission temperature of the planet, to be $T_a = 255$ K.} We are then free to set either the sensible and latent heat fluxes from the surface or the surface temperature $T_s$. \add{On the basis that turbulent dynamics place a strong constraint on the lapse rate in convecting atmospheres \citep{Emanuel1994},} we fix the temperature difference between the surface and atmosphere by setting the surface temperature to roughly match Earth's global-mean surface temperature $T_s = 288$ K,
\add{and we allow the surface fluxes to adjust to satisfy the energy balance equations.}


We now evaluate the planetary entropy budget for the one-layer model.
The entropy fluxes associated with radiation of a given wavelength and angular distribution may be derived from the fundamental statistical mechanics of a Boson gas \citep{Rosen1954}, or through a number of semi-classical methods \citep{Ore1955,Wu2010}. For a blackbody, a formula for the entropy flux $J$
may be derived by integrating the spectral entropy flux distribution over all frequencies to give \citep{Wu2010},
\begin{equation}
    J = \frac{4}{3}\sigma T^3.
    \label{eq:JL}
\end{equation}
Combining \eqref{eq:JL} and \eqref{eq:stefanboltzmann}, the blackbody entropy flux may be expressed in terms of the energy flux $F$ as,
\begin{equation}
    J = \frac{4}{3}\frac{F}{T}.
    \label{eq:blackbody_entropy}
\end{equation}
The entropy flux emitted by a blackbody is larger, by a factor of $4/3$, than the entropy loss of the emitting object $F/T$. This additional entropy transport may be interpreted as the irreversible entropy production associated with emitting radiation into a vacuum \citep{Feistel2011}\footnote{An elegant derivation of \eqref{eq:blackbody_entropy} is presented by \citet{Feistel2011}. Consider two parallel plates held at fixed temperatures exchanging energy through radiation. \add{Assuming they are blackbodies,} the energy flux from each plate may be described by the Stefan-Boltzmann law \eqref{eq:stefanboltzmann}, and it will produce a transfer of heat from the hotter plate to the colder plate. By the second law of thermodynamics, this energy exchange must be associated with positive irreversible entropy production, with the rate of irreversible entropy production tending to zero as the temperature difference between the plates approaches zero. \citet{Feistel2011} showed that the only functional form for the entropy flux associated with the radiation from each plate consistent with this expectation is that described by \eqref{eq:blackbody_entropy}.}. Eq. \eqref{eq:JL} may be used to evaluate the entropy fluxes from the surface and atmosphere in the one-layer model.

Due to the irreversibility associated with reflection, the upward and downward shortwave entropy fluxes must be treated separately. The downward shortwave flux of entropy is given simply by,
\begin{equation}
     J_{SW}^{\downarrow} = \frac{4}{3} \delta_\text{sun} \sigma T_\text{sun}^3, 
     \label{eq:JSdn}
\end{equation}
representing the sun's blackbody entropy flux reduced by the factor $\delta_\text{sun}$. The same approximation cannot be used for the upward flux because of the change of its angular distribution upon reflection.
\add{Instead, we 
assume that the reflection is diffuse (Lambertian), so that the radiance of the reflected beam is independent of direction.
\citet{StephensOBrien1993} solved for this case, finding that the entropy flux of the reflected radiation could be approximated as,
\begin{equation}
   J^{\uparrow}_{SW}= \frac{4}{3}\chi(\alpha \delta_\text{sun})\sigma T_\text{sun}^3,
   \label{eq:JSup}
\end{equation}
where $\chi (u) = u[a_0 +  a_1\ln(u)]$, and
 $a_0 = 0.9652$ and $a_1 = -0.2777$ are empirical constants. }
\citet{Wu2010} provide a detailed evaluation of this approximation and a number of other analytic formulae for entropy fluxes associated with non-blackbody radiation.

\begin{table*}
\renewcommand{\arraystretch}{1.2}
\begin{center}
\begin{tabular}{ l l c c c c c } \hline\hline \vspace{3pt}
 & Model & $\tavg{J^{\text{TOA}}_{SW}}$ & $\tavg{J^{\text{TOA}}_{LW}}$ & $-\tavg{\dot{S}_e}$ & $-\tavg{\dot{S}_e^{\text{trans}}}$ & $-\tavg{\dot{S}_e^\text{mat}}$ \\ \hline
 one-layer model (Fig. \ref{fig:simple_budget}) & & 31 & 1247 &  1279 & 109 & 39 \\ \citet{KatoRose2020} & obs. & -55 & 1238 & 1183 & 76 & 49 \\ 
 \citet{StephensOBrien1993} & obs. & 20  & 1230 & 1250 & &  \\
 \citet{Peixotoetal1991} & obs. &  -41$^\text{a}$ & 925$^\text{a}$ & 884$^\text{a}$ & & 41\\
 \citet{Lembo2019} & CMIP6$^\text{b}$ &  & & & & 58 \\
 \citet{Pascaleetal2011} & HadCM3 &  & & 911$^\text{a}$ & 101 & 52 \\
 \citet{FraedrichLunkeit2008} & PlanetSim &  & & 880$^\text{a}$ & 69 & 35 \\
 \citet{Goody2000} & GISS &  & & & & 52 \\ 
 \hline 
\end{tabular}
\end{center}
\caption{Observational (obs.) and model-based estimates of the top of atmosphere entropy fluxes of shortwave $\tavg{J_{SW}^\text{TOA}}$ and longwave $\tavg{J^\text{TOA}_{LW}}$ radiation and the planetary $-\tavg{\dot{S}_e}$, transfer $-\tavg{\dot{S}^\text{trans}_e}$ and material $-\tavg{\dot{S}_e^\text{mat}}$ entropy export rates of the climate system (per unit area of Earth's surface) from various prior studies and for the one-layer model summarized in Fig. \ref{fig:simple_budget}.  Fluxes toward outer space are defined as positive, and units are mW m$^{-2}$ K$^{-1}$.\smallskip\\
\footnotesize $^\text{a}$Entropy fluxes estimated using \eqref{eq:Jwrong}.\smallskip\\
\footnotesize $^\text{b}$Mean over 7 models participating in the sixth phase of the Coupled Model Intercomparison Project (CMIP6) \citep{Eyring2016}.
} 
\label{table:planetary_budget}

\end{table*}

Eqs. \eqref{eq:JL}, \eqref{eq:JSdn} and \eqref{eq:JSup} may be used to evaluate the planetary entropy budget for the one-layer model (table \ref{table:planetary_budget}).  According to this model, the TOA entropy fluxes are dominated by longwave radiation; the longwave entropy flux at the TOA is factor of 40 larger than the net shortwave entropy flux. Furthermore, the net shortwave entropy flux at the TOA is directed upwards, despite the net shortwave energy flux being downwards. This counter-intuitive result is possible because the entropy associated with photons in the diffuse radiation reflected from the Earth's surface is much higher than the entropy of photons in the beam of radiation incident on the Earth \citep{StephensOBrien1993}. Summing the net shortwave and longwave fluxes and using the steady-state entropy budget \eqref{eq:entropy_balance}, the one-layer model gives an estimate of the total irreversible entropy production by the climate system $\tavg{\dot{S}_i} = 1279$ mW m$^{-2}$ K$^{-1}$. 

Despite the simplicity of the one-layer model, its entropy budget is
similar to more detailed
estimates of Earth's planetary entropy budget based on observations (table \ref{table:planetary_budget}). For example, \citet{StephensOBrien1993} used satellite observations of TOA radiation to estimate the planetary entropy budget, finding a similar dominance of the longwave fluxes and a value of the total entropy \add{export $-\tavg{\dot{S}_e}$} roughly 2\% smaller in magnitude than that of the one-layer model. This difference is partially accounted for by the one-layer model's neglect of temperature variations within the atmosphere. \citet{Lesins1990} showed that the outgoing longwave entropy flux is maximized for an isothermal atmosphere, with meridional and vertical temperature variations reducing the flux by a factor of the order of 1\%.



\citet{KatoRose2020} also provided estimates of the entropy flux at the TOA based on satellite observations. They applied the simple blackbody formula \eqref{eq:blackbody_entropy} to both incoming and outgoing shortwave radiation. This neglects the irreversible entropy production associated with diffuse reflection, resulting in an underestimate of the entropy flux by reflected solar radiation. But the overall planetary entropy budget is nevertheless broadly similar to that of the one-layer atmosphere model. 

On the other hand, the observational study of \citet{Peixotoetal1991} and a number of studies based on global climate models \citep{FraedrichLunkeit2008,Pascaleetal2011} present planetary entropy budgets that are inconsistent with the one-layer model, with estimates of the entropy \add{export $-\tavg{\dot{S}_e}$}  30-40\% smaller in magnitude. The reason for this discrepancy is that these studies evaluate the flux of entropy by radiation as
\begin{equation}
    J' = \frac{F}{T},
    \label{eq:Jwrong}
\end{equation}
where $F$ is the radiative energy flux and $T$ is the temperature of the emitting object. This definition is appealing, because it gives the radiant entropy flux as being equal to the loss of entropy by the emitting object, but it fails to account for the irreversible nature of spontaneous emission and absorption represented by the $4/3$ factor in \eqref{eq:blackbody_entropy} \citep{Feistel2011}. \add{For non-blackbody radiation, \eqref{eq:Jwrong} also neglects differences between the temperature of the emitting object and the spectrally-varying emission temperature of the emitted radiation. As a result, the use of \eqref{eq:Jwrong} can imply irreversible entropy production rates owing to radiative processes that are locally negative, violating the second law \citep{Callies1988}.}

Studies such as \citet{Peixotoetal1991} therefore do not fully account for the irreversible entropy production associated with radiative processes and they underestimate the planetary entropy production rate \citep{Essex1984,Essex1987,StephensOBrien1993}.
Such studies nevertheless remain relevant, because, 
as will be shown below, entropy production associated with radiative transfer does not affect the material entropy budget, and conclusions regarding material entropy production are unaffected \add{by} whether one uses \eqref{eq:Jwrong} or \eqref{eq:blackbody_entropy} to estimate radiant entropy fluxes.



\subsubsection{The material climate system}
\label{sec:material}

To motivate study of the material entropy budget, consider a thought experiment in which the heating and cooling owing to radiative absorption and emission within the climate system is replaced by identical heating and cooling rates produced by a reversible mechanism. Such a change would have no effect on 
 the matter within the climate system;
 the equations governing the fluid dynamics of the atmosphere and oceans are only concerned with radiation insofar as it heats or cools the fluid \citep{Goody2000,Pascaleetal2011}. The atmospheric circulation, hydrological cycle, and ocean currents would behave exactly as before. For this reason, \citet{Goody2000} 
 advocated
 for a view of the entropy budget that focuses exclusively on matter rather than radiation. 




The material climate system includes all matter within the Earth, atmosphere, and oceans but considers radiation as part of the surroundings \citep{Goody2000,Bannon2015}. The import of entropy into the material system $\dot{S}_e^\text{mat}$ occurs through
the heating and cooling of matter that absorbs and emits radiation, and 
it may be written \citep{Goody2000},
\begin{equation}
    \dot{S}^\text{mat}_e = \frac{1}{A}\int_\Omega \frac{\rho\dot{q}_\text{rad}}{T} \, dV,
    \label{eq:entropy_export_material}
\end{equation}
where $\dot{q}_\text{rad}$ is the net radiative heating rate per unit mass, $\rho$ is the density, and the integral is over $\Omega$, representing the entire climate system. \add{Once again, we divide the integral on the right-hand side by the surface area of the Earth $A$ in order to express \add{$\dot{S}_e^\text{mat}$} per unit area.}
We may also define the irreversible material entropy production $\dot{S}_i^\text{mat}$, which by the steady-state material entropy budget satisfies $\tavg{\dot{S}_i^\text{mat}} = -\tavg{\dot{S}_e^\text{mat}}$. 
The steady-state material entropy budget may then be written,
\begin{equation}
            \tavg{\dot{S}^\text{mat}_i} = - \frac{1}{A}\int_\Omega \tavgb{\frac{\rho\dot{q}_\text{rad}}{T} }\,  dV. \label{eq:entropy_material}
        \end{equation}
\add{The requirement that $\dot{S}_i^\text{mat}$ be positive is an application of the second law known as the Clausius-Dunhem inequality \citep{Pelkowski2014}. As we shall show below, for the simple one-layer model introduced in the previous subsection, $\dot{S}_i^\text{mat}\ge0$ if turbulent fluxes transport heat down the temperature gradient.}

Since the material entropy budget incorporates the effect of radiation as an external (and reversible) heating, the material irreversible entropy production rate $\dot{S}_i^\text{mat}$ accounts for only a portion of the total irreversible entropy production of the climate system $\dot{S}_i$, while the remainder $\dot{S}_i^\text{rad}$ is associated with irreversible entropy production owing to radiation.
As we will show in detail in section \ref{sec:irr_procs}, the material entropy production includes irreversible processes such as frictional dissipation, molecular heat diffusion, irreversible mixing, and irreversible chemical reactions.

\add{Consider again the
simple one-layer model illustrated in Fig. \ref{fig:simple_budget}.} 
\add{We may define the \add{time-mean} net radiative heating \add{rate} of the atmosphere per unit area
\begin{linenomath}\begin{equation*}
    \tavg{\dot{Q}_\text{rad}} = \frac{1}{A}\int_{\Omega_A} \tavgb{\rho\dot{q}_\text{rad}} \, dV,
\end{equation*}\end{linenomath}
where $\Omega_A$ is the volume of the atmosphere. The heating rate} $\tavg{\dot{Q}_\text{rad}}$ may be evaluated based on the radiative energy fluxes shown in Fig. \ref{fig:simple_budget},
\begin{equation}
    \tavg{\dot{Q}_\text{rad}} = \sigma T_s^4 - 2\sigma T_a^4.
    \label{eq:Qrad}
\end{equation}
\add{Further, using the energy balance equations for the one-layer model \eqref{eq:one_layer}, one may show that the net radiative heating of the surface is given by $-\tavg{\dot{Q}_\text{rad}}$, and that this is equal to the transport of energy from the surface to the atmosphere via turbulent fluxes,
\begin{linenomath}\begin{equation*}
    -\tavg{\dot{Q}_\text{rad}} = \tavg{F_{LH}+F_{SH}}.
\end{equation*}\end{linenomath}
Using the previous two equations and noting that the temperatures of the atmosphere and surface are assumed to be constant,}
the steady-state material entropy budget of the one-layer model may be expressed,
\begin{linenomath}\begin{equation*}
    \tavg{\dot{S}^\text{mat}_i} = \tavgb{F_{LH}+F_{SH}} \left(\frac{1}{T_a} - \frac{1}{T_s}\right).
\end{equation*}\end{linenomath}
\add{This equation demonstrates that, in steady state, the irreversible entropy production $\tavg{\dot{S}_i^\text{mat}}$ is positive provided the surface fluxes move energy from high to low temperature. In Earth-like climates, $T_s > T_a$, and the surface \addr{turbulent} fluxes transport energy from the surface to the atmosphere, so that $\tavg{\dot{S}^\text{mat}_i}$ is positive as required by the second law. 
It is sometimes suggested that the greenhouse effect is incompatible with the second law of thermodynamics. The one-layer model shows this is not true; the greenhouse effect associated with the absorption of longwave radiation allows the model to maintain a surface temperature higher than the effective emission temperature of the planet while it maintains positive irreversible entropy production.}

For the parameters chosen, the one-layer model gives a material entropy production rate of $\tavg{\dot{S}_i^\text{mat}} = 39$ mW m$^{-2}$ K$^{-1}$. Given the assumptions of the model, \add{this provides only a rough} estimate of the material entropy production in the climate system, but it highlights the small magnitude of $\tavg{\dot{S}_i^\text{mat}}$ compared to the total irreversible production rate  $\tavg{\dot{S}_i}$, implying that the bulk of the irreversible entropy production in the climate system occurs due to radiative processes 
\citep{Essex1984,Essex1987,Li1994,StephensOBrien1993,Goody2000}.


More detailed estimates of the material entropy budget of Earth's climate system confirm the picture above (table \ref{table:planetary_budget}). Studies based on observations \citep{Peixotoetal1991,KatoRose2020} as well as global climate model simulations \citep{Goody2000,FraedrichLunkeit2008,Pascaleetal2011,Lembo2019} have estimated the material entropy export $-\tavg{\dot{S}_e^\text{mat}}$, finding values in the relatively broad range of 35-60 mW m$^{-2}$ K$^{-1}$. The large range in such estimates is partly due to methodological differences across studies.  For example, \citet{Peixotoetal1991} considered only global- and annual-mean radiative fluxes and temperature profiles in order to estimate $\tavg{\dot{S}_e^\text{mat}}$, while 
\citet{KatoRose2020} also took into account spatial and temporal variations.
But differences also arise because the spatial distribution of radiative heating $\dot{q}_\text{rad}$ is strongly dependent on properties such as the surface albedo and the distribution of clouds and water vapor in the atmosphere, and it is therefore difficult to estimate accurately from observations and dependent on uncertain parameterizations in models. As a result, even among studies applying similar methodologies to climate model output \citep{FraedrichLunkeit2008,Pascaleetal2011}, the estimated value of $\tavg{\dot{S}_e^\text{mat}}$ can vary substantially. 



Under steady-state conditions, estimation of the \add{net} entropy \add{export} leads directly to an estimate of the material entropy production of the climate system. However, differences between estimates of $-\tavg{\dot{S}_e^\text{mat}}$ and $\tavg{\dot{S}_i^\text{mat}}$ of up to 30\% have been reported in the literature \citep{Lembo2019}. In principle, such differences may result from imbalances in the entropy budget due to climate variation \citep{Goody2000}, but estimates of the imbalance in the planetary energy \add{budget} \citep{Trenberth2014} suggest that such differences are likely to be on the order of a few mW m$^{-2}$ K$^{-1}$. Rather, differences between estimates of the entropy \add{export} and estimates of material entropy production reflect the difficulty of diagnosing irreversible processes using available observations or using standard model outputs. We further discuss the issues surrounding the estimation of irreversible processes in the climate system in section \ref{sec:global_ent_bud} and in climate models in section \ref{sec:models}.


\subsubsection{The transfer climate system}
\label{sec:trans}
While the majority of studies of Earth's entropy budget adopt one of the two definitions discussed above, \citet{Gibbins2020} have recently advocated for an alternate definition that is intermediate between the planetary and material perspectives, which they refer to as the transfer climate system \citep[also discussed in][as their case ``MS2"]{Bannon2015}. The transfer climate system is similar to the material climate system, but it additionally includes radiation that is ``internal'' to the climate system (Fig. \ref{fig:system_boundary}). Internal radiation corresponds to photons that transport energy between different material elements of the climate system, in contrast to those photons that are incident on the Earth from the sun or that are emitted to outer space. 
Since the transfer approach includes some, but not all, radiation as part of the climate system, it gives an irreversible entropy production rate $\dot{S}_i^\text{trans}$ whose magnitude is between the planetary and material values.

\citet{Gibbins2020} argue that the transfer climate system provides an entropy budget that is more robust to details of internal heat transport mechanisms within the climate system. In particular, they note that, from the perspective of the transfer entropy budget, heat transport from high to low temperature is associated with the same amount of irreversible entropy production whether it is caused by radiative fluxes or conductive fluxes. But only in the latter case would this entropy production be included in the material entropy budget.

For the transfer climate system, the entropy import rate $\dot{S}^\text{trans}_e$ is equal to the sum of the entropy tendencies associated with the absorption of solar radiation and the emission of longwave radiation directly to space. According to the one-layer model, 
the atmosphere emits an amount of radiation 
$\sigma T_a^4$ directly to space,
while the surface absorbs an equal amount of radiation from the sun.
Assuming steady-state conditions, we may therefore write the time-mean irreversible entropy production rate $\tavg{\dot{S}_i^\text{trans}}$ for the one-layer model as
    \begin{linenomath}\begin{equation*}
        \tavg{\dot{S}^\text{trans}_i} = \sigma T_a^4 \left(\frac{1}{T_a} - \frac{1}{T_s}\right).
    \end{equation*}\end{linenomath}
This may be evaluated with the parameters of the model to give $\tavg{\dot{S}^\text{trans}_i} = 109$ mW m$^{-2}$ K$^{-1}$.

A disadvantage of using the transfer approach is that the transfer entropy import rate $\dot{S}_e^\text{trans}$ depends on the origin and destination of each photon that enters the climate system, rather than just the net radiative heating rate or the TOA fluxes, making its estimation from observations and models more involved.
Nevertheless, \citet{KatoRose2020} have recently estimated the transfer entropy budget from observations, and some of its characteristics may be deduced from the results of previous studies using global climate models \citep{FraedrichLunkeit2008,Pascaleetal2011}. As expected, the magnitude of the time-mean entropy export $-\tavg{\dot{S}_e^\text{trans}}$ 
is between the corresponding values for the planetary and material entropy budgets (table \ref{table:planetary_budget}). Compared to the observational and climate-model based estimates, the simple one-layer model overestimates the transfer entropy production rate. This is likely because of its neglect of solar absorption in the atmosphere, which leads to an artificially high value of the solar absorption temperature \citep[cf.][]{Gibbins2020}. 

But even detailed model-based estimates of the transfer entropy production rate differ from each other considerably. 
This highlights our limited knowledge of the transfer entropy production rate, which has only recently been explicitly defined in the context of the climate system's entropy budget \citep{Bannon2015,Gibbins2020}. 
Better quantification of the transfer entropy budget and further understanding of its relationship to the thermodynamics of the climate system present promising avenues for future research.

\subsection{The climate system as a heat engine}
\label{sec:heat_engine}

The climate system is often described as a heat engine, transporting energy from the warm tropical surface to the cold polar troposphere and producing \add{kinetic energy, in the form of} atmospheric and oceanic circulations, in the process 
\citep[e.g.,][]{Brunt1926,Lorenz1967,Barryetal2002,Pauluis2011,Bannon2015,Laliberteetal2015}. But there are some important differences between the climate system and the classic engineering account of a heat engine. \add{For example, the circulations produced by the climate engine act on, and are dissipated within, the system itself \citep{Johnson2000}. 
This creates an important negative feedback loop, in which the circulations themselves act to reduce the temperature differences that are responsible for their existence \citep[e.g.,][]{Barryetal2002}.}
In the following, we clarify how concepts such as the work output and the thermodynamic efficiency of a heat engine may be \add{meaningfully} applied to the climate system.

\subsubsection{Heat engines and irreversibility}

Consider a heat engine operating between two thermal reservoirs at different temperatures. The engine ingests heat at a rate $\dot{Q}_\text{in}$ from the warm reservoir at a temperature $T_\text{in}$, transporting it to the cool reservoir at temperature $T_\text{out}$ where it is expelled at a rate $\dot{Q}_\text{out}$. In the process, the engine is able to perform work at the rate $\dot{W}_\text{ext}$. Here we include the subscript ext to emphasize that this work is done on an external body. For instance, the engine may be used to drive a piston that accelerates a locomotive. The eventual dissipation of the locomotive's kinetic energy occurs outside of the engine.

The action of a heat engine may be described by combining the first and second laws of thermodynamics under steady-state conditions
to form 
the \add{Gouy-Stodola} theorem \citep[e.g.,][]{Bannon2015},
    \begin{equation}
        \eta_{C} \dot{Q}_\text{in} = \dot{W}_\text{ext}  + \dot{S}^\text{mat}_i T_\text{out},
        \label{eq:GouyStoudola}
    \end{equation}
where $\dot{S}^\text{mat}_i$ represents the irreversible entropy production rate of the engine and
\begin{linenomath}\begin{equation*}
    \eta_C = \frac{T_\text{in}-T_\text{out}}{T_\text{in}}.
\end{equation*}\end{linenomath}
If the engine is perfectly reversible, it produces work with an efficiency given by $\eta_C$, equal to the Carnot efficiency. Irreversible processes decrease the work output relative to this theoretical maximum. In the engineering context, irreversible entropy production \add{results in} ``lost work'', and the aim of the engineer is to reduce $\dot{S}^\text{mat}_i$ as much as possible.
    
The climate system does not have a warm or cold reservoir, and the heating rates $\dot{Q}_\text{in}$ and $\dot{Q}_\text{out}$ and the associated temperatures are more difficult to define. Moreover, the climate system as a whole cannot perform work on any external body. For the climate system, we therefore have that $\dot{W}_\text{ext} = 0$ and any traditionally defined efficiency is also zero. Nevertheless, previous authors have defined an efficiency of the climate system in various ways. \add{In the following, we detail two such definitions. We first construct an equivalent Carnot efficiency of the climate system that relates its irreversible entropy production to the heat input through radiation. We then define a mechanical efficiency of the climate system that relates the work performed in generating the atmospheric and oceanic circulation to the heat input by radiation. This second definition may be considered to be the analogue to the engineering concept of the efficiency of a heat engine applied to the climate system.}

\subsubsection{Carnot efficiency of the climate system}
\label{sec:carnot}

\add{To define the Carnot efficiency of the climate system, we take \eqref{eq:GouyStoudola}, set $\dot{W}_\text{ext}=0$, and replace $\dot{S}_i^\text{mat}$ and $\dot{Q}_\text{in}$ with their time-averaged values to give,
    \begin{linenomath}\begin{equation*}
        \eta_C = \frac{\tavg{\dot{S}^\text{mat}_{i}}T_\text{out}}{\tavg{\dot{Q}_\text{in}}}. \hspace{20pt} (\dot{W}_\text{ext} = 0)
    \end{equation*}\end{linenomath}
The numerator $\tavg{\dot{S}^\text{mat}_i} T_\text{out}$ gives a measure of the strength of irreversible processes within the climate system, which was taken by \citet{Bannon2015} to be a measure of the activity of the atmosphere and oceans.}
The task is then to determine the effective input and output temperatures and heating rate $\tavg{\dot{Q}_\text{in}}$. Since there is no warm or cold reservoir, the heating rates must be defined in an averaged sense. In particular, we may take the energy input as the sum over all regions that experience net \add{radiative} heating \citep[e.g.,][]{Bannon2015},
    \begin{equation}
        \tavg{\dot{Q}_\text{in}} = \frac{1}{A}\int_\Omega \tavgb{\rho\dot{q}_\text{rad}^+} \, dV, \label{eq:heating_material}
    \end{equation}
    where $\dot{q}_\text{rad}^+$ is the net radiative heating rate when it is positive and zero otherwise, \add{and we have divided the integral on the right-hand side by $A$, the surface area of the Earth, to express the heating rate per unit area. We may similarly} define the effective temperature of heat input by,
    \begin{linenomath}\begin{equation}
        \frac{\tavg{\dot{Q}_\text{in}}}{T_\text{in}} = \frac{1}{A}\int_\Omega \tavgb{\frac{\rho\dot{q}_\text{rad}^+}{T}} \, dV.
        \label{eq:effectiveT}
    \end{equation}\end{linenomath}
    Making similar definitions of the heat output and output temperature based on the radiative cooling rate, \eqref{eq:GouyStoudola} may be written
    \begin{equation}
        \tavg{\dot{S}_i^\text{mat}} = \tavg{\dot{Q}_\text{in}}\left(\frac{1}{T_\text{out}} - \frac{1}{T_\text{in}}\right).
        \label{eq:heat_engine_2}
    \end{equation}
     Here we have assumed that $\dot{W}_\text{ext} = 0$, and hence that $\tavg{\dot{Q}_\text{in}}$ = $\tavg{\dot{Q}_\text{out}}$ for the climate system. Comparing with \eqref{eq:entropy_material}, the heat engine relation above may be seen to be  equivalent to the steady-state material entropy budget.
  
For the simple one-layer atmosphere model described by Fig. \ref{fig:simple_budget}, the heat input $\tavg{\dot{Q}_\text{in}}$ is given by $\tavg{\dot{Q}_\text{rad}}$ in \eqref{eq:Qrad}, and the input and output temperatures are those of the surface and the atmosphere, respectively. The model therefore has a Carnot efficiency of $\eta_C = 11\%$.
\citet{Bannon2015} found a slightly lower value of $\eta_C = 8\%$ using a similar single-layer model of the climate system which allows for atmospheric absorption of shortwave radiation, while \citet{Gibbins2020} found Carnot efficiencies in the range 6-12\% also using highly simplified models of the climate system. 
     
It is important to note that the heating rate $\tavg{\dot{Q}_\text{in}}$ is not an external parameter for a given planet (as it would be in traditional heat engine analysis), but it depends on features such as the surface albedo and the cloud and water vapor distribution, which may vary with climate. Furthermore, the definitions given above for the input and output heating rates are non-unique \citep{Bannon2015}. This non-uniqueness has some parallels in the ambiguity of defining the boundaries of the climate system in our discussion of planetary, material, and transfer entropy production rates above. For example, taking $\tavg{\dot{Q}_\text{out}}$ as cooling by longwave emission \add{directly} to space and $\tavg{Q_\text{in}}$ as solar absorption gives an efficiency based on the transfer climate system \citep{Gibbins2020}. This perspective was used in \citet{Bannon:2017aa} to derive approximate upper bounds to the Carnot efficiencies of Earth, Mars, Venus, and Titan (see also section \ref{sec:other_planets}).

\add{Finally, we note that some authors define a Carnot efficiency for the climate system based on a heating rate that includes additional terms such as the heating owing to frictional dissipation and that due to latent and sensible heat fluxes \citep[e.g.,][]{Johnson2000}. This approach was used by \citet{Lucarini2009} to relate the Carnot efficiency to the generation of kinetic energy by the climate system. Here, we instead define a separate mechanical efficiency which relates the generation of kinetic energy by the climate system to the radiative energy input $\tavg{\dot{Q}_\text{in}}$.}
    
\subsubsection{Mechanical efficiency of the climate system}    
\label{sec:mech} 

While the climate system cannot perform work on an external body, the atmosphere and ocean perform work on themselves and each other, and this work drives the winds and ocean currents. 
More specifically, work represents a conversion between kinetic energy and internal or potential energy within the climate system \citep[e.g.,][]{Lucarini2009,Lucarini2010}. This conversion may occur reversibly via motions of the ocean and atmosphere, or irreversibly via dissipative processes.
In the latter case, kinetic energy may be transformed into internal or potential energy\footnote{\add{Dissipation can irreversibly increase potential energy through the thermal expansion associated with frictional heating; see section \ref{sec:APE}.}}, but the reverse transformation is prohibited by the second law of thermodynamics. 

In Earth's climate system, 
kinetic energy dissipation occurs via two processes:
\begin{enumerate}
    \item Frictional dissipation of the winds and ocean currents that occurs as a result of the turbulent cascade of kinetic energy to scales small enough for viscosity to act.
    \item Frictional dissipation in the microscopic shear zones surrounding particles that have appreciable sedimentation velocities relative to the fluid (e.g., raindrops and snowflakes) \citep{Pauluis2000,Pauluis2002}. 
\end{enumerate}
At steady state, the time-mean rate at which the climate system performs reversible work $\tavg{\dot{W}_\text{rev}}$ must be equal to the total dissipation rate owing to these two processes. We may therefore write the
steady-state mechanical energy budget of the climate system as \citep{Pauluis2002,Romps2008},
\begin{equation}
    \tavg{\dot{W}_\text{rev}} = \tavg{\dot{D}_\text{fric}} + \tavg{\dot{D}_\text{sed}},
    \label{eq:work_moist}
\end{equation}
where 
$\tavg{\dot{D}_\text{fric}}$ is the time-mean rate of frictional dissipation of the winds and ocean currents, and $\tavg{\dot{D}_\text{sed}}$ is the time-mean rate of dissipation associated with the sedimentation of precipitation. 
We may also define the time-mean rate of generation of kinetic energy associated with winds and ocean currents by $\tavg{\dot{W}_K}$. \add{In steady state, we must have $\tavg{\dot{W}_K} = \tavg{\dot{D}_\text{fric}}$, a balance often expressed through the Lorenz energy cycle \citep[e.g.,][]{Lorenz1955}, which describes the atmospheric heat engine as a series of conversions between different reservoirs of internal, potential and kinetic energy. We discuss the Lorenz energy cycle in more detail in section \ref{sec:APE}.} 

According to \eqref{eq:work_moist}, $\tavg{\dot{W}_K}$ accounts for only a portion of the reversible work performed by the climate system; 
the remainder is used to lift water upwards through the atmosphere to balance the downward irreversible flux of water owing to precipitation \citep{Pauluis2002}. 
Since $\tavg{\dot{W}_K}$ represents the work responsible for powering the atmospheric and oceanic circulation, it may be considered to be the ``useful'' component of the total reversible work.
This motivates the definition of the mechanical efficiency of the climate system $\eta_M$ by
\begin{equation}
     \eta_M = \frac{\tavg{\dot{W}_K}}{\tavg{\dot{Q}_\text{in}}}.
     \label{eq:mech_eff}
 \end{equation}
The mechanical efficiency refers to the efficiency with which the climate system generates and dissipates kinetic energy of the winds and ocean currents \citep{Pauluis2002,Goody2003}. It is similar to the classic concept of the efficiency of a heat engine, except that the useful work $\tavg{\dot{W}_K}$ is done on, and dissipated within, the system itself. 

To examine the factors affecting the mechanical efficiency, we use the fact that the dissipation rates \add{$\tavg{\dot{D}_\text{fric}}$} and \add{$\tavg{\dot{D}_\text{sed}}$}
are associated with irreversible entropy sources which we denote $\tavg{\dot{S}_i^\text{fric}}$ and \add{$\tavg{\dot{S}_i^\text{sed}}$}, respectively.
This allows the heat-engine relation \eqref{eq:heat_engine_2} to be written
    \begin{linenomath}\begin{equation*}
        \tavg{\dot{Q}_\text{in}}\left(\frac{1}{T_\text{out}} - \frac{1}{T_\text{in}}\right) = \tavg{\dot{S}^\text{fric}_{i}} + \tavg{\dot{S}^\text{sed}_i} + \tavg{\dot{S}^\text{nm}_i}, 
    \end{equation*}\end{linenomath}
    where $\dot{S}^\text{nm}_i$ represents irreversible material entropy production by non-mechanical processes \add{such as heat diffusion, irreversible mixing, and irreversible chemical reactions} \citep{Goody2003}.
    As we will show in more detail in section \ref{sec:irr_procs}, the entropy source 
    $\tavg{\dot{S}_i^\text{fric}}$ may be written 
    in terms of the time-mean dissipation rate $\tavg{\dot{D}_\text{fric}}$ and an effective temperature $T_\text{fric}$ so that,
\begin{linenomath}\begin{equation*}
    \tavg{\dot{S}_i^\text{fric}} = \frac{\tavg{\dot{D}_\text{fric}}}{T_\text{fric}}.
\end{equation*}\end{linenomath}
\add{Here, the effective temperature of frictional dissipation is defined to satisfy
\begin{equation}
    \frac{\tavg{\dot{D}_\text{fric}}}{T_\text{fric}} =  \int_\Omega \frac{\rho\epsilon}{T} \, dV, \label{eq:Tfric}
\end{equation}
where $\epsilon$ is the local rate of frictional dissipation per unit mass.
Generally, $T_\text{fric}$ is weighted near the warm lower boundary of the atmosphere, implying $T_\text{fric}>T_\text{out}$.}

Since $\tavg{\dot{D}_\text{fric}}=\tavg{\dot{W}_K}$ in steady state, we may combine the previous two equations to give
     \begin{equation}
        \tavg{\dot{W}_\text{max}} = \tavg{\dot{W}_K}  + T_\text{fric} \tavg{ \dot{S}_i^\text{sed} + \dot{S}^\text{nm}_i }, \label{eq:max_work}
    \end{equation}
     where 
     \begin{linenomath}\begin{equation*}
         \tavg{\dot{W}_\text{max}} = \left(\frac{T_\text{fric}}{T_\text{out}}\right)\eta_C \tavg{\dot{Q}_\text{in}}
     \end{equation*}\end{linenomath}  
     is the maximum rate at which work can be performed by the system for a given $T_\text{fric}$, achieved when there are no other irreversible entropy sources besides that associated with frictional dissipation of the winds and ocean currents \citep{Pauluis2002}. 
     
     Note that, since we generally expect $T_\text{fric}>T_\text{out}$, the mechanical efficiency implied by the performance of work at a rate $\tavg{\dot{W}_\text{max}}$ is higher than the Carnot efficiency $\eta_C$. 
     This is possible because the work is being performed on the system itself, providing an additional dissipative heat source $\tavg{D_\text{fric}}$. The rate of work $\tavg{\dot{W}_\text{max}}$ 
     therefore does not represent the maximum work that can be done on an external body, which is limited by Carnot's theorem to not exceed that of an ideal Carnot engine \citep{Renno1996,Bister2011,Hewittetal1975}. Rather, $\tavg{\dot{W}_\text{max}}$ represents the maximum rate of work that would be performed by an ideal Carnot engine in which the heat input and output is given by the combination of radiative heating and cooling and dissipative heating experienced by the \add{climate system}. 

 For a given value of $\tavg{\dot{W}_\text{max}}$, the mechanical efficiency of the climate system is determined by the amount of entropy produced by precipitation sedimentation and non-mechanical irreversible processes and the effective temperature of frictional dissipation. 
 We shall see in the following section that processes associated with water, including diffusion of water vapor and irreversible phase change, are responsible for most of the non-mechanical irreversible entropy production in the atmosphere. These processes, coupled with the work required to lift water upward through the atmosphere, reduce the mechanical efficiency of the atmosphere relative to a hypothetical atmosphere that does not contain water, and they exert a strong influence on the dynamics of the atmospheric circulation. 

\section{Irreversible processes in the climate system}
\label{sec:irr_procs}

We now consider in detail the different processes that contribute to irreversible  entropy production in the climate system. Because of its clear relationship to work and kinetic energy generation, we focus on material entropy production in the atmosphere and oceans. We first sketch the derivation of the material entropy budget for both a single-component and multi-component fluid (section \ref{sec:ent_derivation}). Readers familiar with the governing equation for a fluid's entropy \eqref{eq:s_moist} may proceed to the following sections where we consider the application of these results to the atmosphere (section \ref{sec:moist}) and ocean (section \ref{sec:ocean}) more specifically. 
A number of previous authors have provided more detailed treatments focused on the atmosphere \citep[e.g.,][]{Hauf1987,Pauluis2000b,Gassmann2015} and ocean \citep[e.g.,][]{Gregg1984}, and in a more general context \citep[e.g.,][]{deGroot1984}.



\subsection{Derivation of the material entropy budget}
\label{sec:ent_derivation}

\subsubsection{Single-component fluids}
\label{sec:single}

\add{For simplicity, we begin by considering the entropy budget of a fluid made of a single chemical component, or equivalently, a mixture with a fixed composition. The atmosphere and ocean both have variable composition, and they must be treated as multi-component fluids; the more complex multi-component case is discussed in section \ref{sec:multi} below.}

\add{Consider the second law of thermodynamics applied to a fluid element of unit mass.} If the fluid's interactions are purely reversible, we have,
\begin{linenomath}\begin{equation*}
        \frac{ds}{dt} = \frac{\dot{q}_\text{rev}}{T},
\end{equation*}\end{linenomath}
where $s$ is the entropy of the fluid element and $\dot{q}_\text{rev}$ is the reversible heating rate, both expressed per unit mass. We also have the first law of thermodynamics,
    \begin{linenomath}\begin{equation*}
        \frac{du}{dt} = \dot{q} + \dot{w},
    \end{equation*}\end{linenomath}
 where $u$ is the internal energy of the fluid element, $\dot{w}$ is rate of work done on the fluid element by its environment, and $\dot{q}$ is the heating rate. Assuming reversible conditions, $\dot{q}=\dot{q}_\text{rev}$, and the work is given by $\dot{w}_\text{rev} = -p\frac{d\alpha}{dt}$, where $p$ is the pressure and $\alpha = \rho^{-1}$ is the specific volume of the fluid element. Combining the above equations, we have
        \begin{equation}
        \frac{du}{dt} = T\frac{ds}{dt} - p\frac{d\alpha}{dt}.
        \label{eq:fund_therm_rel}
    \end{equation}
    This is the fundamental thermodynamic relation linking entropy to other state variables for any substance of fixed composition \citep[see e.g.,][]{deGroot1984,Landau1987}. 
    
    While \eqref{eq:fund_therm_rel} was derived for reversible conditions and under the assumption of thermodynamic equilibrium, it may be applied under more general circumstances provided an approximation known as \textit{local thermodynamic equilibrium} is valid \citep[e.g.,][]{deGroot1984}. This approximation allows for thermodynamic functions such as temperature, pressure, and entropy, to be defined locally within a fluid as a function of space and time. In the bulk of the atmosphere and ocean, local thermodynamic equilibrium is a very good approximation. The exception is at very high altitudes ($\gtrsim 80 $ km), where the density of the gas becomes so low that molecular collisions become infrequent \citep[e.g.,][]{Houghton2002}. But this region accounts for a trivial fraction of the atmosphere's mass, and it is therefore reasonable to assume \eqref{eq:fund_therm_rel} is valid when considering the entropy budget of the atmosphere or climate system as a whole \citep[e.g.,][]{Lesins1990,Lucarini2009}.

    

    
    
    

\add{To use \eqref{eq:fund_therm_rel}, we require a separate expression for the rate of change of the fluid's internal energy.} The equation governing the specific internal energy $u$ of a \add{single-component} fluid  under the influence of radiation may be written \citep{deGroot1984},
\begin{equation}
    \frac{\partial \rho u}{\partial t} + \nabla \cdot \left(\rho \mathbf{v} u + \mathbf{D}_u\right)
    = -p \nabla \cdot \add{\mathbf{v}} + \rho\dot{q}_\text{rad}  + \rho \epsilon,
    \label{eq:u_flux}
\end{equation}
where $\rho$ is the fluid density, $\mathbf{v}$ is the fluid velocity, $\mathbf{D}_u$ is the heat flux owing to molecular diffusion, and $\dot{q}_\text{rad}$ and $\epsilon$ are the heating rates owing to radiation and frictional dissipation, respectively, expressed per unit mass of the fluid. 
The first term on the right-hand side gives the rate of work done on the fluid by its surroundings. 
Eq. \eqref{eq:u_flux} is an Eulerian equation for the internal energy as a function of space and time, while \eqref{eq:fund_therm_rel} is a Lagrangian equation valid for a given element of fluid. These viewpoints may be related to one another by expressing the \add{Lagrangian} derivative $d/dt$, representing the rate of change following a given fluid element\footnote{The Lagrangian derivative is sometimes given the special notation $\frac{D}{Dt} = \frac{\partial}{\partial t} + \mathbf{v}\cdot\nabla$.}, in terms of its Eulerian  counterpart,
\begin{linenomath}\begin{equation*}
    \frac{du}{dt} = \frac{\partial u}{\partial t} +\mathbf{v}\cdot \nabla u.
\end{equation*}\end{linenomath}
The internal energy equation may then be written in Lagrangian form as,
\begin{equation}
    \rho\frac{du}{dt} =-\rho p\frac{d\alpha}{dt} + \rho\dot{q}_\text{rad} - \nabla \cdot \mathbf{D}_u + \rho\epsilon,
    \label{eq:u}
\end{equation}
where we have used the equation for mass continuity,
\begin{equation}
    \frac{\partial \rho}{\partial t} + \nabla \cdot \left(\rho \mathbf{v} \right) = 0,
    \label{eq:mass}
\end{equation}
and we have rewritten the work term in terms of the Lagrangian rate of change of specific volume $\alpha$.

%

Combining the internal energy equation \eqref{eq:u} with the fundamental thermodynamic relation \eqref{eq:fund_therm_rel} and using mass continuity, we may write an explicit equation governing the fluid's entropy,
\begin{equation}
    \frac{\partial \rho s}{\partial t} + \nabla \cdot \left( \rho \mathbf{v} s \right) + \nabla \cdot \left(\frac{\mathbf{D}_u}{T}\right) - \frac{\rho\dot{q}_\text{rad}}{T} =
       \frac{\rho\epsilon}{T} - \frac{\mathbf{D}_u\cdot \nabla T}{T^2}.
      \label{eq:budget_local}
\end{equation}
This represents the local Eulerian entropy budget of a single-component fluid. Terms on the left-hand side give, from left to right, the local rate of change of entropy, the flux divergence of entropy by fluid motions, the flux divergence of entropy owing to molecular heat diffusion, and the entropy tendency due to radiative heating and cooling. The right-hand side contains the entropy production due to irreversible processes.

The connection between the local Eulerian entropy budget and the material entropy budget of the climate system may be readily seen by integrating \eqref{eq:budget_local} in space and averaging in time. Since there are no advective and molecular fluxes at the top of the atmosphere, the flux divergences on the left-hand side vanish on integration over the entire climate system. Further considering steady-state conditions, the time tendency also vanishes, and we have
\begin{equation}
    -\frac{1}{A}\int_\Omega \tavgb{\frac{\rho\dot{q}_\text{rad}}{T} }\,  dV =
    \tavg{\dot{S}_i^\text{fric}} + \tavg{\dot{S}_i^\text{heat}},
    \label{eq:dS_simple}
\end{equation}
where we have defined
\begin{linenomath}\begin{align}
    \dot{S}_i^\text{fric} &= \frac{1}{A}\int_\Omega \rho\dot{s}_i^\text{fric} \, dV = \frac{1}{A}\int_\Omega \frac{\rho\epsilon}{T}  \,  dV, \label{eq:s_fric} \\
    \dot{S}_i^\text{heat} &= \frac{1}{A}\int_\Omega \rho \dot{s}_i^\text{heat} \, dV = -\frac{1}{A}\int_\Omega \frac{\mathbf{D}_u\cdot \nabla T}{T^2} \, dV. \label{eq:s_heat}
\end{align}\end{linenomath}
Comparing \eqref{eq:dS_simple} to the material entropy budget \eqref{eq:entropy_material}, we find that the material entropy production is given by $\dot{S}_i^\text{mat} = \dot{S}_i^\text{fric} + \dot{S}_i^\text{heat}$.
For a planet\add{ary atmosphere} comprised of a single-component fluid, there are two processes that lead to irreversible material entropy production\add{:} molecular heat diffusion and viscous dissipation.

It is important to note that our derivation of \eqref{eq:budget_local} required only the internal energy equation and the fundamental thermodynamic relation, which may be taken as the relation defining entropy. Eq. \eqref{eq:budget_local} contains no additional information about the flow that is not already contained in the energy budget \citep{Romps2008}. Rather, the additional information provided by the second law is contained in the requirement for the irreversible entropy production terms on the right-hand side of \eqref{eq:budget_local} to be positive \add{definite} \citep{Gassmann2015}. This puts constraints on the form of the molecular heat flux $\mathbf{D}_u$ and the viscous dissipation $\rho\epsilon$. 

For example, it is easy to show that the second law is satisfied for the simple case of Fickian diffusion of temperature, for which
$\mathbf{D}_u = -\kappa \nabla T$, for some $\kappa \ge 0$. The associated irreversible entropy production due to heat diffusion is \citep{Peixotoetal1991},
\begin{linenomath}\begin{equation*}
    \rho\dot{s}^\text{heat}_i = \kappa \left(\frac{|\nabla T|}{T}\right)^2,
\end{equation*}\end{linenomath}
which is positive definite as required by the second law. \add{For more complex heat diffusion laws (e.g., those applicable to anisotropic materials), the requirement of positive definite entropy production may be used to constrain the functional form of $\mathbf{D}_u$ and ensure it is consistent with the second law.} 



\subsubsection{Multi-component fluids}
\label{sec:multi}

The budget equation \eqref{eq:dS_simple} is valid for a fluid whose composition is invariant in time and space. But the dynamics of Earth's atmosphere and ocean are both strongly influenced by their variable composition.
In particular, 
irreversible entropy production in the atmosphere is dominated by processes associated with water in all its phases \citep{Pauluis2002}.
We therefore must consider Earth's atmosphere as a multi-component fluid when discussing its entropy budget.



We consider a fluid that is a mixture of $N$ species, 
and we denote the density of species $x$ by $\rho_x$. 
The continuity equation for each species may be written
\begin{linenomath}\begin{equation*}
        \frac{\partial \rho_x}{\partial t} + \nabla \cdot \left(\rho_x \mathbf{v} + \mathbf{D}_x \right) =\rho\dot{\chi}_x.
    \end{equation*}\end{linenomath}
    where the velocity $\mathbf{v}$ is the barycentric velocity, given by the mass-weighted mean velocity over all species \citep{deGroot1984}, and $\mathbf{D}_x$ is the non-advective flux of species $x$, representing processes such as Brownian motion of molecules and, in the atmosphere, sedimentation of hydrometeors such as raindrops and snowflakes \citep{Gassmann2015}. 
    
    The quantity $\dot{\chi}_x$ represents the mass source of species $x$ per unit mass of the mixture due to chemical reactions. Mass conservation requires that 
    \begin{linenomath}\begin{equation*}
       \sum_x \dot{\chi}_x = 0.
    \end{equation*}\end{linenomath}
    For example, in the atmosphere, condensation represents a source of liquid water and a sink of water vapor of equal magnitude. Since by definition the barycentric velocity gives the velocity of the center of mass of an element of the fluid mixture, mass conservation also requires \add{that the} non-advective mass fluxes \add{of all species} sum to zero:
    \begin{equation}
        \sum_x \mathbf{D}_x = 0. \label{eq:mass_diff}
    \end{equation} 
    Combining the previous three equations, it may be shown that density of the mixture $\rho = \sum_x \rho_x$ satisfies the continuity equation \eqref{eq:mass}.
    
    It is useful to define the mass fraction of a species $q_x = \rho_x/\rho$. The mass fraction of water vapor, $q_v$, is known as the specific humidity. The mass continuity equation for each species may be rearranged into a Lagrangian equation for $q_x$,
    \begin{equation}
        \rho \frac{dq_x}{dt} = \rho\dot{\chi}_x - \nabla \cdot \mathbf{D}_x.
        \label{eq:q}
    \end{equation}
    We may also define the specific internal energy of the mixture by
    \begin{equation}
        u = \sum_x q_x u_x,
        \label{eq:umean}
    \end{equation}
    where $u_x$ is the specific internal energy of species $x$. Finally, we may write the fundamental thermodynamic relation for each species,
    \begin{equation}
        \frac{du_x}{dt} = T\frac{ds_x}{dt} - p_x\frac{d\alpha_x}{dt},
    \end{equation}
    where $s_x$ is the specific entropy of species $x$, $p_x$ is the partial pressure of species $x$ and $\alpha_x = \rho_x^{-1}$. Combining the previous two equations, we may write a thermodynamic relation for the mixture given by \citep{Pauluis2011},
        \begin{equation}
        \frac{du}{dt} = T\frac{ds}{dt} - p\frac{d\alpha}{dt} + \sum_x g_x \frac{dq_x}{dt}.
        \label{eq:fund_therm_rel_moist}
    \end{equation}
    Here, the entropy of the mixture $s$ is defined analogously to \eqref{eq:umean}, $\alpha = \rho^{-1}$, and the total pressure $p$ is the sum of the partial pressures of each species. The quantity $g_x = u_x + p_x \alpha_x - Ts_x$ is the specific Gibbs free energy \add{for each species $x$; it is equal to the chemical potential divided by the molar mass.}

    Use of \eqref{eq:fund_therm_rel_moist} involves some approximation that should be noted. In addition to the assumption that the fundamental thermodynamic relation is valid for each species, we have assumed that each species has the same temperature $T$. Furthermore, by expressing the internal energy of the mixture as the mass-weighted sum of the internal energy of each of species in \eqref{eq:umean}, we have neglected interfacial effects between the different species. Such effects are important for understanding the formation of clouds and precipitation in the atmosphere \citep{Pruppacher2010}, but they are typically neglected when considering its bulk thermodynamics. 
    Note, however, that we have made no assumption of chemical equilibrium between species; as we shall see, phase changes outside of equilibrium are an important irreversible entropy source in the atmosphere.
    
    As for a single-component fluid, we may derive an equation for the entropy tendency of a multi-component fluid by substituting the thermodynamic relation \eqref{eq:fund_therm_rel_moist} into the equation governing the internal energy, which for a multi-component fluid may be written
    \citep[e.g.,][]{deGroot1984,Gassmann2015},
\begin{equation}
    \rho \frac{d u}{d t} + \nabla \cdot \left(\mathbf{D}_u + \sum_x \mathbf{D}_x h_x \right)= -\rho p \frac{d\alpha}{dt} + \rho\dot{q}_\text{rad} + \rho\epsilon.
    \label{eq:u_moist}
\end{equation}
This equation is identical to the single-component case \eqref{eq:u} but for the appearance of the flux divergence of enthalpy $h_x = u_x+p_x\alpha_x$ owing to molecular motions on the left-hand side\footnote{The heat flux $\mathbf{D}_u$ is sometimes defined to include these molecular enthalpy fluxes \citep{deGroot1984}.}. 
\add{Substituting \eqref{eq:fund_therm_rel_moist} into \eqref{eq:u_moist}, using \eqref{eq:q}, and rearranging, one may write an equation for the Lagrangian rate of change of entropy, given by
\begin{linenomath}\begin{align*}
     \rho \frac{d s}{d t} + &\nabla \cdot \left(\frac{\mathbf{D}_u}{T}\right) + \sum_x \left[ \tfrac{\nabla\cdot\left(\mathbf{D}_x h_x\right)}{T}  - \tfrac{g_x\nabla \cdot{\mathbf{D}_x}}{T} \right] - \frac{\rho\dot{q}_\text{rad}}{T}= \\
     &  -\frac{\mathbf{D}_u\cdot\nabla T}{T^2} -\sum_x \frac{\rho g_x \dot{\chi}_x}{T} +  \frac{\rho\epsilon}{T}.
\end{align*}\end{linenomath}
Using the fact that $h_x -g_x = Ts_x$, the terms involving the molecular flux of species mass may be written, 
\begin{equation}
    \tfrac{\nabla\cdot\left(\mathbf{D}_x h_x\right)}{T}  - \tfrac{g_x\nabla \cdot{\mathbf{D}_x}}{T} = \nabla\cdot\left(\mathbf{D}_x s_x\right) +\mathbf{D}_x\cdot \left(\tfrac{1}{T}\nabla h_x - \nabla s_x\right).
\end{equation}
Finally, using mass conservation \eqref{eq:mass},}
we may write a local Eulerian budget equation for the entropy, given by
\begin{linenomath}\begin{align}
    \frac{\partial \rho s}{\partial t} + \nabla \cdot \left(\rho \mathbf{v} s \right) + \nabla \cdot \left(\frac{\mathbf{D}_u}{T} + \sum_x \mathbf{D}_x s_x\right) -  \frac{\rho\dot{q}_\text{rad}}{T} = \nonumber\\ \rho\dot{s}_i^\text{fric} + \rho\dot{s}_i^\text{heat} +   \rho\dot{s}_i^\text{diff} + \rho\dot{s}_i^\text{chem},
    \label{eq:s_moist}
\end{align}\end{linenomath}
\add{where
\begin{linenomath}\begin{align*}
    \rho \dot{s}_i^\text{diff} &= -\sum_x \mathbf{D}_x\cdot \left(\tfrac{1}{T}\nabla h_x -  \nabla s_x \right), \\
    \rho \dot{s}_i^\text{chem} &= - \sum_x \frac{\rho g_x\dot{\chi}_x}{T}.
\end{align*}\end{linenomath}
The budget \eqref{eq:s_moist}} is similar to the entropy budget derived for a single component fluid, but it has an additional term associated with entropy transport owing to the non-advective flux of mass $\mathbf{D}_x$ on the left-hand side, and additional irreversible sources $\rho \dot{s}_i^\text{diff}$ and $\rho \dot{s}_i^\text{chem}$ associated with non-advective transport of mass and chemical \add{reactions}, respectively, on the right-hand side \citep[e.g.,][]{deGroot1984}. 

The entropy production associated with non-advective transport $\rho \dot{s}_i^\text{diff}$ may be simplified further using the fundamental thermodynamic relation written in terms of the enthalpy $h_x$,
\begin{equation}
    dh_x = T ds_x + \frac{1}{\rho_x} dp_x, \label{eq:fundamental_enthalpy}
\end{equation}
which allows one to write,
\begin{equation}
    \rho\dot{s}_i^\text{diff} = -\sum_x \frac{\mathbf{D}_x\cdot \nabla p_x}{\rho_x T}. \label{eq:s_diff}
\end{equation}
Molecular diffusion of a species $x$ induces positive irreversible entropy production when it transports the species from high partial pressure to low partial pressure. This corresponds to the entropy production associated with the molecular mixing of the species within the fluid.


\subsection{Thermodynamics of a moist atmosphere}
\label{sec:moist}

We now discuss more specifically the irreversible entropy sources in the atmosphere. To do so, we introduce an approximate thermodynamic treatment of a moist atmosphere following \citet{Gassmann2008,Gassmann2015} \add{and consistent with other treatments in the literature \citep[e.g.,][]{Hauf1987,Bannon2002}. We note that while some numerical models of the atmosphere employ similar thermodynamic treatments \citep[e.g.,][]{Romps2008,Bryan2002}, many models employ simplified equation sets that neglect some of the irreversible processes discussed. Moreover, most atmospheric models include numerical sources and sinks of entropy in addition to the irreversible physical sources discussed here, and diagnosing the entropy budget in such models must be done with care. We further discuss the issue of numerical production of entropy in models of the climate system in section \ref{sec:models}.
}

The atmosphere is taken to be a mixture of water vapor ($v$), liquid water ($l$), solid water ($s$) and ``dry'' air $(d)$ containing all the well-mixed gases. In Earth's atmosphere, dry air is by far the most abundant component; typical mass fractions of water vapor $q_v$ are no larger than 3-4\%, while typical mass fractions of condensed water species are orders of magnitude smaller. Dry air and water vapor are assumed to be ideal gases governed by an equation of state of the form,
\begin{linenomath}\begin{equation*}
    p_x = \rho_x R_x T,
\end{equation*}\end{linenomath}
with $R_x = R^*/m_x$ being the gas constant for species $x$ expressed in terms of the molar mass $m_x$ and the universal gas constant $R^*$. Condensed water species are assumed to be incompressible, with negligible specific volume. The specific entropy $s$ is then given by $s = \sum_x q_x s_x$ where $x = (d,v,l,s)$ and the specific entropy of each constituent is defined \citep{Hauf1987}
\begin{subequations}
\begin{linenomath}\begin{align}
      s_d &= c_{pd} \ln(T/T_0) - R_d \ln(p_d/p_0) + s_{d0}, \\
      s_v &= c_{pv} \ln(T/T_0) - R_v \ln(p_v/p_0) + s_{v0}, \label{eq:s_vap}\\
      s_l &= c_{pl} \ln(T/T_0) + s_{l0},  \\
      s_s &= c_{ps} \ln(T/T_0) + s_{s0}. 
\end{align}\end{linenomath}
\end{subequations}
Here $c_{px}$ is the isobaric specific heat capacity of each constituent, which we take to be constant, and $T_0$, $p_0$, and $s_{x0}$ are the temperature, pressure, and entropy of each constituent, respectively, at a reference point which we take to be the triple point of water. 
The reference entropies of water phases are related by,
\begin{linenomath}\begin{align*}
    s_{v0} &= s_{l0}+\frac{L_{v0}}{T_0}, \\
    s_{s0} &= s_{l0}-\frac{L_{f0}}{T_0},
\end{align*}\end{linenomath}
where $L_{v0}$ and $L_{f0}$ are the latent heats of vaporization and freezing, respectively, at the triple point of water. Specification of the entropy is completed by setting the value of \addr{$s_{d0}$ and} $s_{l0}$. 
\addr{Because we will not consider chemical reactions between dry air and water substance, the values of these parameters do not affect the calculation of irreversible entropy production in the atmosphere, 
and many authors take $s_{l0}=s_{d0}=0$ \citep[e.g.,][]{Romps2008}. But care must be taken to account for the effects of this choice when interpreting entropy changes of open systems \citep{Marquet2017b}.}
As noted previously, this formulation also neglects interfacial effects between species, and the resultant entropy budget therefore neglects certain irreversible processes associated with the spatial distribution of different phases within an air parcel (e.g., the irreversibility of cloud droplets coalescing together). 

With the \add{above} caveats in mind, we now consider the irreversible entropy sources in a moist atmosphere. \add{Like a single-component fluid, entropy production in the atmosphere includes that due to frictional dissipation $\dot{s}_i^\text{fric}$ and heat diffusion $\dot{s}_i^\text{heat}$. In subsections \ref{sec:moist}.1-3 below, we focus on the remaining entropy production terms $\dot{s}_i^\text{chem}$ and $\dot{s}_i^\text{diff}$ that owe their existence to the multi-component nature of a moist atmosphere.} 

\subsubsection{Irreversible phase changes}

The irreversibility of phase changes is included within the entropy source $\dot{s}_i^\text{chem}$, which we divide into components associated with evaporation/condensation $\dot{s}_i^\text{evap}$ and melting/freezing $\dot{s}_i^\text{melt}$. Sublimation may be considered to be a combination of melting and evaporation. 

 Consider first evaporation and condensation of a liquid droplet in the air. Denoting the evaporation rate per unit mass by $e$, we have that $e = \dot{\chi}_v = -\dot{\chi}_l$, and hence,
\begin{linenomath}\begin{equation*}
    \dot{s}^\text{evap}_i = \frac{e (g_l - g_v)}{T}.
\end{equation*}\end{linenomath}
Positive irreversible entropy production requires $g_v<g_l$ for net evaporation ($e>0$) and $g_l<g_v$ for net condensation ($e<0$), and the phase change is reversible if $g_v = g_l$, a condition known as saturation. Defining $g^{*l}_v$ as the value of the Gibbs free energy of water vapor at saturation with respect to liquid, we have
\begin{equation}
   \dot{s}^\text{evap}_i = \frac{e (g_v^{*l} - g_v)}{T}.    
\end{equation}
Using the definitions of the entropy of water vapor \eqref{eq:s_vap} and the enthalpy of water vapor $h_v = c_{pv}(T-T_0) + T_0s_{v0}$, this may be written \citep[e.g.,][]{Pauluis2002},
\begin{equation}
\dot{s}^\text{evap}_i = - eR_v\ln(\mathcal{R}), \label{eq:s_cond}   
\end{equation}
where $\mathcal{R} = p_v/p^{*l}_v$ is the relative humidity and $p^{*l}_v(T)$ is the saturation vapor pressure over a liquid surface and is only a function of temperature. When the relative humidity is less than one, the air is subsaturated with respect to liquid water, and evaporation is irreversible. When the relative humidity is greater than one, the air is supersaturated, and condensation is irreversible. 

\add{The above discussion neglects the effects of surface tension and impurities on phase equilibrium. Both of these effects are important for determining the conditions under which cloud droplets may form and grow \citep{Pruppacher2010}. Were it not for the abundance of aerosols that act as nuclei for the formation of cloud droplets, supersaturation may be a common occurrence in the atmosphere. In actual fact, substantial supersaturation with respect to liquid water is rare,} and condensation generally occurs close to phase equilibrium. Evaporation, on the other hand, often occurs at relative humidities well below 100\% and is an important source of irreversible entropy production in the climate system. \add{Contributors to $\dot{s}_i^\text{evap}$ include the evaporation of precipitation falling through subsaturated air, and evaporation from the Earth's surface, particularly over water bodies. Surface evaporation is driven by the thermodynamic disequilibrium between the surface and the atmosphere, but it is strongly modulated by kinetic effects; empirically, the evaporation rate from a saturated surface is found to scale with the square of the windspeed. This modulation gives rise to interesting feedbacks that can amplify atmospheric circulations such as tropical cyclones (see section \ref{sec:TCs})}.

A similar derivation \add{to that given above} may be performed for the solid/liquid phase transition leading to \citep{Romps2008} 
\begin{equation}
\dot{s}^\text{melt}_i = - mR_v\ln\left(\frac{p_v^{*l}}{p_v^{*s}}\right), 
\label{eq:s_frz}
\end{equation}
where $m$ is the melting rate per unit mass and $p_v^{*s}$ is the saturation vapor pressure over a solid ice surface. Note that both $p_v^{*l}$ and $p_v^{*s}$ are functions of temperature only, and they coincide at the freezing point\footnote{By neglecting the specific volumes of liquid and solid water, we have assumed that the freezing point is independent of pressure and equal to the triple point temperature. This is a very good approximation except at very high pressures not experienced in Earth's atmosphere.} $T_0$. For temperatures below $T_0$, $p_v^{*l}>p_v^{*s}$ and freezing is irreversible, while for temperatures above $T_0$, $p_v^{*l}<p_v^{*s}$ and melting is irreversible. In the atmosphere, melting and freezing can occur tens of kelvins from the freezing point, leading to a substantial irreversible production of entropy. \add{Furthermore, in regions of the atmosphere with temperatures below freezing, the relative humidity with respect to ice $p_v/p_v^{*s}$ ranges from a few percent to values approaching 200\% \citep{Gettelman2006}. Both sublimation at subsaturation and deposition at supersaturation may therefore be important irreversible sources of entropy in the atmosphere, contributing to both $\dot{s}_i^\text{evap}$ and $\dot{s}_i^\text{melt}$. }

\subsubsection{Diffusive mixing}
\label{sec:qv_diff}

In the atmosphere, the non-advective flux of species mass occurs as a result of two processes: 1) diffusive molecular mixing as a result of the random Brownian motions of molecules of each species and 2) sedimentation of condensed water particles massive enough to have appreciable terminal velocities. We consider entropy production by molecular mixing $\dot{s}_i^\text{mix}$ here and return to the sedimentation flux in the following subsection.

According to \eqref{eq:s_diff}, the entropy source due to molecular mixing is proportional to the specific volume of each constituent. \add{The specific volume of condensed water species has been assumed to be negligible, and so we must only consider diffusive fluxes of gaseous species.}
\add{Consider the diffusive mixing of dry air and water vapor. The mass conservation condition \eqref{eq:mass_diff} requires that such mixing involves equal and opposite mass fluxes of dry air and water vapor. But in the atmosphere,} fractional gradients in the partial pressure of water vapor can be orders of magnitude larger than those of dry air, and the dominant component of $s_i^\text{mix}$ is that due to the diffusion of water vapor down its partial pressure gradient,
given by \citep{Pauluis2002,Romps2008},
\begin{equation}
    \rho \dot{s}_i^\text{mix} \approx  R_v\mathbf{D}_v \cdot \nabla \ln\left(\frac{p_v}{p_0}\right). \label{eq:s_mix}
\end{equation}
Diffusive mixing of water vapor \add{and dry air} is particularly strong at the boundaries between clouds and the clear-air environment. Here, the combination of diffusive mixing and evaporation of cloud and precipitation particles produces a transport of water vapor into the environment that plays an important role in governing the tropical relative humidity \citep{Romps2014,Singh2019}. 

In fact, the entropy source owing to the \add{mixing of dry air and water vapor} $\dot{s}_i^\text{mix}$ has a close connection to that of evaporation $\dot{s}_i^\text{evap}$.
Consider a cloud droplet suspended in air with a relative humidity $\mathcal{R}$. Evaporation from this droplet into the subsaturated environment results in an irreversible entropy source given by \eqref{eq:s_cond}. Alternatively, suppose the evaporation from the droplet occurs reversibly in a molecular boundary layer surrounding the droplet that is at saturation. \add{Diffusive mixing then transports this water vapor} from the droplet to the far-field which has relative humidity $\mathcal{R}$. \add{Neglecting the small contribution owing to the diffusion of dry air,} this process results in an irreversible source of entropy given by \eqref{eq:s_mix}. As pointed out by \citet{Pauluis2002}, the entropy production in these two cases is the same. Evaporating water and transporting it from the droplet to its environment results in the same irreversible entropy production regardless of the microscopic details of the transport process.

\subsubsection{Irreversible sedimentation of precipitation}
\label{sec:precip_sed}

Consider an air parcel containing a mass fraction of precipitation $q_p$ with a sedimentation velocity $-v_t\mathbf{k}$ relative to air. Here $\mathbf{k}$ is a unit vector pointing upwards (antiparallel to the gravitational vector). Generally, it is reasonable to assume that the fall speed $v_t$ is equal to the terminal velocity of precipitation set by a balance between the downward gravitational force \add{on hydrometeors (precipitation particles)} and the upward drag force on hydrometeors owing to friction with the surrounding air \citep{Pruppacher2010}.

The barycentric velocity of the air-precipitation mixture may be written,
\begin{equation}
   \mathbf{v} = (1-q_p)\mathbf{v}_a + q_p\left(\mathbf{v}_a  -v_t\mathbf{k}\right),
   \label{eq:barycentric}
\end{equation}
where $\mathbf{v}_a$ is the velocity of air.
Since $\mathbf{v}_a \ne \mathbf{v}$, 
sedimentation of precipitation is coupled with a compensating upward non-advective transport of air with a velocity,
\begin{linenomath}\begin{equation*}
\mathbf{v}_a - \mathbf{v} = q_p v_t \mathbf{k}.
\end{equation*}\end{linenomath}
The above equation implies an upward non-advective flux of dry air $\mathbf{D}_d = \rho q_d q_p v_t \mathbf{k}$ and water vapor $\mathbf{D}_v = \rho q_v q_p v_t \mathbf{k}$. On substitution into \eqref{eq:s_diff}, these non-advective fluxes give an irreversible entropy source totalling \citep{Gassmann2015},
\begin{linenomath}\begin{equation*}
    \rho \dot{s}_i^\text{sed} = - \frac{q_p v_t}{T} \frac{\partial p}{\partial z}.
\end{equation*}\end{linenomath}
Assuming hydrostatic balance, this may be written,
\begin{equation}
    \rho \dot{s}_i^\text{sed} \approx \frac{\rho {g_{\earth}} q_p v_t}{T} \label{eq:s_sed},
\end{equation}
where ${g_{\earth}}$ is Earth's gravitational acceleration.  
The entropy source $\dot{s}_i^\text{sed}$ is therefore positive when precipitation falls to the surface ($v_t>0$). 

The term
$\dot{d}_\text{sed} = {g_{\earth}} q_p v_t$ that appears on the right-hand side of \eqref{eq:s_sed} represents a dissipation rate associated with the loss of gravitational potential energy by falling precipitation. Physically, this manifests as frictional dissipation due to the upward drag force acting on precipitation particles sedimenting relative to the air \citep{Pauluis2000,Pauluis2002}. In our derivation, however, we do not consider in detail the interface between air and condensed water species, and $\rho\dot{d}_\text{sed}$ appears instead as the hydrostatic approximation to an irreversible pressure work \citep{Gassmann2015}.

As discussed in section \ref{sec:mech}, the integrated dissipation $\tavg{\dot{D}_\text{sed}} =  \int_\Omega \tavg{\rho \dot{d}_\text{sed} }\, dV$ is a sink in the \add{time-averaged} mechanical energy budget.
This may be seen explicitly by considering the \add{time-averaged} rate of work performed by the barycentric flow, which may be written,
\begin{equation}
    \tavg{\dot{W}_K} = -\int_\Omega \tavg{\mathbf{v}\cdot \nabla p} \, dV,
    \label{eq:work_revirr}
\end{equation}
\add{where the angle brackets refer to a time mean.}
Rearranging the barycentric velocity \eqref{eq:barycentric} 
into a reversible component $\mathbf{v}_\text{rev} = \mathbf{v}_a$, associated with fluid motions, and an irreversible component $\mathbf{v}_\text{irr} = -q_p v_t\mathbf{k}$, associated with the sedimentation of precipitation,
we may write,
\begin{eqnarray*}
 \tavg{\dot{W}_K} =
\underbrace{-\int_\Omega\tavg{ \mathbf{v}_\text{rev}\cdot \nabla p} \, dV}_{\tavg{\dot{W}_\text{rev}}} \,\,\,
- \,\,\, \underbrace{\int_\Omega \tavg{\rho {g_{\earth}} q_p v_t} \, dV}_{\tavg{\dot{D}_\text{sed}}}.
\end{eqnarray*}
Here we have used hydrostatic balance to express the second term on the right-hand side in terms of the gravitational acceleration.
The first term on the right-hand side gives the rate of work performed by reversible fluid motions $\tavg{\dot{W}_\text{rev}}$, and the second term gives the precipitation-induced dissipation rate $\tavg{\dot{D}_\text{sed}}$.
Comparing this equation to \eqref{eq:work_moist}, we see that, in steady state, \add{$\tavg{\dot{W}_K} = \tavg{\dot{D}_\text{fric}}$. We} may therefore identify the rate of work performed by the barycentric flow as being the rate of generation of kinetic energy of fluid motions. The total rate at which reversible work is performed $\tavg{\dot{W}_\text{rev}}$ is larger and represents the work required not just to generate kinetic energy of fluid motions, but also to lift water against the Earth's gravitational field \citep{Pauluis2000,Pauluis2002}. 

\add{In general it is the work $\tavg{\dot{W}_K}$ that is of most interest in studies of the climate system, as it represents the work performed in driving the atmospheric and oceanic circulations. In particular, $\tavg{\dot{W}_K}$ represents the source of kinetic energy in computations of the transformations between internal, potential, and kinetic energy that make up the Lorenz energy cycle \citep[see section \ref{sec:APE} and][]{Lorenz1955}. However, most discussions of the Lorenz energy cycle to date treat moist processes in a simplified way, and they do not explicitly consider the work required to lift water through the Earth's gravitational field \citep{Pauluis2007}.}

\subsubsection{The entropy budget of a moist atmosphere}

Combining the results from the previous subsections, we may write the approximate local Eulerian material entropy budget \eqref{eq:s_moist} for a moist atmosphere,
\begin{equation}
    \frac{\partial \rho s}{\partial t} + \nabla \cdot \left(\rho \mathbf{v} s \right) + \nabla \cdot \left(\frac{\mathbf{D}_u}{T} + \sum_x \mathbf{D}_x s_x\right) -  \frac{\rho\dot{q}_\text{rad}}{T} = \rho\dot{s}_i^\text{mat} 
    \label{eq:ent_bud_local_moist}
\end{equation}
where the material irreversible entropy production is given by,
\begin{equation}
    \dot{s}_i^\text{mat} = \dot{s}_i^\text{fric} + \dot{s}_i^\text{heat} +
    \dot{s}_i^\text{evap} + \dot{s}_i^\text{melt}
     + \dot{s}_i^\text{mix} + \dot{s}_i^\text{sed}.
     \label{eq:entropy_terms}
\end{equation}    
The terms on the right-hand side represent, from left to right, frictional dissipation of the winds \eqref{eq:s_fric}, molecular heat diffusion \eqref{eq:s_heat}, irreversible evaporation and condensation \eqref{eq:s_cond}, irreversible melting and freezing \eqref{eq:s_frz}, irreversible molecular mixing \eqref{eq:s_mix}, and dissipation associated with the sedimentation of precipitation \eqref{eq:s_sed}.  

Integrating the above equation \add{in space}, dividing by the Earth's surface area $A$, \add{and averaging in time}, one obtains the steady-state material entropy budget for the entire atmosphere \citep{Romps2008},
\begin{linenomath}\begin{align}
    \frac{1}{A}\int_{\partial\Omega_A}  \tavgb{\frac{\mathbf{D}_u}{T} + \sum_x \mathbf{D}_x s_x } \cdot d\mathbf{A} -  &\frac{1}{A} \int_{\Omega_A}\tavgb{\frac{\rho\dot{q}_\text{rad}}{T}} \,  dV \nonumber\\ &= \tavg{\dot{S}_i^\text{mat}},
    \label{eq:ent_bud_atm}
\end{align}\end{linenomath}
where $\Omega_A$ represents the volume of the atmosphere, $\partial\Omega_A$ represents its boundary (the top of the atmosphere and the surface) and $d\mathbf{A}$ is a surface element oriented with outward pointing normal. \add{The first integral on the left-hand side gives the boundary fluxes of entropy owing to molecular heat and water transport. Specifically, $\int_{\partial\Omega} \tfrac{\mathbf{D}_u}{T} \cdot d\mathbf{A}$ may be interpreted as the flux of entropy out of the atmosphere carried by the molecular flux of heat, while $\int_{\partial\Omega} \mathbf{D}_x s_x \cdot d\mathbf{A}$ may be interpreted as the flux of entropy out of the atmosphere carried by the molecular flux of species $x$.} The right hand side $\tavg{\dot{S}_i^\text{mat}}$ gives the total material irreversible entropy production in the atmosphere defined
\begin{linenomath}\begin{equation*}
    \tavg{\dot{S}_i^\text{mat}} =\frac{1}{A} \int_{\Omega_A} \tavgb{\rho \dot{s}_i^\text{mat} } \, dV,
\end{equation*}\end{linenomath}
\add{and expressed per unit of the Earth's surface area $A$}.
The total entropy production rate associated with each process given in \eqref{eq:entropy_terms} may be defined analogously.

Consider the atmosphere over a saturated liquid surface (e.g., the ocean)\footnote{For a surface with relative humidity less than one, the surface evaporative flux of vapor also involves an irreversible source of entropy that should be added to $\dot{S}_i^\text{mat}$ on the right-hand side.}. Using the relationship $s_v^{*l} - s_l = L_v/T$, where $s_v^{*l}$ is the saturation entropy of water vapor over a liquid surface and $L_v$ is the latent heat of vaporization,
we may write \eqref{eq:ent_bud_atm} as 
\begin{equation}
    -\frac{1}{A}\int_{\partial\Omega_A}  \tavgb{\frac{F_{LH} + F_{SH}}{T} } dA -  \frac{1}{A}\int_{\Omega_A}\tavgb{\frac{\rho\dot{q}_\text{rad}}{T}} \,  dV = \tavg{\dot{S}_i^\text{mat}}
    \label{eq:ent_atm_bud_flux}
\end{equation}
where $F_{LH}$ and $F_{SH}$ are the surface latent and sensible heat fluxes from the surface to the atmosphere, and $dA = |d\mathbf{A}|$. In steady state, conservation of total energy in the atmosphere requires the sensible and latent heat flux to balance the net radiative cooling of the atmosphere. \add{We may therefore write \add{the} above equation in a simpler form \citep[e.g.,][]{Pauluis2002,SinghOGorman2016},
\begin{equation}
-\tavg{\dot{Q}_\text{rad}}\left(\frac{1}{T_a}- \frac{1}{T_s}\right) = \tavg{\dot{S}_i^\text{mat}},
    \label{eq:ent_bud_atm2}
\end{equation}
where $\dot{Q}_\text{rad} = \frac{1}{A}\int_{\Omega_A} \rho\dot{q}_\text{rad} \, dV$ is the net radiative heating of the atmosphere, and we have defined the characteristic temperatures
\begin{linenomath}\begin{equation*}
    \frac{1}{T_s} = \frac{\int_{\partial\Omega_A}  \tavg{\left(\frac{1}{T}\right)(F_{LH} + F_{SH})} dA} {\int_{\partial\Omega_A}  \tavgb{F_{LH} + F_{SH} } dA}
\end{equation*}\end{linenomath}
and
\begin{linenomath}\begin{equation*}
    \frac{1}{T_{a}} = \frac{1}{\tavg{\dot{Q}_\text{rad}}}\int_{\Omega_A}  \tavgb{\frac{\rho\dot{q_\text{rad}}}{T}} dV.    
\end{equation*}\end{linenomath}
Here $T_s$ is a characteristic surface temperature representing the temperature at which the atmosphere is heated by surface fluxes, and $T_a$ is a characteristic atmospheric temperature representing the temperature at which the atmosphere is cooled by radiation. With these definitions, \eqref{eq:ent_bud_atm2} is identical to the material entropy budget of the one-layer atmosphere model presented in section \ref{sec:material}.}


Eq. \eqref{eq:ent_bud_atm2} provides an intuitive perspective on the second law as applied to the atmosphere. The atmosphere is heated by surface fluxes at the relatively warm surface and cooled by radiation in the relatively cold troposphere, thus creating an entropy sink that, \add{in steady state,} is balanced by irreversible processes\footnote{\add{If one instead defines $T_s$ as the characteristic temperature at which the surface is heated by radiation, \eqref{eq:ent_bud_atm2} becomes the entropy budget for the entire climate system. Since the ocean transports energy from warm regions to cooler regions, this definition results in a higher $T_s$, and a larger entropy sink. In steady state, this larger sink is balanced by the additional irreversible entropy production that occurs in the ocean (see section \ref{sec:ocean})}.}.
\add{A key task is then to determine} relative importance of the different processes contributing to $\dot{S}_i^\text{mat}$. As highlighted by \citet{Pauluis2002,Pauluis2002b}, the irreversible entropy production in the atmosphere is dominated by moist processes, including the \add{molecular mixing} of water vapor $\dot{S}_i^\text{mix}$, irreversible phase change $\dot{S}_i^\text{cond}$ \& $\dot{S}_i^\text{melt}$, and dissipation associated with precipitation sedimentation $\dot{S}_i^\text{sed}$. 
In the next three sections, we will discuss in detail the role played by irreversible moist processes in the dynamics of convective clouds (section \ref{sec:conv}), tropical cyclones (section \ref{sec:TCs}) and the general circulation (section \ref{sec:global_atmosphere}). 


An important limitation of the derivation leading to  \eqref{eq:ent_bud_atm2} is the assumption that all chemical species within the atmosphere have the same temperature. This is a good approximation for dry air, water vapor, and clouds, but it is not accurate for hydrometeors with appreciable sedimentation velocities, which often differ in temperature from their surroundings by several kelvins. While the assumption of uniform temperature is made commonly in studies of the atmosphere's entropy budget, it does not allow for consideration of irreversible entropy production associated with heat diffusion between \add{precipitation and the surrounding atmosphere}. \citet{Bannon2002} and \citet{Goody2003} derived order-of-magnitude estimates to suggest that such heat diffusion may contribute significantly to the entropy budget of moist convection. \citet{Bannon2002} and \citet{Raymond2013} derived equation sets suitable for numerical modeling that include the relevant irreversible production terms, \add{but, to our knowledge, there are no detailed modeling studies of the entropy budget that include entropy production associated with heat diffusion between hydrometeors and their surroundings. The magnitude of the errors induced by neglecting this entropy source, and whether it is larger than other errors in simulated entropy budgets due to, for example, numerical truncation \citep[see e.g.,][and section \ref{sec:models}]{WoollingsThuburn2006}, remains unknown.}

\subsubsection{The role of latent heating}
\label{sec:latent}

Latent heat release does not appear explicitly in the atmospheric entropy budget. This is because phase changes that occur at equilibrium do not cause a change in entropy. Rather, the effects of latent heat in \eqref{eq:ent_bud_atm} are included through the entropy input into the atmosphere owing to molecular fluxes of water species. \add{Nevertheless, many studies of the entropy budget make explicit mention of entropy production by latent heat release and find that it is a major contributor to the global atmospheric budget \citep[e.g.,][]{Pascaleetal2011,Peixotoetal1991,FraedrichLunkeit2008}}. 
\add{This is because
such studies consider an ``external'' view of the effects of moisture, in which latent heating is treated as an additional external heat source, similar to radiative heating, but the atmosphere is otherwise treated largely as a single-component fluid (see section \ref{sec:single}). Our derivation above, in contrast, considers an ``internal'' view, in which the atmosphere is treated as a true multi-component fluid, with phase changes being cast as mass exchanges between the different components. As we discuss below, both of these perspectives are mathematically valid, but only the internal approach \addr{provides a} direct expression of the second law of thermodynamics applied to the atmosphere.}

The connection between the internal and external viewpoints was first elucidated by \citet{Pauluis2002b}, who showed that 
the total irreversible entropy production associated with phase change and water vapor mixing could be approximately written as the sum of two terms, one related to the latent heating rate, and the other related to the work performed by the expansion of water vapor in the atmosphere. Using this relationship, the entropy budget \eqref{eq:ent_bud_atm} may be written approximately as,
\begin{linenomath}\begin{align}
    \int_{\partial\Omega_A}  \tavgb{\frac{\mathbf{D}_u}{T}} \cdot d\mathbf{S} -  \int_{\Omega_A}\tavgb{\frac{\rho(\dot{q}_\text{rad}+ \dot{q}_\text{lat})}{T} }\,  dV \approx \nonumber \\ \tavg{\dot{S}_i^\text{fric}}+\tavg{\dot{S}_i^\text{heat}}+\tavg{\dot{S}_i^\text{sed}}+ \int_{\Omega_A} \tavgb{\frac{1}{T}\frac{dp_v}{dt}} \, dV,
    \label{eq:ent_bud_dry}
\end{align}\end{linenomath}
where $\dot{q}_\text{lat}$ is the net heating rate due to phase changes\footnote{This relationship neglects the temperature dependence of the latent heats of vaporization and freezing \citep{Pauluis2002b,Romps2008}.}.
This equation is similar to the single-component entropy budget [cf. \eqref{eq:dS_simple}] with the latent heating rate included as an external heating in addition to that due to radiation, but it includes the entropy production associated with sedimentation of hydrometeors \addr{$\tavg{\dot{S}_i^\text{sed}}$} and a term related to the work performed by water vapor expansion (last term on the right-hand side). 

\citet{Romps2008} explicitly showed that \eqref{eq:ent_bud_dry} is an approximation of the budget for a quantity he referred to as the ``dry'' entropy, the entropy of a single-component gas with the same pressure, temperature, heat capacity, and gas constant as the moist atmosphere. A number of authors have considered budgets similar to \eqref{eq:ent_bud_dry} but neglected the terms associated with hydrometeor sedimentation and water vapor work \citep{Peixotoetal1991,FraedrichLunkeit2008,Pascaleetal2011}. The neglect of these terms is consistent with the treatment of moist thermodynamics in many climate models, which often neglect the effects of water substance \add{on the specific heat capacity and density of air,} except in the calculation of buoyancy \citep{Pauluis2002}. Indeed, \citet{Pascaleetal2011} obtained accurate closure of a budget similar to \eqref{eq:ent_bud_dry} in a comprehensive global climate model without considering precipitation sedimentation or water vapor work.



The dry-entropy budget and its variants provide a view of atmospheric thermodynamics in which the latent heating and cooling associated with phase changes is treated as an external heat source to an otherwise dry fluid. Such a view has proven useful as a means of estimating terms within the entropy budget \citep{Pauluis2002,Pauluis2002b}, or as an analysis tool in and of itself \citep{Peixotoetal1991,FraedrichLunkeit2008,Pascaleetal2011,Romps2008}. 
But in contrast to the entropy budget, the dry-entropy budget includes source terms that may be locally negative and do not represent irreversible processes, and its connection to the second law of thermodynamics is less direct. 

\subsection{Irreversible entropy production in the ocean}
\label{sec:ocean}

Comparatively few studies have investigated the entropy budget of the ocean relative to that of the atmosphere, but 
the formalism developed in section \ref{sec:multi} is equally applicable to the ocean. 
The oceanic entropy budget includes irreversible entropy sources owing to frictional dissipation $\dot{S}_i^\text{fric}$, heat diffusion $\dot{S}_i^\text{heat}$, diffusion of mass $\dot{S}_i^\text{diff}$, and phase changes $\dot{S}_i^\text{chem}$. In the ocean, $\dot{S}_i^\text{diff}$ accounts for irreversible entropy production owing to molecular \add{mixing between regions of high and low salinity}, while $\dot{S}_i^\text{chem}$ accounts for irreversible entropy production associated with the melting and freezing of sea ice outside of phase equilibrium. Estimates of the entropy production by these processes indicate that, while salt diffusion can be a significant source of entropy in certain regions \citep{Gregg1984}, both salt diffusion \citep{Shimokawa2001} and sea-ice melt \citep{Bannon2018} contribute only a small portion of the total irreversible entropy production of the ocean as a whole. In contrast to the atmosphere, the ocean's entropy budget may be approximated by that of a single-component fluid, with irreversible entropy production occurring primarily through heat diffusion and frictional dissipation. 

A further difference between the atmosphere and the ocean is that the ocean is heated and cooled almost exclusively at its upper surface (the exception is the heating owing to the geothermal heat flux, which on Earth is quantitatively small). Heat transport through sensible and latent heat fluxes occurs at the air-sea interface, while the penetration depth of shortwave and longwave radiation through ocean water is no more than 100 m and a few mm, respectively. The temperature at which the ocean is heated or cooled by radiation and turbulent fluxes is therefore very nearly equal to the surface temperature. The steady-state material entropy budget of the ocean may therefore be written \citep{Tailleux2015,Bannon2018},
\begin{equation}
     \frac{1}{A_O}\int_{\partial\Omega_O}  \tavgb{\frac{F_\text{rad} + F_{LH} + F_{SH}}{T_s} }\, dA = \tavg{\dot{S}_i^\text{mat}},
    \label{eq:ent_bud_oce}
\end{equation}
where $F_\text{rad}$ is the upward radiant energy flux, $F_{SH}$ and $F_{LH}$ are the sensible and latent heat fluxes from the ocean to the atmosphere, the integral is over the ocean surface $A_O$, {and the angle brackets refer to a time mean}. For simplicity, we have neglected the geothermal heat flux into the ocean and we have assumed that the difference between the effective temperature of precipitation and runoff and the effective temperature of evaporation is negligible; \citet{Bannon2018} show that making these approximations has a relatively minor effect on the ocean's entropy budget.

Building on the work of \citet{Tailleux2015}, \citet{Bannon2018} derived an observational estimate of the left-hand side of \eqref{eq:ent_bud_oce}, finding a rate of entropy export $-\tavg{\dot{S}_e} = 1.7$ mW m$^{-2}$ K$^{-1}$ averaged over the ocean surface area. This estimate is \add{somewhat higher than the material entropy production rate of the ocean of 1.2-1.4 W m$^{-2}$ K$^{-1}$ per unit area of the ocean surface found in the modeling study of \citet{Pascaleetal2011}. However, it should be noted that} the model's thermodynamic formulation did not include frictional heating, and so this was left out of the estimate of irreversible entropy production \add{and may influence the result.} Recall that the total material entropy production of the climate system is of the order 35-60 mW m$^{-2}$ K$^{-1}$ averaged over the Earth's surface; irreversible entropy production by the ocean is a very small fraction of this total, implying that the atmosphere accounts for the bulk of the irreversble entropy production in the climate system.

\citet{Bannon2018} also derived an independent estimate of the material entropy production in the ocean using estimates of the thermal diffusivity and temperature structure of the global ocean. According to this estimate, $\tavg{\dot{S}_i^\text{heat}} \approx 0.86$ mW m$^{-2}$ K$^{-1}$ and $\tavg{\dot{S}_i^\text{fric}} \approx 0.64$ mW m$^{-2}$ K$^{-1}$ with small contributions from salt diffusion and ice melt. Given the difficulty in measuring diffusivities in the ocean, the resultant estimate of $\tavg{\dot{S}_i^\text{mat}}$ is in rather remarkable agreement with the estimate of the entropy export $-\tavg{\dot{S}_e}$ given above.


The ocean, unlike the atmosphere, is forced both thermodynamically, through surface heat and freshwater fluxes (so-called buoyancy fluxes), and mechanically, through the work done on the ocean by the winds and tides\footnote{The ocean also performs work on the atmosphere. but this a negligible term in the atmosphere's mechanical energy and entropy budgets.}. \add{Together, these two forcings produce an overturning circulation that spans the depth of the ocean and governs the exchange of heat and chemical species between the atmosphere and ocean on timescales from hundreds to thousands of years \citep[e.g.,][]{Cessi2019}. An application of the oceanic entropy budget is in understanding the role of each type of forcing in determining the ocean's deep overturning circulation. 
}

%


It is well known that the circulation produced in a fluid heated and cooled at the same geometric level is substantially weaker than if the heating occurs below the cooling \citet{Sandstrom1908}\footnote{See \citet{Kuhlbrodt2008} for an English translation.}. \add{Mechanically-induced turbulent mixing may therefore play an important role in amplifying the ocean's overturning circulation by allowing the surface buoyancy flux from the atmosphere to penetrate a finite depth into the ocean, thereby allowing the ocean to be heated and cooled at different levels. Indeed, \citet{Munk1998} and \citet{Wunsch2004} argued that mechanically-induced mixing is critical to the observed ocean stratification, and that 
mechanical forcing plays a dominant role in governing the ocean's overturning circulation.} But more recent work notes that buoyancy forcing is an important source of available potential energy for the global ocean \citep[e.g.,][see also section \ref{sec:APE}]{Hughes2009,Tailleux2009,Tailleux2013a}, and numerical evidence suggests that the ocean's overturning circulation responds both to changes in buoyancy fluxes and changes in the atmospheric wind field \citep[e.g.,][]{Morrison2011}. 

A key question is how buoyancy fluxes act to alter the kinetic energy generation rate of the ocean. The estimate of the entropy source owing to frictional dissipation $\tavg{\dot{S}_i^\text{fric}}$ provided by \citet{Bannon2018} is based on the assumption that the frictional dissipation rate is equal to the work input by the winds and tides. To the extent that this budget is closed, it is therefore consistent with the notion that
buoyancy forcing results in no net increase in kinetic energy generation of the oceanic circulation.
We note, however, that the estimates of entropy production given in \citet{Bannon2018} are necessarily crude due to the lack of detailed observations, and they are not sufficiently accurate to strongly constrain the kinetic energy dissipation rate. Moreover, as pointed out by \citet{Hughes2009}, buoyancy forcing may facilitate a release of kinetic energy at large scales even if does not provide a net increase in the total kinetic energy generation rate of the ocean. Constraining the magnitude of such energy transfers between scales using either the entropy or mechanical energy budgets remains an observational challenge.

\section{The entropy budget of atmospheric convection}
\label{sec:conv}

In the field of meteorology, convection refers to fluid motions that transport heat in the vertical direction. This is primarily accomplished through clouds and their associated circulations, from shallow boundary-layer clouds over the subtropical ocean, to explosive continental convection that spans the depth of the troposphere and can potentially produce lightning, hail and other severe weather. Despite its ubiquity in the atmosphere, our fundamental understanding of atmospheric convection and its interaction with planetary-scale flows remains limited. Basic questions such as what physical factors determine cloud updraft velocities are still not completely resolved. A key reason for this is the importance of moist processes, including evaporation, latent heating, and precipitation, in atmospheric convection. The complex interaction of moist processes with atmospheric fluid dynamics presents a challenging theoretical problem. 

In this section, we describe how analysis of the second law of thermodynamics has provided a range of insights into the dynamics of moist convection. We begin by introducing a common idealized conceptual and modeling framework for studying moist convection known as radiative-convective equilibrium (section \ref{sec:RCE}),
before reviewing theories of moist convective updraft velocities developed based on analysis of the second law
 (section \ref{sec:convtheory}). Finally, we introduce some new results concerning the effect of the ``organization'' of moist convection on its mechanical efficiency
 with the aim of motivating further work in this area (section \ref{sec:org}). 

\subsection{Radiative-convective equilibrium (RCE)}
\label{sec:RCE}

Radiative-convective equilibrium (RCE) describes a hypothetical state in which a surface of infinite extent, usually assumed to consist of liquid water, is held at a fixed temperature, and the atmosphere above is allowed to cool under the influence of radiation. The cooling destabilizes the atmosphere, eventually leading to convection. ``Equilibrium'' is achieved when the convective heat flux from the surface balances the integrated radiative cooling rate \citep{Robe1996}. \add{Note that, while we retain the term RCE for consistency with the corresponding literature, it is somewhat of a misnomer, as the state of RCE is far from thermodynamic equilibrium. Rather, RCE is a non-equilibrium, statistically steady, but turbulent, state, involving the continuous irreversible production of entropy.} RCE is similar to the canonical fluid mechanics problem of Rayleigh-B\'enard convection between two plates, but the upper plate is absent and replaced by bulk cooling of the fluid. 

RCE provides a starting point for thinking about vertical heat transport in an atmosphere without horizontal variations. The first studies to calculate RCE solutions used it as a model for the tropical-mean or global-mean climate \citep{Manabe1964,Manabe1967}. While recent studies have constructed analytical approximations for mean temperature and humidity profiles in RCE \citep{Romps2014}, solutions to the full turbulent cloud field are only accessible through numerical models. But with increased availability of computational resources, RCE has become a popular numerical and theoretical framework for studying the dynamics of moist convection \citep[e.g.,][]{Held1993,Bretherton2005,Wing2014,Singh2014,Tompkins1998}, and in particular, for developing and testing theories of moist convective updraft velocities \citep[e.g.,][]{Robe1996,Pauluis2002,Singh2013,Seeley2015,Singh2015}. Recent work has also considered the problem of rotating RCE, in which the atmosphere is assumed to exist on a planet with a finite rotation rate \citep[e.g.,][]{Nolanetal2007,KhairoutdinovEmanuel2013}. In this section, we will consider only nonrotating RCE, but the rotating case will be relevant to the discussion of tropical cyclones in section \ref{sec:TCs}.

A disadvantage of RCE is that it does not exist on Earth, and so there are no direct observations with which to compare numerical solutions. Our discussion in this section will therefore remain theoretical and modeling based, but we will discuss estimates of the entropy budget of Earth's atmosphere in section \ref{sec:global_atmosphere}.


\begin{figure*}
\centering
\includegraphics[width=14cm]{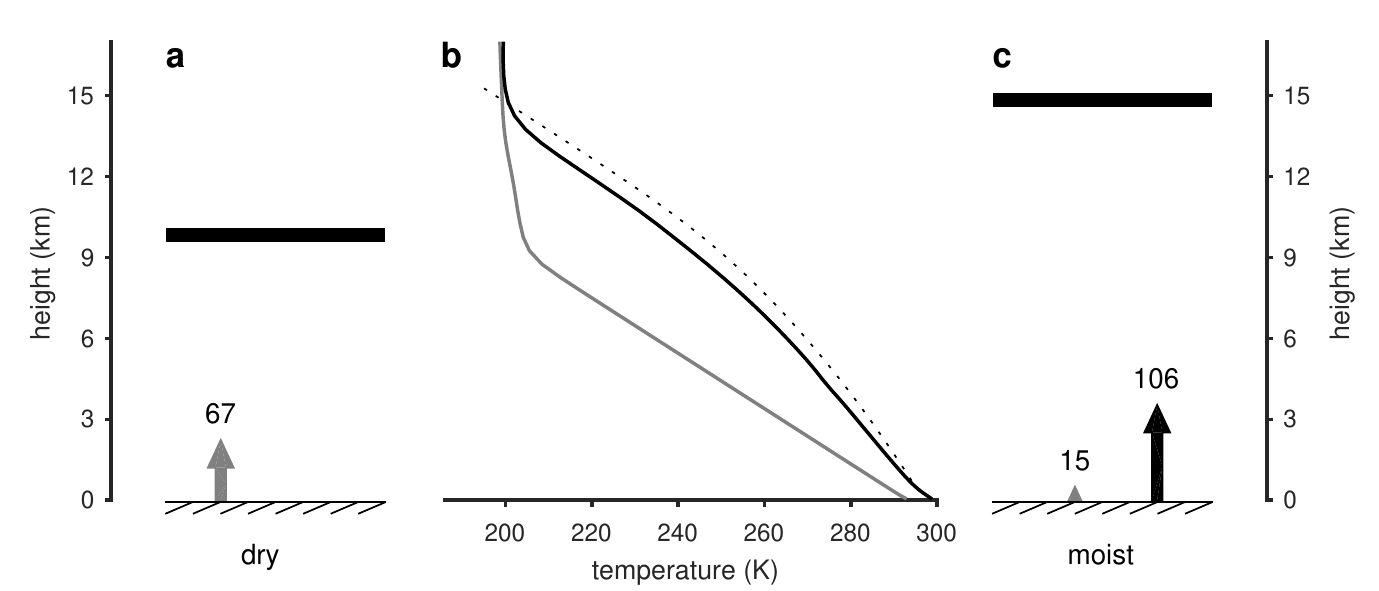}
\caption{Schematic of (a) dry and (c) moist radiative-convective equilibrium. Horizontal black lines denote the tropopause in each case and vertical arrows show the mean sensible (gray) and latent (black) heat fluxes into the atmosphere in W m$^{-2}$. Panel (b) shows mean temperature profiles in the dry (gray) and moist (black) cases, and temperature of a parcel initialized with the mean temperature and humidity of the lowest model level of the moist simulation and lifted adiabatically while assuming all condensate is immediately removed from the parcel by precipitation (dotted). Mean quantities are calculated as horizontal averages over the domain and over days 90-100 in each simulation.}
		\label{fig:RCE}
\end{figure*}

\subsubsection{Dry radiative-convective equilibrium}
\label{sec:dryRCE}

While our primary interest is in understanding moist convection, it is instructive to begin by considering the simpler case of ``dry'' convection, in which the phase change of water plays no role in the dynamics, latent heating is absent, and the enthalpy transport from the surface is entirely composed of sensible heat fluxes.
We further simplify the problem by assuming that the radiative cooling rate may be expressed as a function of the temperature,
\begin{equation}
    \dot{q}_\text{rad} = -\frac{c_{pd}(T - T_0)}{\tau},
    \label{eq:qrad}
\end{equation}
where $c_{pd}$ is the specific heat capacity of dry air, $T_0 = 200$ K, and $\tau = 40$ days \citep{Pauluis2002}.  


To demonstrate the RCE state, we ran a simulation of the above configuration using a \add{``cloud-permitting''} model of the atmosphere. \add{Cloud-permitting models are numerical models with grid spacings of $\mathcal{O}$(1 km), allowing 
them to explicitly represent convective clouds on the model grid. \add{Though more highly resolved} than global climate models, cloud-permitting models remain too coarse to properly \emph{resolve} all but the largest cloud systems, and they must rely on parameterizations for turbulence and other subgrid-scale processes.
We discuss the challenges in modeling the second law using cloud-permitting models further in section \ref{sec:CRMs}}.



Our RCE simulation was run with the Bryan Cloud Model \citep[CM1 version 13;][]{Bryan2002} with the lower boundary fixed at a temperature of 301.5 K and without moisture. The domain was $200\times200$ km$^2$ horizontally, with periodic boundary conditions in both $x$ and $y$ dimensions, and with a horizontal grid spacing of 500 m and 64 vertical levels. The model was run for 100 days, and statistics were accumulated over days 90-100 representing a statistically steady state. Further details of the model configuration are as specified in \citet{Singh2013}. 

A schematic of the dry RCE simulation is presented in Fig. \ref{fig:RCE}a. The radiative cooling \eqref{eq:qrad} produces a troposphere with a depth of roughly 10 km in which vigorous convection occurs and the domain-mean temperature decreases with height at a rate close to the dry-adiabatic lapse rate\footnote{This is the rate at which temperature decreases with height as an air parcel is lifted adiabatically and without phase change.} ($\sim 10 $ K km$^{-1}$). Above this level, radiative cooling is not sufficient to cause convection, and only weak overturning is present. 

\begin{figure*}
\centering
\includegraphics[width=12cm]{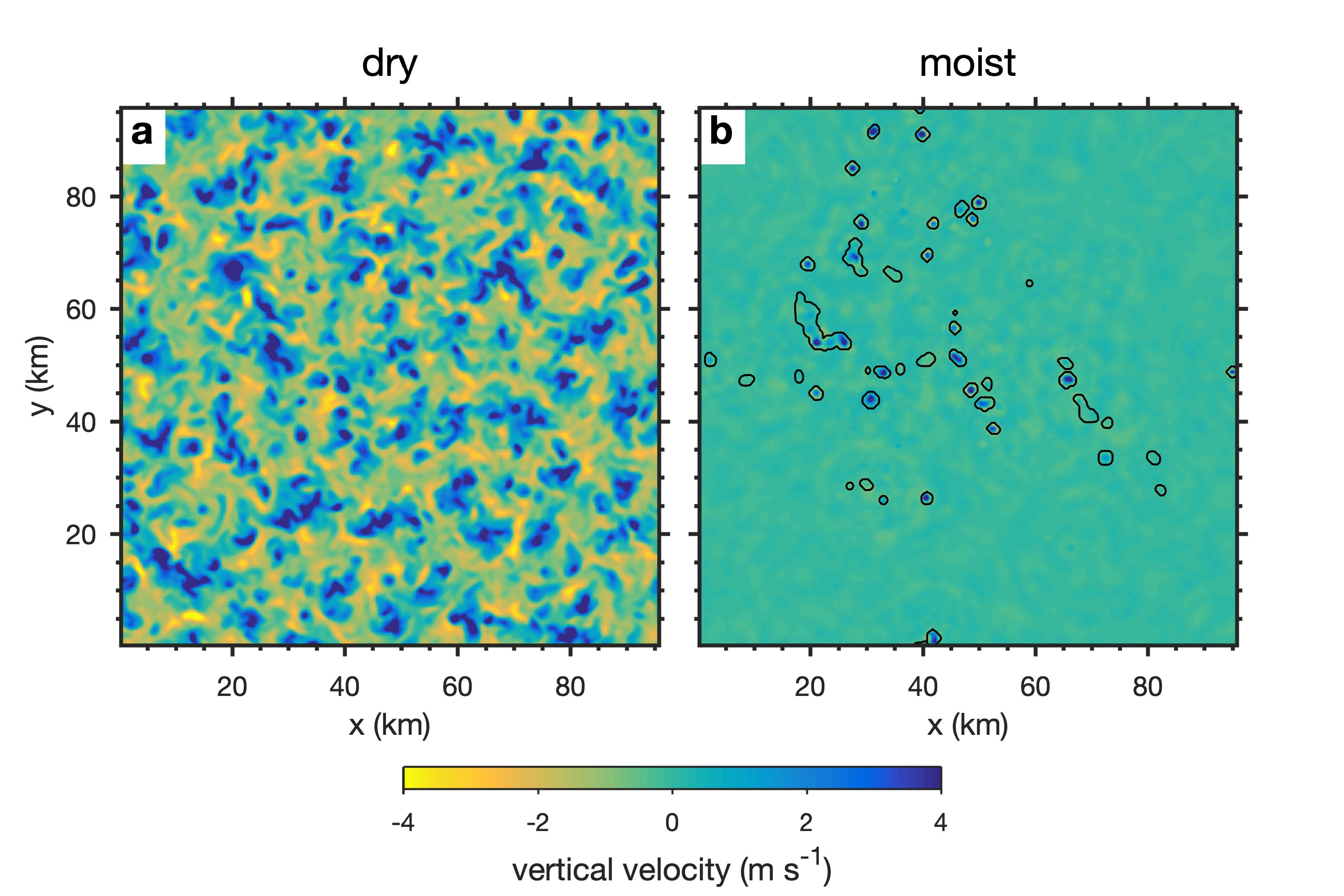}
\caption{Snapshot of vertical velocity at a height of 4 km and at day 90 in simulations of (a) dry and (b) moist radiative-convective equilibrium. Black contour in (b) shows regions within clouds (cloud water content greater than 0.01 g kg$^{-1}$).}
		\label{fig:w_snapshot}
\end{figure*}

Fig. \ref{fig:w_snapshot}a shows a snapshot of the vertical velocity distribution at 4 km; updrafts and downdrafts fill the domain, with magnitudes in the range 2-4 m s$^{-1}$.  What sets the magnitude of these updrafts, and how does it depend on the radiative cooling rate? The entropy budget provides a useful perspective.

The steady-state entropy budget for the dry RCE case may be written (section \ref{sec:irr_procs}),
    \begin{linenomath}\begin{equation*}
        \tavg{\dot{Q}_\text{rad}} \left(\frac{1}{T_a} - \frac{1}{T_s} \right) = \tavg{\dot{S}^\text{fric}_i} + \tavg{\dot{S}^\text{heat}_i},
    \end{equation*}\end{linenomath}
where $\dot{Q}_\text{rad} = A^{-1}\int_{\Omega_A} \rho\dot{q}_\text{rad}\, dV$ is the integrated radiative heating rate, expressed in units of W m$^{-2}$ by dividing by the area of the domain $A$, \add{the angle brackets refer to a time average,} $T_s$ is the surface temperature, and $T_a$ is the effective \add{temperature at which the atmosphere is cooled by radiation}, defined,
\begin{equation}
    \frac{\tavg{\dot{Q}_\text{rad}}}{T_a} =\frac{1}{A} \int_{\Omega_A} \tavgb{\frac{\rho\dot{q}_\text{rad}}{T} }\,  dV.
    \label{eq:eff}
\end{equation}
In the atmosphere, we expect frictional dissipation to dominate over molecular heat diffusion as an irreversible entropy source except in a molecular boundary layer immediately adjacent to the surface \citep{Pauluis2002,SinghOGorman2016}. The entropy production owing to molecular heat diffusion may therefore be neglected entirely if we replace the surface temperature $T_s$ with the temperature of the air above this molecular boundary layer, $T_{sa}$ \citep{Romps2008}. The entropy budget of the dry RCE state then becomes,
    \begin{equation}
        \tavg{\dot{Q}_\text{rad}} \left(\frac{1}{T_{a}} - \frac{1}{T_{sa}} \right) = \tavg{\dot{S}^\text{fric}_i}.
        \label{eq:ent_bud_dry_2}
    \end{equation}
Identifying $T_{sa}$ as the input temperature, $T_{a}$ as the output temperature, and $|\tavg{\dot{Q}_\text{rad}}|$ as the heat input, \eqref{eq:ent_bud_dry_2} may be used to define the Carnot efficiency (section \ref{sec:carnot}) and mechanical efficiency (section \ref{sec:mech}) of convection in RCE.

In a statistically steady state, we expect the rate of generation of kinetic energy by the atmosphere to equal its dissipation. We may write this balance,
\begin{equation}
    \tavg{\dot{W}_K} = T_\text{fric}\tavg{\dot{S}^\text{fric}_i},
    \label{eq:ke_rel}
\end{equation}
where $T_\text{fric}$ is the effective temperature of frictional dissipation defined \add{as in \eqref{eq:Tfric}}.
The left-hand side of \eqref{eq:ke_rel} gives the work done by the pressure gradient force in producing kinetic energy of the winds, and the right-hand side gives the frictional dissipation rate.

Combining the entropy budget \eqref{eq:ent_bud_dry_2} and the mechanical energy budget \eqref{eq:ke_rel} allows one to derive an estimate of the vigor of convection (as measured by its rate of work) given the heat input $|\tavg{\dot{Q}_\text{rad}}|$ and estimates of the effective temperatures $T_{sa}$, $T_a$, and $T_\text{fric}$ \citep{Emanuel2001},
\begin{equation}
    \tavg{\dot{W}_K} = \tavg{\dot{Q}_\text{rad}}T_\text{fric} \left(\frac{1}{T_a} - \frac{1}{T_{sa}} \right).
    \label{eq:work_dry}
\end{equation}
The pressure-work $\dot{W}_K$ may be related, using hydrostatic balance, to the upward buoyancy flux, which, in turn, may be used to derive a scale for the vertical velocity \citep[see e.g.,][]{Emanuel1994}. For dry RCE, the mechanical efficiency is close to its maximum value, and, \add{for fixed values of the temperatures $T_{sa}$, $T_{a}$, and $T_\text{fric}$, \eqref{eq:work_dry} gives that} the rate of work done by the convective heat engine scales linearly with the heat input $|\tavg{\dot{Q}_\text{rad}}|$.

\subsubsection{Moist radiative-convective equilibrium}
\label{sec:moistRCE}

Let us now consider the case of moist RCE. The dry RCE simulation is rerun, but with the lower boundary condition assumed to be a saturated surface of water (Fig. \ref{fig:RCE}c). The enthalpy flux from the surface now includes evaporative fluxes of water vapor in addition to sensible heat, and the transport and interaction of the three phases of water in the atmosphere is accounted for. \addr{Once steady state is reached}, the precipitation rate through the lower boundary is equal to the rate at which water vapor is evaporated into the atmosphere. 

The simulation highlights a number of important differences between dry and moist RCE:
\begin{itemize}
    \item In moist RCE, rising air parcels rapidly become saturated, leading to condensation and latent heating. As a result, the temperature lapse rate in moist RCE is reduced from dry adiabatic to nearly moist adiabatic\footnote{The moist adiabatic lapse rate is the rate at which temperature decreases with height for a saturated air parcel lifted adiabatically \add{accounting for the latent heat released by condensation.}} \citep[Fig. \ref{fig:RCE}b;][]{Singh2013}.
    \item The irreversible fallout of precipitation reduces the water content of air and allows descending air parcels to be unsaturated. This leads to an asymmetry in upward motion; moist convection favors narrow rapid updrafts and broad weak descent \citep[Fig. \ref{fig:w_snapshot}b;][]{Bjerknes1938}.
    \item  Turbulence in moist RCE is weaker than in the dry case \citep{Pauluis2002}; in our simulations, the mid-tropospheric vertical velocity distribution is substantially narrower in moist RCE compared to dry RCE (Fig. \ref{fig:w_pdf}), despite the fact that the moist case has a larger heat input $|\tavg{\dot{Q}_\text{rad}}|$ (see Fig. \ref{fig:RCE}).
    \item Finally, \citet{Robe1996} showed that cloud updraft velocities in moist RCE are virtually independent of the heat input $|\tavg{\dot{Q}_\text{rad}}|$ \citep[see also][]{Craig1996}, in contrast to the case of dry RCE, for which updraft velocities increase with the heat input \citep{Emanuel1994}. In moist RCE, increased radiative cooling is balanced by an increased area fraction of clouds rather than any change in their updraft velocities.
\end{itemize}
These differences suggest that the effects of moisture fundamentally change the dynamics of moist convection compared to its dry counterpart. A theory of moist convection must therefore account for these differences; it is the search for such a theory toward which we now turn.

\subsection{Theories of moist convection}
\label{sec:convtheory}

\subsubsection{The moist convective heat engine}

\add{\citet{Emanuel1996} and \citet{Renno1996} attempted to use the entropy budget to derive a theory for moist convective updraft velocties. As we shall see below, however, both theories are limited in their utility because they do not properly account for the irreversible entropy production associated with moist processes.}

\add{The theory of \citet{Emanuel1996} focuses} on the integrated vertical buoyancy flux associated with convection, given by,
\begin{linenomath}\begin{equation*}
   \tavg{F_b} = \frac{1}{A}\int_{\Omega_A} \tavgb{\rho w b} \, dV,
\end{equation*}\end{linenomath}
where $w$ is the vertical velocity of the fluid, and \begin{linenomath}\begin{equation*}
   b = -{g_{\earth}} \left(\frac{\rho - \overline{\rho}}{\overline{\rho}}\right)    
\end{equation*}\end{linenomath}
 is the buoyancy, defined using the horizontal mean density $\overline{\rho}$. If the pressure field is in approximate hydrostatic balance \add{with the mean density}, the integrated buoyancy flux is approximately equal to the rate at which the atmosphere performs work in order to generate the kinetic energy of the winds $\tavg{F_b} \approx \tavg{\dot{W}_K}$ \citep[e.g.,][]{Romps2008}. \add{\citet{SinghOGorman2016} show that this is a very good approximation in simulations of RCE.} 
 
 \citet{Emanuel1996} assumed that the \add{buoyancy flux at a given level could be approximated by the buoyancy of an air parcel lifted adiabatically from the subcloud layer multiplied by the total cloud mass flux. Integrating vertically, this allows the kinetic energy generation rate to be written,}
 \begin{linenomath}\begin{equation*}
      \tavg{\dot{W}_K} \approx \tavg{F_b}  \approx \tavg{M_c} (\text{CAPE}),
 \end{equation*}\end{linenomath}
where $M_c$ is the upward cloud mass flux from the subcloud layer, and CAPE is the convective available potential energy, defined as the kinetic energy produced by buoyancy forces as an air parcel of unit mass rises adiabatically to the tropopause (the temperature of such a parcel is shown in the dotted line on Fig. \ref{fig:RCE}.). Further assuming that frictional dissipation is the dominant source of irreversible entropy production in the atmosphere, the above equation may be combined with the entropy budget to give a scaling relation for the CAPE,
\begin{equation}
    \text{CAPE} = \frac{\tavg{\dot{Q}_\text{rad}}T_\text{fric}}{\tavg{M_c}}\left(\frac{1}{T_a} - \frac{1}{T_{sa}}\right).
    \label{eq:Emanuel_scaling}
\end{equation}
This provides a velocity scaling for cloud updrafts given by $w\sim\sqrt{2\text{CAPE}}$. A similar scaling was derived by \citet{Renno1996} based on a heat engine analysis of an air parcel completing a cycle rising through a cloud and descending through the environment. Since the cloud mass flux $\tavg{M_c}$ scales with the integrated radiative cooling rate $\tavg{\dot{Q}_\text{rad}}$ in RCE, \eqref{eq:Emanuel_scaling} implies that the cloud updraft velocity is only weakly dependent on $\tavg{\dot{Q}_\text{rad}}$, as seen in numerical simulations \citep{Robe1996}. 

\add{The flaw in the theories of \citet{Emanuel1996} and \citet{Renno1996}} is that they assume that the dominant irreversible entropy production mechanism in the atmosphere is that of frictional dissipation, thereby neglecting irreversible moist processes. As we show below, this is a poor assumption for moist RCE, and it is almost certainly a poor assumption for convection on Earth. 
A full theory for moist convection based on the entropy budget requires careful consideration of the irreversible entropy production associated with moist processes.

\begin{figure}
\centering
\includegraphics[width=8.6cm]{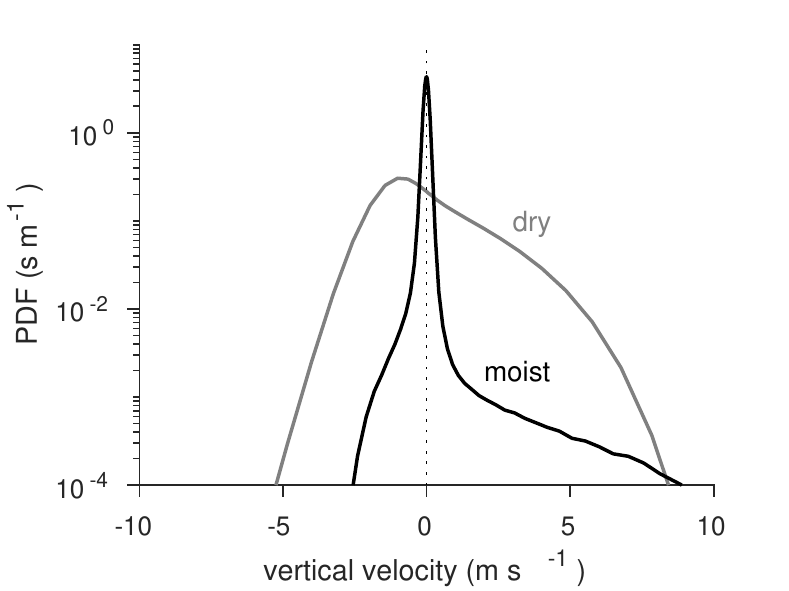}
\caption{Empirical probability distribution functions (PDFs) of the vertical velocity at a height of 4 km in simulations of dry (gray) and moist (black) RCE. PDFs are constructed from hourly snapshots from days 90-100 of each simulation. }
		\label{fig:w_pdf}
\end{figure}


\subsubsection{Moist irreversible processes in radiative-convective equilibrium}

\citet{Pauluis2002} were the first to provide a comprehensive estimate of the material entropy budget of both dry and moist RCE. The authors used simulations of RCE with a two-dimensional cloud-permitting model in which only the liquid-vapor phase transition was considered to provide a detailed evaluation of the relative importance of frictional dissipation compared to other irreversible sources of entropy. Since then, \citet{Romps2008} and \citet{SinghOGorman2016} have confirmed these results using three-dimensional simulations that include the ice phase. 

Table \ref{table:RCE_budget} shows an example of the entropy budget in a simulation of moist RCE over a liquid water surface held at a temperature of 301.5 K. The simulation differs from the examples shown in Figs. \ref{fig:RCE}-\ref{fig:w_pdf} in that it allows for the interactive effects of radiation by including an explicit parameterization of radiative transfer.  The simulation is identical to the control simulation in \citet{SinghOGorman2016} except that it is run with a horizontal grid spacing of 2 km, and on a doubly periodic domain $288\times288$ km$^2$ in size. We include interactive radiation and use a larger domain in order to allow the simulation to undergo the phenomenon of ``convective self-aggregation'', which is described in the next subsection.

The key result of \citet{Pauluis2002}, also evident in table \ref{table:RCE_budget}, is that the entropy source associated with frictional dissipation $\tavg{\dot{S}_i^\text{fric}}$ is a relatively small component of the irreversible entropy production of RCE, accounting for \add{less than $\sim$ 15\% of the total}. This is in stark contrast to dry RCE, in which frictional dissipation is the main mechanism by which entropy is produced irreversibly.

\begin{table*}
\renewcommand{\arraystretch}{1.2}
\begin{center}
\begin{tabular}{ l r c r r  } \hline\hline \vspace{3pt}
  & symbol & units & disaggregated & aggregated  \\ \hline
 entropy budget &         &          &   \\
 \hspace{10pt} export & $-\tavg{\dot{S}^\text{mat}_e}$ & mW m$^{-2}$ K$^{-1}$& $37.4 \pm 2.1$ & $36.5 \pm 3.4$ \\
 \hspace{10pt} irreversible source & $\tavg{\dot{S}^\text{mat}_{i}}$ & mW m$^{-2}$ K$^{-1}$ & $37.7  \pm 0.2$ & $34.9 \pm 0.2$ \\
 \hspace{20pt} heat diffusion & $\tavg{\dot{S}_i^\text{heat}}$& mW m$^{-2}$ K$^{-1}$ & $-0.24 \pm 0.0$ & $-0.12 \pm 0.0$ \\
 \hspace{20pt}  frictional dissipation & $\tavg{\dot{S}_i^\text{fric}}$& mW m$^{-2}$ K$^{-1}$& $6.0 \pm 0.1$ & $2.4 \pm 0.1$ \\
 \hspace{20pt} hydrometeor sedimentation & $\tavg{\dot{S}_i^\text{sed}}$& mW m$^{-2}$ K$^{-1}$& $11.0 \pm 0.1$ & $9.1 \pm 0.1$ \\
 \hspace{20pt} mixing \& phase change                 & $\tavg{\dot{S}_i^\text{mem}}$& mW m$^{-2}$ K$^{-1}$& $21.0 \pm 0.1$ & $23.5 \pm 0.1$ \\
 radiative cooling rate & $-\tavg{\dot{Q}_\text{rad}}$ & W m$^{-2}$ & $106.4 \pm 0.5$ & $122.6 \pm 0.8$ \\
 mechanical efficiency & $\eta_M$ & \% & 1.5 & 0.5 \\
 Carnot efficiency & $\eta_C$ & \% & 9.2 & 7.5 \\

 precipitation efficiency & $\epsilon$ & \% & 25 & 49 \\
 \hline 
\end{tabular}
\end{center}
\caption{Entropy budget and other statistics calculated from a simulation of moist radiative-convective equilibrium with interactive radiation. Simulation is run following the configuration of \citet{SinghOGorman2016},  but on a $288\times288$ km horizontal domain and with a horizontal grid spacing of 2 km. Lower boundary condition is an ocean surface held fixed at 301.5 K, and simulation is initialized from a state of rest. ``Disaggregated'' corresponds to mean over day 20-40 of the simulation, when convection remains scattered, and humidity variations across the domain are weak (Fig. \ref{fig:CWV_snapshot}a). ``Aggregated'' corresponds to a mean over day 150-230 of the simulation, when the domain consists of a single moist region containing convection surrounded by a dry region devoid of clouds (Fig. \ref{fig:CWV_snapshot}b). Uncertainties associated with estimating the steady-state budget using a finite timeseries are quantified using a block-bootstrap method and given as the 90$^\text{th}$ percentile confidence interval.}
		\label{table:RCE_budget}

\end{table*}

The physical reason for the small value of $\tavg{\dot{S}_i^\text{fric}}$ is that moist RCE includes a number of additional irreversible sources of entropy, including entropy production associated with the sedimentation flux of precipitation $\tavg{\dot{S}_i^\text{sed}}$ \citep{Pauluis2000} and that associated with \add{the mixing of water vapor and dry air} $\tavg{\dot{S}_i^\text{mix}}$ and irreversible phase changes $\tavg{\dot{S}_i^\text{evap}}$ and $\tavg{\dot{S}_i^\text{melt}}$ \citep{Pauluis2002}, which we have combined in table \ref{table:RCE_budget} into a single term $\tavg{\dot{S}_i^\text{mem}}$,
\begin{linenomath}\begin{equation*}
    \tavg{\dot{S}_i^\text{mem}} = \tavg{\dot{S}_i^\text{mix} + \dot{S}_i^\text{evap} + \dot{S}_i^\text{melt}}.
\end{equation*}\end{linenomath}
The sum of all irreversible entropy sources must, in steady state, balance the total radiative sink of entropy  $-\tavg{\dot{S}_e^\text{mat}}$. The additional sources of entropy associated with phase change, mixing, and precipitation sedimentation must therefore occur at the expense of entropy production associated with frictional dissipation. In fact, the sources $\tavg{\dot{S}_i^\text{mem}}$ and $\tavg{\dot{S}_i^\text{sed}}$  are the two largest irreversible entropy sources in RCE, accounting for the vast majority of the irreversible entropy production (table \ref{table:RCE_budget}). 

\add{Before discussing the implications of the above results, we briefly mention some caveats regarding the interpretation of our numerically simulated entropy budget.}
\add{In most models, spurious numerical sources and sinks of entropy exist which can induce errors in the budget. 
Moreover,} irreversible molecular processes that produce entropy in the atmosphere are not \add{explicitly} resolved in cloud-permitting simulations, \add{and the parameterizations that attempt to account for their effects}
do not always faithfully represent the second law of thermodynamics. For example, the entropy production associated with heat diffusion is small but negative in our simulation of RCE, appearing to violate the second law. We further discuss the reasons for such apparently unphysical entropy sinks, and other issues relating to accurately modeling the entropy budget, in section \ref{sec:models}. 

\add{Despite the above caveats, 
the overall dominance of moist irreversible processes and the small value of the entropy production owing to frictional dissipation found in our simulation of RCE is consistent with previous numerical \citep{Pauluis2002,Romps2008} and theoretical  \citep{Pauluis2002,Goody2003} analyses and is therefore likely to be robust.}
Since, in steady state, the frictional dissipation of winds in the atmosphere must balance their generation by mechanical work, \add{this implies} that moist convection has a low mechanical efficiency
$\eta_M$ compared to dry convection and compared to the efficiency of an ideal Carnot heat engine (table \ref{table:RCE_budget}). Here we define the mechanical efficiency similarly to \eqref{eq:mech_eff}, with $\tavg{\dot{Q}_\text{in}} = |\tavg{\dot{Q}_\text{rad}}|$ the radiative cooling rate, so that,
\begin{linenomath}\begin{equation*}
    \eta_M = \frac{\tavg{\dot{W}_K}}{|\tavg{\dot{Q}_\text{rad}}|}.
\end{equation*}\end{linenomath}

\citet{Pauluis2002} performed a nondimensional analysis of their RCE simulations to show that this mechanical efficiency depends on the relative importance of latent heat transport compared to sensible heat transport and the relative importance of the work done by water vapor compared to the total work performed by moist convection. In moist RCE at temperatures characteristic of Earth's tropics, latent heat transport is the dominant vertical heat transport mechanism, and water vapor expansion accounts for a substantial fraction of the total work. Both of these factors reduce the efficiency of moist convection. 

\citet{Romps2008} pointed out that irreversible melting and freezing further contributes to the low mechanical efficiency of moist convection. He also noted that the reduced lapse rate in moist convection compared to dry convection (Fig. \ref{fig:RCE}b) results in a smaller temperature difference between the input and output temperatures given the same radiative cooling profile. \add{For given vertical profile of $\dot{q}_\text{rad}$,} the moist convective heat engine therefore has a lower Carnot efficiency than its dry convective counterpart.



An alternative perspective on the reduced mechanical efficiency of moist convection compared to dry convection is to consider the atmosphere as a combination heat engine and steam cycle. As described by \citet{Pauluis2011}, the thermodynamic action of atmospheric convection is to transport heat vertically in the atmosphere, but also to dehumidify the atmosphere. These two thermodynamic operations are in competition with each other, such that the dehumidification process reduces the work available for the atmospheric heat engine. The magnitude of the reduction in available work associated with this dehumidification depends on the relative importance of latent heat transport compared to sensible heat transport, and on the relative humidity at which the mixed-cycle engine operates.

The implications of the small mechanical efficiency of moist convection are profound. Firstly, it provides an explanation for the reduced kinetic energy of moist convection relative to dry convection as highlighted in the snapshots in Fig. \ref{fig:w_snapshot} and the vertical velocity distributions in Fig. \ref{fig:w_pdf}. Furthermore, a small mechanical efficiency is inconsistent with the theories of \citet{Emanuel1996} and \citet{Renno1996}; such theories are predicated on the dominance of entropy production associated with frictional dissipation within the entropy budget, and they cannot account for the case in which the entropy budget is dominated by moist processes. Finally, the presence of moist irreversible sources of entropy means that the entropy \addr{sink $-\tavg{\dot{S}_e^\text{mat}}$} no longer places a \addr{direct} constraint on the work done by atmospheric convection; a change in the heat input $|\tavg{\dot{Q}_\text{rad}}|$ or its effective temperature $T_{a}$ may be balanced by changes in the entropy sources associated with moisture, rather than those associated with frictional dissipation.

\subsubsection{The role of mixing and microphysics}
\label{sec:updrafts}


Despite the challenges described above, the entropy budget may nevertheless provide guidance toward a theory of moist convective intensity. In particular, \citet{Pauluis2002} argued that the importance of the moist processes for the entropy budget implies that moist convective updraft velocities may depend strongly on the effects of condensate on the buoyancy of air and ultimately on the microphysical processes that determine cloud and precipitation formation. Additionally, the importance of vapor diffusion and irreversible phase change in the entropy budget potentially suggests that the mixing of cloudy and non-cloudy air parcels may be of central importance to any theory for convective vigor in moist RCE. 

Indeed, current theories for moist convective updraft velocities suggest that they are limited by the sedimentation velocity of precipitation \citep{Parodi2009}, or that they are determined by the efficiency of mixing between clouds and their environment \citep{Singh2013,Seeley2015,Singh2015}. In the latter case, it is argued that the import of subsaturated air from the environment allows the profile of temperature in moist RCE to decrease with height faster than that of a moist adiabat, thereby allowing for finite CAPE. At present, estimates of the rate at which this import of subsaturated air occurs must rely on detailed simulations with high-resolution models such as those reported above. Such simulations do not resolve the molecular mixing processes directly, but rely on turbulence closures whose validity in the vicinity of clouds is difficult to establish. Whether the entropy budget may be used to constrain the rate at which cloudy and non-cloudy air is mixed is a potentially important area of future work.

\begin{figure*}
\centering
\includegraphics[width=12cm]{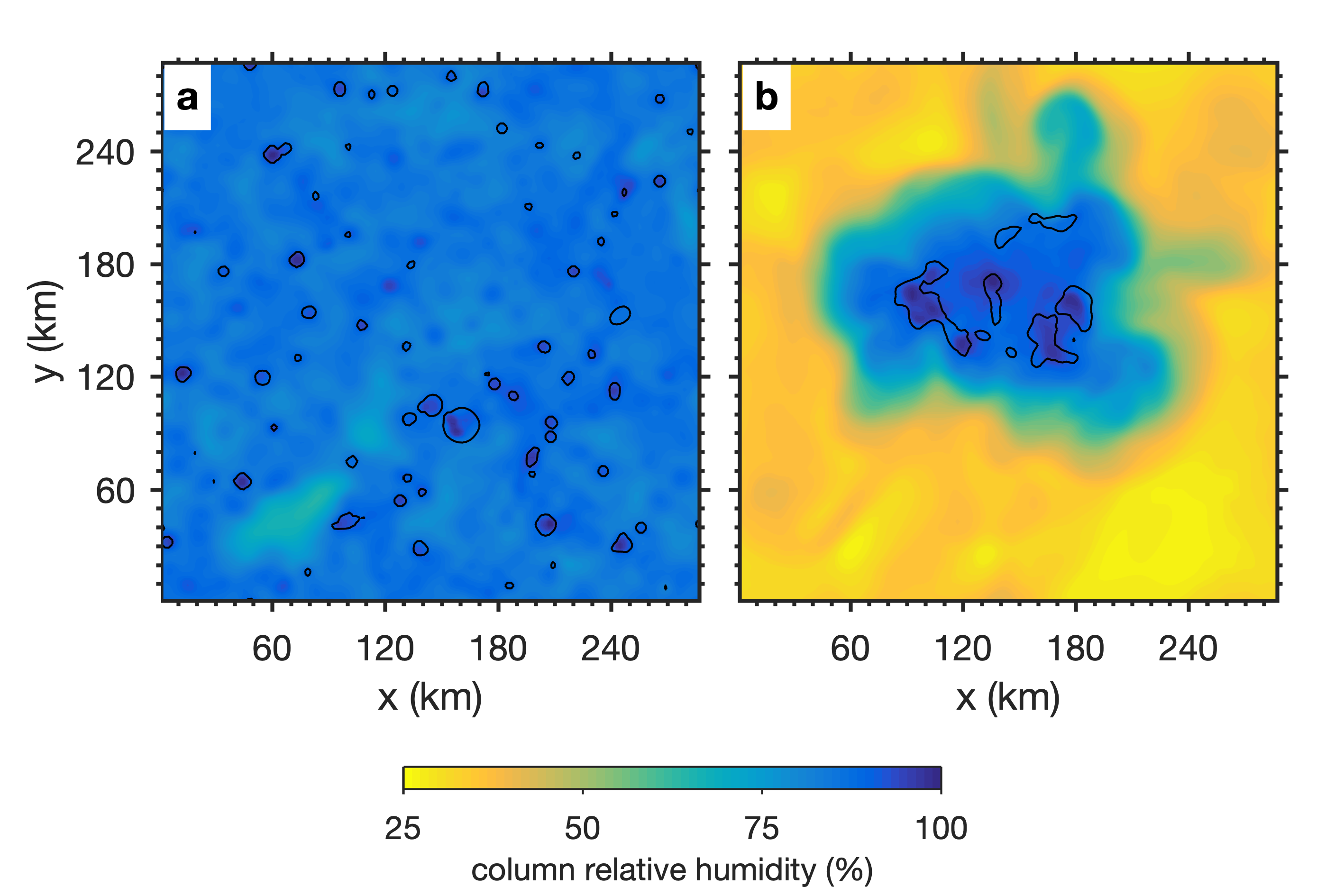}
\caption{Snapshots of column relative humidity (colors) and contour of column cloud water content of 0.5 kg m$^{-2}$ (black) in a simulation of moist RCE with interactive radiation at (a) day 30 and (b) day 230. Simulation is run following the configuration of \citet{SinghOGorman2016},  but on a $288\times288$ km horizontal domain and with a horizontal grid spacing of 2 km. Lower boundary condition is an ocean surface held fixed at 301.5 K, and simulation is initialized from a state of rest.}
		\label{fig:CWV_snapshot}
\end{figure*}

\subsection{Convective organization and the mechanical efficiency of moist convection}
\label{sec:org}

Previous studies of the entropy budget of RCE have been limited to cases of disorganized convection,  characterized by a statistical equilibrium in which convective clouds grow and decay quasi-randomly within the domain \citep[Fig. \ref{fig:w_snapshot};][]{Pauluis2002,Romps2008,SinghOGorman2016}. However, convection in Earth's atmosphere is often organized into larger-scale structures such as squall lines, mesoscale convective complexes, or tropical cyclones  \citep{Houze1982}. While the heat engine characteristics of tropical cyclones have been the subject of considerable research (see section \ref{sec:TCs}), the broader characteristics of the entropy budget of organized moist convection remain relatively unexplored. In this section, we provide a preliminary analysis of the entropy budget of a particular type of organized convection to highlight the potential for future research in this area.

Under certain conditions, simulations of RCE are known to spontaneously develop organization in a process termed ``self-aggregation''. In the aggregated state, clouds and convective activity become confined to a small region that remains moist, while the rest of the domain is characterized by a dry troposphere and subsiding air \citep[e.g.,][]{Bretherton2005,Wing2014,Emanuel2014}. Studies of convective self-aggregation have shown that it results from feedbacks between clouds, water vapor, and radiation that lead to an instability of the disaggregated state to perturbations in tropospheric humidity \citep{Emanuel2014}. This instability is sensitive to the imposed surface temperature \citep{Wing2014} as well as other details of the model formulation, including the resolution and domain size \citep{Muller2012,Jeevanjee2013}. Given its sensitivity to model formulation, the idealization of the framework of RCE, and the long timescale taken for convection to self-aggregate (see below), the importance of self-aggregation as a mechanism for convective organization on Earth remains debated \citep{Jakob2019}. It is nevertheless of interest to understand how the thermodynamic characteristics of the aggregated state differ from those of the disaggregated state. For example, does self-aggregation increase or decrease the mechanical efficiency of moist convection?

The simulation of moist RCE described in the previous section initially develops a state in which clouds occur randomly and relatively uniformly throughout the domain (Fig. \ref{fig:CWV_snapshot}a). After roughly 50 days, the simulation begins to aggregate, and after roughly 150 days, a new state in which clouds are clustered into a small region is obtained (Fig. \ref{fig:CWV_snapshot}b). 

We focus our analysis on two quasi-steady periods within this simulation in which convection is disaggregated (days 20-40) and aggregated (days 150-230), respectively. By comparing the entropy budget in these periods, table \ref{table:RCE_budget} presents, for the first time, an analysis of how the entropy budget is affected by convective organization. \add{In each period, the statistics of the flow are relatively steady, and the entropy budget is roughly closed (table \ref{table:RCE_budget}), while in the $\sim100$ day period in between, the process of self aggregation occurs, and we do not expect the statistics to be steady. The initiation of the self-aggregation process therefore limits the length of the period over which we analyze the disaggregated state. However,} the disaggregated state is less variable than the aggregated state, and the uncertainty in the steady-state budget introduced by such variability is relatively small in both \add{the aggregated and disaggregated} cases (table \ref{table:RCE_budget}).



In our simulation, convective self-aggregation is associated with a large fractional reduction in the entropy production associated with frictional dissipation, implying a reduction in the rate of work performed by moist convection. Indeed, the mechanical efficiency of moist convection decreases by a factor of three between the aggregated and disaggregated state. The reduction in $\tavg{\dot{S}_i^\text{fric}}$ is balanced by an increase in entropy production associated with mixing and phase change $\tavg{\dot{S}_i^\text{mem}}$ and a slight decrease in the total irreversible entropy production of the atmosphere. The entropy source owing to the sedimentation of precipitation $\tavg{\dot{S}_i^\text{sed}}$ also decreases with aggregation, and this is associated with a doubling of the precipitation efficiency, defined as the ratio of total condensation in the atmosphere to the surface precipitation, in the aggregated state compared to the disaggregated state.

The results above suggest that the RCE atmosphere is a less efficient heat engine when convection is aggregated compared to when it is not, but that the efficiency by which condensation in the atmosphere is converted to precipitation at the surface increases under aggregation.
The reasons for these differences in efficiency are likely to be related to the large-scale reorganization of the humidity distribution that occurs under aggregation. For example, the humidity of the near-cloud environment is likely to be higher when convection is aggregated, contributing to the higher precipitation efficiency of the aggregated state. On the other hand, the variability of humidity within the domain also increases under aggregation \citep{Wing2014}, and this may lead to a larger entropy source associated with vapor diffusion, contributing to the lower mechanical efficiency of aggregated convection.

\add{The idealizations inherent in a simulation of RCE on a finite domain limit the direct applicability of our results to Earth's atmosphere. Previous studies of aggregation in RCE have shown that domain geometry affects the number of aggregated regions produced \citep{Wing2016}, while \citet{Jakob2019} noted that a state of RCE is only observed in the atmosphere on spatial scales substantially larger than our domain. Moreover, a variety of additional mechanisms are present in Earth's atmosphere, including large-scale waves and other large-scale circulations \citep[e.g.,][]{Houze1982}, inhomogeneities in the surface \citep[e.g.,][]{Rieck2014}, and background wind shear \citep[e.g.,][]{Rotunno1988} which may organize convection more efficiently than the mechanisms that cause the $\sim100$-day development of aggregation in our simulation. Further work is required to determine if our RCE results are relevant to these more general forms of convective organization. 
Nevertheless,} we note that the tendency for tropical convection to aggregate has been hypothesized to increase in a warmer climate \citep[see e.g.,][]{Wing2019}. Our results suggest that such an increase in aggregation may have implications for the global atmospheric heat engine under future climate change \citep[see section \ref{sec:future} and][]{Laliberteetal2015}.

\section{The thermodynamics of tropical cyclones}
\label{sec:TCs}

When planetary rotation is included, simulations of RCE spontaneously generate one or more rapidly rotating storms analogous to tropical cyclones (TCs) on Earth \citep{Nolanetal2007,KhairoutdinovEmanuel2013,CarstensWing2020,Ramsayetal2020}.
TCs are stunning examples of organized deep convection. \add{They are variously called hurricanes, typhoons or cyclones, depending on their intensity and the ocean basin in which they occur.
TCs} are characterized by a primary circulation, consisting of a rapidly rotating vortex around a low-pressure center, and a secondary circulation, consisting of a (mostly) thermally direct overturning circulation. The small exception comes from the enigmatic, dry TC eye, in which a thermally indirect flow is produced as buoyant dry air is mechanically forced to descend. Outside the dry central eye, high-entropy moist air ascends in the saturated eyewall until it reaches the upper troposphere, where it spreads out laterally to large radii. At intermediate radii, deep thunderstorms comprise one or more asymmetric rainbands that spiral away from the TC center. Precipitation within the eyewall and the rainbands is a rapid, irreversible sink of rising water mass and occurs primarily within 200 km from the TC center, while radiative cooling slowly removes energy and entropy from the air at larger radii. 

Like all storms, TCs are transient; they make an interesting thermodynamic system in part due to their genesis and death. TCs are open systems, continuously exchanging energy, mass and entropy laterally with the rest of the atmosphere, \add{as well as vertically via fluxes through the surface and TOA} (Fig. \ref{fig:TCdomain}, right side). These characteristics contrast with the idealized RCE simulations discussed above, in which transient effects and lateral fluxes from a larger environment are not considered. In RCE, the atmosphere is an open system only to the extent that water may enter and leave it through surface evaporation and precipitation.


The idealization of TCs as being steady and closed to lateral fluxes (Fig. \ref{fig:TCdomain} left side) has nevertheless been a rewarding model for understanding their basic physics, and we will consider an idealized steady-state TC in a closed domain as a starting point for our discussion. First we review steady-state theory for the potential intensity (highest-achievable surface wind speed, or lowest-achievable central surface pressure deficit) of a TC given the thermodynamic parameters of the environment (section \ref{sec:PItheory}). We next review how the secondary (overturning) circulation of TCs as a whole can be viewed as a combination heat/steam engine, and we discuss recently developed techniques to calculate the integrated work production of a TC (section \ref{sec:TC_work}). Finally, we consider TCs as open systems, capable of genesis and extinction, and mention areas for future research (section \ref{sec:OpenTCs}).

\subsection{Potential intensity theory}\label{sec:PItheory}


It is generally agreed upon that the primary energy source of a TC is the flux of enthalpy from the sea surface \citep{Byers1944,Kleinschmidt1951,MalkusRiehl1960,Riehl1950,Riehl1954,Emanuel1986}\footnote{For a translation of \citet{Kleinschmidt1951} see the appendix of \citet{Gray1994}.}. This flux is driven by the thermodynamic disequilibrium between the sea surface and the subsaturated air immediately above it. Acting like a heat engine, the TC transports energy from the warm surface to the cooler troposphere, producing potentially catastrophic winds in the process. \add{\citet{Riehl1950} first identified the enthalpy disequilibrium as being responsible for the energy flux from the sea, and \citet{Kleinschmidt1951}} provided the first estimate of the maximum wind speed of a TC. That study and a later attempt by \citet{MalkusRiehl1960} laid the groundwork for quantifying TC thermodynamics, preceding the celebrated modern-day potential intensity (PI) theory \citep{Emanuel1986,Emanuel1988,BisterEmanuel1998} for the maximum attainable surface wind speed (or minimum central surface pressure) of an inviscid, axisymmetric TC embedded in a given thermodynamic environment.

\begin{figure*}
\includegraphics[width=16cm]{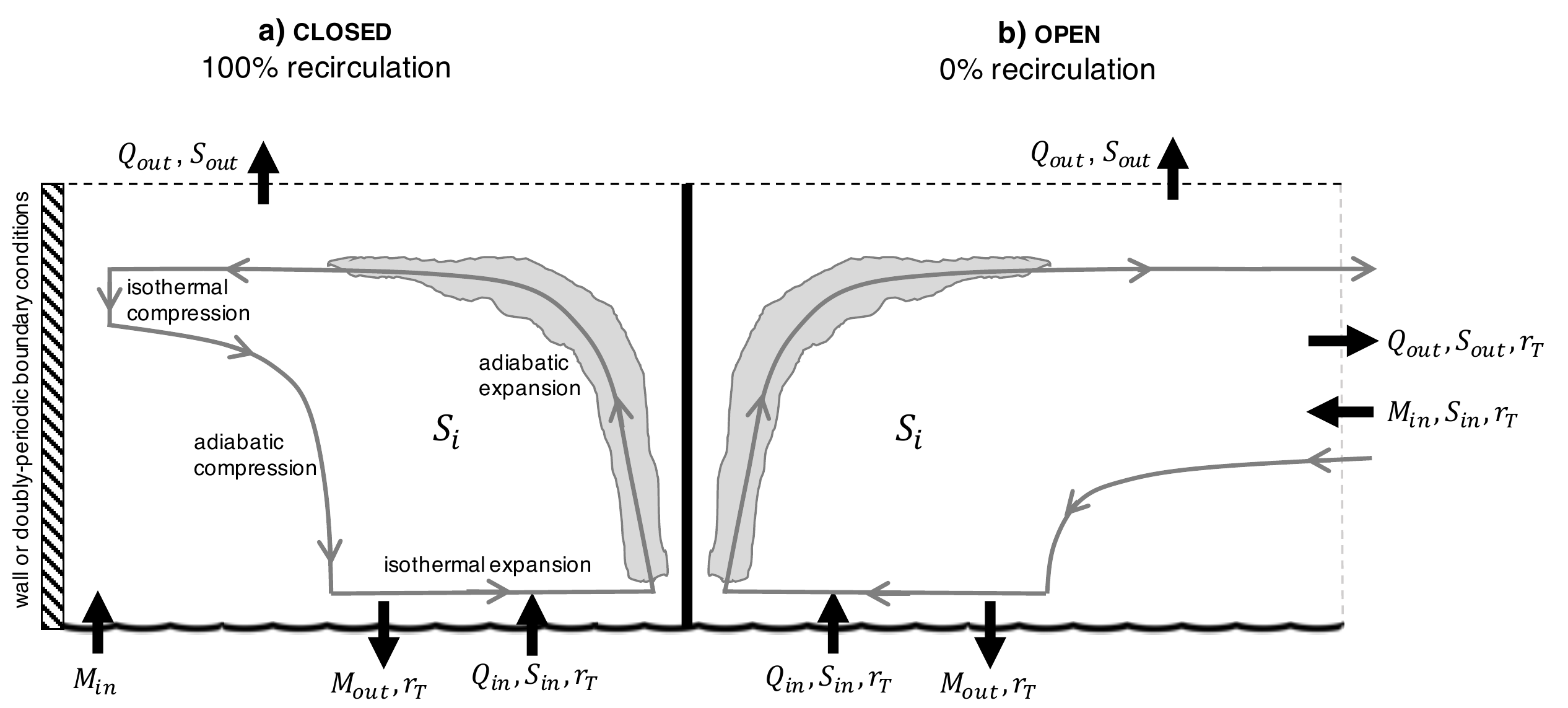}
\caption{A schematic of two idealized tropical cyclone overturning circulations (radial cross-section). Grey arrows indicate dry air mass fluxes. Black arrows indicate \add{boundary fluxes of }heat $Q$, entropy $S$, water content $r_T$, and momentum $M$ into and out of the domain\add{, and arrows indicate the characteristic direction of the fluxes in particular regions}. (a) A closed circulation bounded by an outer wall or doubly periodic boundary conditions\add{, with the idealized Carnot cycle legs indicated}. (b) A TC in lateral contact with the rest of the atmosphere.}
		\label{fig:TCdomain}
\end{figure*}

\citet{Emanuel1986} explicitly likened a TC to a Carnot heat engine, demonstrating that the same analytic result for potential intensity can be derived from either the equations of motion and the first law of thermodynamics, or from consideration of an idealized thermodynamic cycle performed by an air parcel within a TC. 
The thermodynamic cycle consists of four legs as follows (Fig. \ref{fig:TCdomain}): \textit{Isothermal expansion:} boundary layer air that converges toward the low pressure center expands but stays relatively isothermal, heated by sensible heat fluxes from the sea surface. Simultaneously, surface latent heat fluxes dramatically increase the air's moist entropy. \textit{Adiabatic expansion:} frictional convergence of boundary layer air near the surface is balanced by ascent in the deep, cloudy eyewall of the TC. This process is roughly slantwise moist neutral: saturated rising parcels do not ascend strictly vertically but rather along sloping surfaces of constant angular momentum $M$. During sloping ascent away from the TC core, parcels initially approximately conserve their moist entropy $s$ (neglecting the mass loss due to precipitation, which is a small fraction of the air mass), such that surfaces of constant $M$ and $s$ are parallel. \textit{Isothermal compression:} as air is exhausted farther radially over time, it loses energy radiatively as it starts to descend, and finally (\textit{adiabatic compression}) the parcel slowly subsides back to the boundary layer along a vortex line. These last two legs do not occur separately in real TCs; rather, air parcels lose both entropy and \add{energy} as air descends at large radii. However, the artificial separation of the two legs aids in mathematical comparison to a Carnot cycle.


The outer environment restores a parcel's $M$ and $s$ back to environmental conditions, ultimately through contact with the \add{saturated,} frictional sea surface. With assumptions about the storm structure in the free troposphere above the boundary layer, chiefly the assumption of slantwise moist neutrality, one can determine the energetics of the storm needing only to specify the TC's boundary layer radial structure of $s$ as a function of $M$ and environmental factors like surface air temperature $T_{s}$ and outflow air temperature $T_{o}$ \citep{Emanuel1986}. Having specified these conditions, a theoretical closed loop describing a parcel undergoing cyclic changes in pressure, volume, temperature, and entropy that is closely analogous to a heat engine may be defined. This parcel travels the outermost loop of the entire TC circulation, allowing it to experience the largest temperature range and heating/cooling range, and thus achieve the maximum wind speed $v_\text{max}$:
\begin{equation}
v^2_\text{max} = \frac{T_s - T_o}{T_s}\frac{C_k}{C_D}(h^*_{s}-h_{BL}).\label{eq:Vmax}
\end{equation}
This expression is the outcome of PI theory. Note that the first factor on the right-hand side $(T_s-T_o)/T_s$ is identical to the Carnot efficiency of a heat engine operating between thermal reservoirs at temperatures $T_s$ and $T_o$. The constants $C_k$ and $C_D$ are bulk surface exchange coefficients for enthalpy and momentum, respectively. The atmospheric specific enthalpy $h$ here is typically approximated as $h=q_d h_d + q_v h_v \approx c_{pd} (T-T_0) + L_v q_v$ for dry air heat capacity at constant pressure $c_{pd}$, latent heat of vaporization $L_v$, mass fraction of water vapor $q_v$, and reference temperature $T_0$. The relevant enthalpy disequilibrium that drives flux from the ocean to the atmosphere is the difference between the saturated enthalpy of the sea surface $h_{s}^*$ and the enthalpy of the subsaturated boundary layer $h_{BL}$.

According to PI theory, the positive feedback responsible for TC intensification is wind-induced surface heat exchange \citep[WISHE,][]{Emanuel1986}, whereby stronger surface winds induce stronger sea-to-air energy fluxes, which intensify the vortex and lead to stronger surface winds. This feedback is eventually arrested and balanced by frictional drag at the surface. TC structure and PI theory were reviewed by \citet{Emanuel1991,CampMontgomery2001,Emanuel2004,Wang2012}; recently \citet{Emanuel2018} comprehensively reviewed the full breadth of tropical cyclone research.

PI theory was further developed by \citet{Emanuel1988,Emanuel1991} and a complementary approach was provided by \citet{Holland1997}. In these theories, heating due to kinetic energy dissipation was not considered, so the thermodynamic efficiency in (\ref{eq:Vmax}) includes the surface temperature $T_{s}$ in the denominator. \citet{BisterEmanuel1998} showed that, \add{when frictional dissipation is taken into account, the potential intensity is increased, 
with
the outflow temperature $T_{o}$ replacing} $T_s$ in the denominator of \eqref{eq:Vmax}. In numerical models that don't include this frictional heating term, the original formulation is still appropriate \citep[e.g.,][]{CroninChavas2019}. Recent work also shows that the outflow temperature $T_o$ is not a constant \citep{EmanuelRotunno2011,Emanuel2012,Fangetal2019}, so the constant Carnot efficiency-like term in \eqref{eq:Vmax} does not accurately reflect the range of temperatures at which the atmosphere loses heat.
    
The application of heat engine theory to cyclones is even simpler for dry fluids. \citet{Rennoetal1998} undertook a heat engine analysis for nearly-dry dust devils (for which planetary rotation is unimportant), where the driving temperature difference was defined as that between the surface and the top of the boundary layer. \citet{Mrowiecetal2011} showed that the original \citet{Emanuel1986} PI theory, formulated assuming a saturated core neutral to slantwise convection, was also valid for purely dry TCs that only receive sensible heat from the surface (provided that there is a sufficient WISHE-type feedback). This works because PI theory incorporates the role of moisture only in the definition of entropy.

Potential intensity theory is useful for bounding the upper limit of peak TC wind speeds. However, the expression for $v_\text{max}^2$ is not equivalent to a calculation of work produced by the extremely dissipative TC heat engine \citep{Bister2011}. \citet{Hakim2011} demonstrated that, even for the most extremal closed path considered by PI theory, the Carnot cycle analogy is only partially appropriate; while the isothermal expansion and adiabatic expansion legs were decent approximations of the inflowing and ascending/outflowing air, respectively, the return flow was neither isothermal nor adiabatic. And of course, in a highly dissipative moist TC, $\oint T ds$ does not equal the work produced in a cycle, as it would for a reversible Carnot cycle.

A recent development of PI theory considers a ``differential Carnot cycle'' \citep{RousseauRizziEmanuel2019} in which the closed parcel path is taken as an infinitesimally wide region bounding the extremal overturning streamline only for the inflowing and ascending/outflowing branches of the circulation. This approach avoids the somewhat unrealistic description of the TC descending branch as two separate legs of adiabatic compression and isothermal cooling, and it better illustrates that the thermodynamic efficiency of the TC's overall overturning circulation is unaccounted for in the calculation of $v_\text{max}$. For a calculation of TC work and entropy production as a whole, the entire circulation must considered. \add{\citet{Renno2008} proposed that one could predict the pressure drop anywhere within a realistic vortex by taking into account the irreversibility of the vortical heat engine due to the presence of moisture or dust.
Very recent work has shown that, indeed, a TC's core wind structure is closely related to the sources and magnitude of entropy production available in the system \citep{WangLin2020,WangLin2021}.}

\subsection{Work and entropy budgets of TCs}
 \label{sec:TC_work}   
    
Consider a steady-state TC, far from equilibrium with its environment. Let the TC system be defined by a cylinder centered on the TC center, bounded below by the ocean surface and above by the top of the atmosphere (Fig. \ref{fig:TCdomain}). Though a developing TC does grow in volume of dry air at the expense of the environment (as defined perhaps by something like the size of the expanding outflow region), we shall neglect the cyclogenesis/cyclolysis stages and consider a steady-state mature TC with a fixed amount of dry air in RCE. 



To evaluate the work performed by TCs in RCE, we apply a procedure of isentropic averaging \add{to analyze the overturning circulation of an axisymmetric TC simulation described in \citet{ONeillChavas2020}}
\citep[see also][]{Rossby1937,PauluisMrowiec2013,Mrowiecetal2016}. To do so, we introduce the potential temperature
\begin{equation}
\theta=T (p_\text{ref}/p)^{R_d/c_{pd}}\label{eq:theta}
\end{equation}
where $p_\text{ref}=1000$~hPa is a reference pressure. The potential temperature is conserved for a parcel of dry air that experiences adiabatic changes in pressure, and it is related to the entropy of dry air by the approximate equation $s_d\approx c_{pd}\ln(\theta/\theta_0)$, for some reference potential temperature $\theta_0$. We also introduce the equivalent potential temperature $\theta_e$, which is a similar adiabatic invariant for moist air. The equivalent potential temperature $\theta_e$ may be related to the entropy by $s=c_{pd}\ln(\theta_e/\theta_{e0})$ for some reference equivalent potential temperature $\theta_{e0}$ \add{\citep{Pauluis2010}}, \addr{provided the liquid water reference entropy $s_{l0}$ is taken to be zero. While other adiabatic invariants may be defined that take into account the physical value of $s_{l0}$ \citep{Marquet2011,Marquet2016c}, here we will proceed with the equivalent potential temperature $\theta_e$ as it is commonly used in the tropical meteorology literature. } 

\begin{figure*}
\centering
\includegraphics[width=16cm]{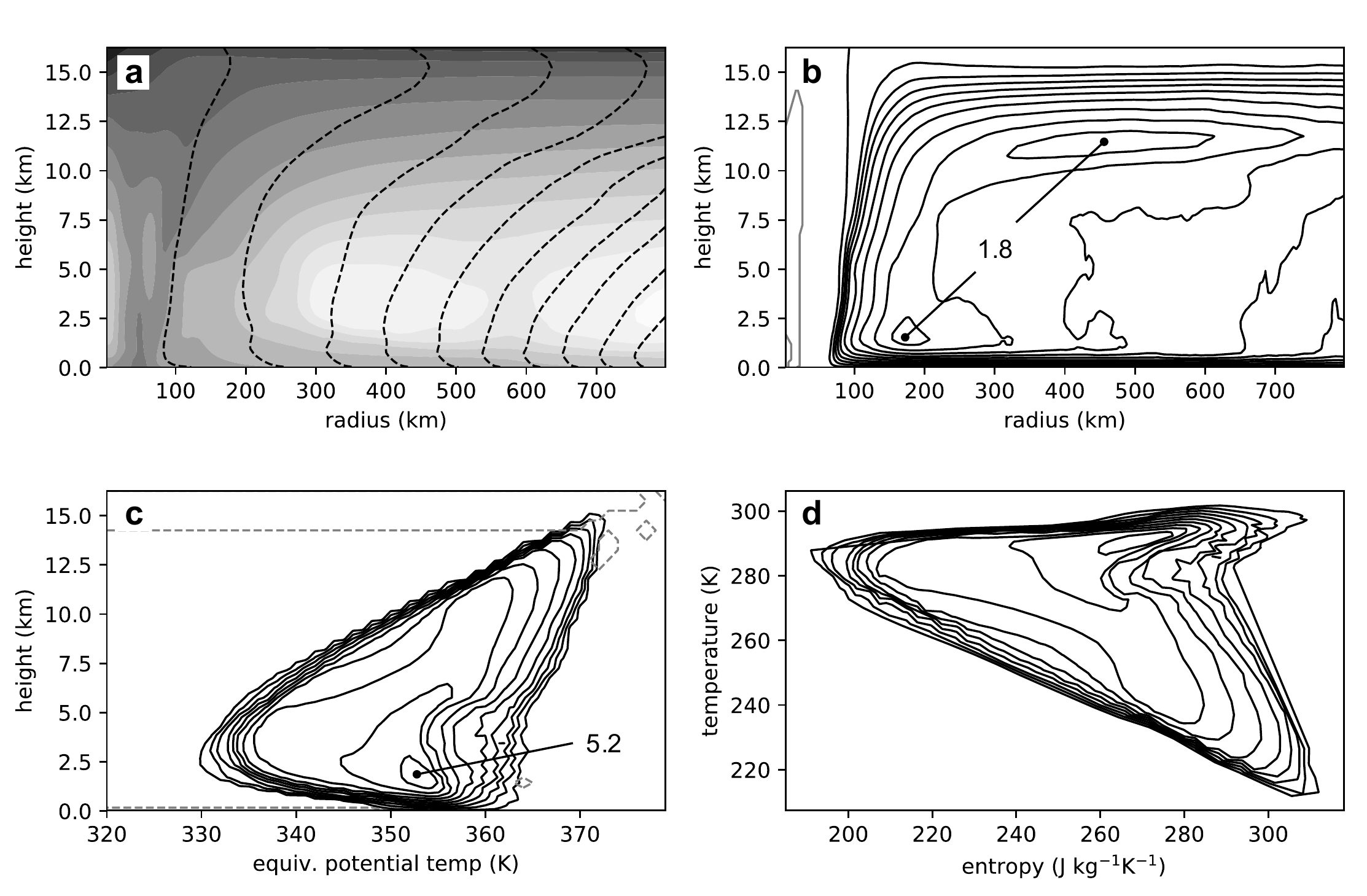}
\caption{Structure and overturning circulation of an axisymmetric tropical cyclone simulation at steady state [data from a 30-day window of a TC in a 6,000 km radius domain at $40^\circ$ latitude as described in \citet{ONeillChavas2020}]. (a) Equivalent potential temperature ($\theta_e$ [K], darker shades for higher values) and absolute angular momentum (dashed; contour value increasing with radius). (b) Eulerian mass overturning streamfunction $\Psi$ [kg s$^{-1}$] integrated radially outward to 800 km; each contour represents 10\% of the mass flux. (c) Isentropic mass overturning streamfunction $\Psi_{e}$ [kg s$^{-1}$] integrated radially outward to 800 km; each contour represents 10\% of the mass flux. (d) Temperature-entropy ($T$-$s$) diagram corresponding to parcel paths in panel (c) using the MAFALDA technique \citep{Pauluis2016}, where data gaps present in the contouring procedure in panel (c) have been filled via linear interpolation. Numbers in panels (b) and (c) give maximum streamfunction magnitude in units of $10^9$ kg s$^{-1}$.}
\label{fig:TCstreamfunction}
\end{figure*}

Being thermally direct warm-core vortices, TCs are characterized by rising high-$\theta_e$ air and subsiding low-$\theta_e$ air. Isentropic averaging \addr{[where the word ``isentropic'' here is more historical and less accurate \citep[see ][]{Marquet2017b}]} allows one to recast spatial data into a spatial dimension and a thermodynamic dimension. The vertical mass flux at each height is binned by a discretized thermodynamic variable such as $\theta_e$. These isentropic mass fluxes may then be used to define a streamfunction $\Psi_e(\theta_e,z)$ as a function of equivalent potential temperature and altitude that provides a thermodynamic perspective on the overturning circulation. \add{To ensure closure of the streamlines,  calculations of $\Psi_e$ are performed by first subtracting the average vertical velocity at each vertical level of the analysis domain}. \citet{PauluisMrowiec2013} first applied this technique to statistically-steady disorganized convection in RCE. \citet{Mrowiecetal2016} subsequently applied the procedure to a TC, where the $\theta_e$ structure is a rather natural radial coordinate (Fig. \ref{fig:TCstreamfunction}a). \add{\citet{Mrowiecetal2016}} showed that the bulk of the upward mass flux occurred in asymmetric convective regions that don't appear in an Eulerian streamfunction. As a result, the extremal isentropic streamfunction mass flux is always higher than the extremal Eulerian streamfunction mass flux, by a factor of three or so. This is also true when the TC is axisymmetric, as a comparison of the peak mass flux in the Eulerian and isentropic streamfunctions given in Fig. \ref{fig:TCstreamfunction}b and c shows. This demonstrates that the TC upward mass flux does not exclusively occur in the central eyewall.

The isentropic streamfunction may be considered to be an approximation of the thermodynamic temperature-entropy ($T$-$s$) diagram; $T$ decreases monotonically with altitude between the surface boundary layer and the tropopause, and moist entropy increases with increasing $\theta_e$, so the isentropic streamfunction is quite similar to a $T$-$s$ diagram if you flip it upside-down (compare panels \ref{fig:TCstreamfunction}c and d). \citet{Pauluis2016} introduced a more formal method to approximate thermodynamic cycles based on Eulerian data by assuming that each closed circuit in the isentropic streamfunction is the path of a parcel of air (named the Mean Air Flow As Lagrangian Dynamics Approximation, MAFALDA). Isentropically-averaged variables of interest may then be interpolated along each closed path. This is a strong idealization of real parcel motion within a TC, which is not expected to exhibit closed parcel paths in the vicinity of the eyewall. However it allows for an approximate calculation of thermodynamic cycles experienced by the overturning circulation (Fig. \ref{fig:TCstreamfunction}d), including the motions in randomly distributed convective towers. The isentropic streamfunction suggests a complicated, distinctly not-Carnot-like spectrum of closed paths in $T$-$s$ space, \add{and the bulk of the TC convection in fact occurs radially outward of the eyewall. This can be confirmed} using the MAFALDA procedure [Fig. \ref{fig:TCstreamfunction}d, and see \citet{Pauluis2016}].

The total work produced by an air parcel containing a unit mass of dry air traversing a reversible Carnot cycle is equal to the net heating: $\widetilde{W}_\text{max}=-\oint \alpha_d dp=\oint T ds$. Here $s$ is the entropy expressed per unit mass of dry air, $\alpha_d = \rho_d^{-1}$ is the specific volume of dry air, and the $\widetilde{()}$ indicates an integral over a closed parcel path. We follow \citet{Pauluis2016} in considering integrals along paths following the trajectory of dry air as the working fluid rather than the barycentric flow as was done in previous sections. \add{Furthermore}, unlike the definition of $\dot{Q}_\text{in}$ in \eqref{eq:heating_material}, the net heating in the following development \add{includes that due to frictional dissipation and diffusion}, so the Carnot efficiency is an upper bound of the mechanical efficiency. \add{We denote this total positive heating as $Q_\text{tot}^+$. Because the relevant system here is a single air parcel traversing a closed cycle, frictional heating and diffusion can be considered as external heat sources \citep{PauluisZhang2017}.}

In an irreversible MAFALDA loop within a TC, the net heating is partitioned into the ``Gibbs penalty'' and the reversible work $\widetilde{W}_\text{rev}$. The Gibbs penalty \citep{Pauluis2011,Pauluis2016,PauluisZhang2017} is due to the systematic removal of Gibbs free energy from the TC system because water enters the open system via evaporation in subsaturated conditions (low Gibbs free energy) and leaves via precipitation under generally saturated conditions (high Gibbs free energy)\add{, and it corresponds to a loss of work associated with irreversible entropy production}. The reversible work $\widetilde{W}_\text{rev}$ itself is partitioned between the production of kinetic energy $\widetilde{W}_{K}$ and the work used to increase the geopotential of water of any phase $\widetilde{W}_\text{H$_2$O}$, $\widetilde{W}_\text{rev}=\widetilde{W}_{K}+\widetilde{W}_\text{H$_2$O}$. 
 Thus $\widetilde{W}_\text{max}$ can be decomposed in the following way \citep{Pauluis2011,Pauluis2016,PauluisZhang2017,Fangetal2019}:
\begin{eqnarray}
\underbrace{\oint T ds}_{\widetilde{W}_\text{max}} \,\,\, = \,\,\,
\underbrace{-\oint\alpha_d dp}_{\widetilde{W}_K+\widetilde{W}_\text{H$_2$O}}
\,\,\, - \,\,\, \add{\underbrace{\oint \sum\limits_{x=v,l,s} g_x dr_x}_{-\widetilde{\Delta g}}}\label{eq:TCworkterms}
\end{eqnarray}
where $g_x$ is the Gibbs free energy of water in phase $x$, expressed here per unit mass of dry air, and $r_x$ is the mixing ratio, equal to the mass of water in phase $x$ per unit mass of dry air. This is a closed line integral of the fundamental \add{thermodynamic relation \eqref{eq:fund_therm_rel_moist}, written in terms of enthalpy and expressed per unit mass of dry air.}

The most efficient parcel path (`inner core path') in the TC is the most extremal one that travels from the bottom of the boundary layer, up the eyewall and to the top of the outflow, experiencing the maximum gradient in both temperature and entropy.
The mechanical efficiency of a given closed MAFALDA trajectory $\eta_M=\widetilde{W}_K/\widetilde{Q}_\text{tot}^+$ may be approximately expressed as \citep{PauluisZhang2017,Fangetal2019}:
\begin{equation}
    \eta_M = \eta_C - \frac{\widetilde{W}_\text{H$_2$O}}{\widetilde{Q}_\text{tot}^+} - \frac{\widetilde{\Delta G}}{\widetilde{Q}_\text{tot}^+},
\end{equation}
\add{where $\eta_C = (T_\text{in} -T_\text{out})/T_\text{in}$ is the Carnot efficiency and
\begin{equation*}
    \widetilde{\Delta G} = -T_\text{out}\oint \sum\limits_{x=v,l,s}\frac{g_x}{T}dr_x.
\end{equation*}
The temperatures $T_\text{in}$ and $T_\text{out}$ are the effective temperatures experienced by the parcel during the heating and cooling legs of the closed cycle, respectively. They are defined analogously to (\ref{eq:effectiveT}) but based on the total heating rate $Q_\text{tot}$, and evaluated for a closed line integral.}
\citet{PauluisZhang2017} numerically simulated an idealized three-dimensional TC and calculated a remarkable mechanical efficiency of $\eta_M\approx 0.7\eta_C$ for the inner core parcel path. Similar values of the mechanical efficiency as a fraction of the Carnot efficiency were reported by \citet{Fangetal2019} in a more realistic simulation of Hurricane Edouard (2014). They showed that both \add{$\widetilde{Q}_\text{tot}^+$} and $\widetilde{W}_K$ of the inner core path increased sharply during a period of strong intensification. The intensification occurred as the storm grew and axisymmetrized, while $\widetilde{W}_\text{H$_2$O}$ and $\widetilde{\Delta G}$ remained relatively stationary. In the numerical TC experiments of \citet{PauluisZhang2017} and \citet{Fangetal2019}, MAFALDA trajectories that extend to the upper troposphere have a high efficiency, with $\widetilde{W}_{K}>\widetilde{W}_{H_2O},\widetilde{\Delta G}$. 
These results highlight the potential utility of calculations of TC work and entropy budgets for understanding TC dynamics. \addr{A limitation of the above framework is that the magnitudes of $\widetilde{W}_\text{max}$ and $\widetilde{\Delta g}$ are sensitive to the choice of the liquid water reference entropy $s_{l0}$ \citep{Marquet2017b}.
But whether this sensitivity matters for useful interpretation is a matter of debate \citep[see][]{Pauluis2018,MarquetDauhut2018}.
}

The values found for TC mechanical efficiencies given above cannot be directly compared with the efficiencies found in aggregated vs. disaggregated RCE states in section \ref{sec:conv}, because the RCE calculations given in table \ref{table:RCE_budget} are integrated over the domain, while the above TC efficiencies are only measured for the parcel path with the highest efficiency. A comparison of the efficiency of convection in rotating and non-rotating RCE has not yet been carried out in the literature, but the MAFALDA procedure could easily be employed in both cases for a direct comparison.


\add{\citet{WangLin2020,WangLin2021} compared simulations of a dry TC, a reversible moist TC (no precipitation and attendant entropy production), and a highly irreversible, realistic moist TC. The dry and reversible moist storms displayed very similar structure with a deep lower inflow layer, a wide and weak eyewall ascent region and no entropy minimum, in contrast to the realistic TC. The realistic TC was also much smaller, even though it exhibited much higher peak wind intensity. \citet{WangLin2021} calculated full entropy budgets for the same simulations, exploiting the heat engine nature of a TC to develop an analytical expression that relates wind intensity and wind structure. These results suggest that irreversibility due to water phase changes and precipitation substantially reduces the overall mechanical efficiency of realistic moist TCs and is responsible for their small, compact core as compared to dry TC counterparts.}

\subsection{Open TCs}\label{sec:OpenTCs}

Though the isentropic averaging procedure produces a closed streamfunction amenable to a work calculation, the TCs studied in the research described above are all open to the farther environment. Dry air and water substance are exchanged at the system's lateral boundaries, causing the system to import or export energy and entropy laterally. Note that the upper outflow of the TC in Fig. \ref{fig:TCstreamfunction}b clearly flows beyond the right-hand boundary, and boundary layer air enters from the outer edge, but the isentropic streamfunction in panel c still appears as a closed circulation. In fact, it has been shown that most of the condensed water in a TC comes from the lateral convergence of water vapor, rather than locally from the sea surface \citep[e.g.,][]{Kurihara1975,Zhangetal2002}. But the MAFALDA procedure can only operate on closed thermodynamic cycles. In order to close an isentropic streamfunction under substantial lateral exchange with the environment, the average vertical velocity is removed from the subdomain containing the TC at each vertical level. The integrals are calculated out to a rather limited radius away from the TC center (500-800 km) in order for the signature of the mass flux in the eyewall to remain appreciable in spite of its small volume. An isentropic average that integrates outward to the deformation radius of the storm would be dominated by weaker convection occurring far from the TC eye.

TCs have been treated as open systems in thermodynamic studies \citep[e.g.,][]{LiuLiu2004,TangEmanuel2010,TangEmanuel2012b,JuracicRaymond2016}, with TC-environmental exchange considered alongside the traditional vertical boundary sinks and sources.
\citet{JuracicRaymond2016} calculated a moist entropy budget for TCs using dropsonde data interpolated to a three-dimensional grid and included a calculation of lateral fluxes of entropy between the tropical cyclone (in a $4^{\circ}\times 4^{\circ}$ storm-following domain) and the environment. Irreversible entropy production was estimated as instantaneously balancing the entropy sink due to radiative cooling. This assumption could be inaccurate for TCs that are experiencing a lot of environmental shear (and in general lateral exchange with the environment), which can bring in low-entropy air \citep{TangEmanuel2010} and lead to evaporation underneath the cooling cirrus canopy.
A constraint on the environmental exchange with a highly dissipative TC is that the net entropy out of the TC domain (to space as well as the broader atmosphere/ocean system) must be large and positive in order to keep a TC at steady state. Lateral fluxes of entropy out of the TC domain could be negative if radiative cooling is sufficiently high; \citet{JuracicRaymond2016} found that such fluxes \add{could be of either sign in their observed cases, and showed some dependence on the strengthening or weaking status of the TC}.  


No study has yet attempted a heat-engine analysis of the entire lifecycle of a TC, but \citet{Tang2017a,Tang2017b} constructed a simplified framework that illuminates the role of lateral entropy fluxes before and during tropical cyclogenesis. However, their numerical model omits material production of entropy within the TC domain. The intensification study of \citet{Fangetal2019} is another promising avenue. Heat engine concepts could be further brought to bear in the literature seeking to understand the annual frequency of TCs globally \citep[e.g.,][]{Hoogewindetal2020,Hsiehetal2020}. \add{Use of MAFALDA to estimate closed trajectories, and the thermodynamic analysis of \citet{WangLin2020,WangLin2021}, are among recent approaches that can be leveraged to better understand how TCs evolve in a warming world.}
In short, there are many exciting tools and approaches newly available to probe TCs that exploit the second law of thermodynamics.


%


\section{The global circulation of the atmosphere}
\label{sec:global_atmosphere}


In this section, we consider the global atmospheric circulation from a thermodynamic perspective.
We first describe the material entropy budget of the global atmosphere and we compare it to the entropy budget of RCE (section \ref{sec:global_ent_bud}). Next, we consider the global atmospheric heat engine, and we review theories for its meridional energy transport and its response to global climate change \ref{sec:global_heat_engine}). Finally, we broaden our perspective to consider the heat engines of other planets in the Solar System and beyond (section \ref{sec:other_planets}).





\subsection{The material entropy budget of the global atmosphere}
\label{sec:global_ent_bud}

Estimating the entropy budget of the global atmosphere is challenging; observational studies often employ relatively crude estimates of effective temperatures \citep{Peixotoetal1991} that limit the accuracy of the resultant estimates of irreversible entropy production. Global climate models are able to provide more detailed diagnostics than those available from observations, but they present difficulties of their own.
In particular, global climate models are typically run at horizontal grid spacings of the order of 100 km and they are therefore unable to resolve convective clouds. Irreversible entropy production associated with moist convection, which was described in detail in the previous section and is known to account for a large fraction of the total irreversible entropy production in the atmosphere \citep{Pascaleetal2011}, must be wholly parameterized within a global climate model. The extent to which parameterizations of convection accurately represent this entropy production remains unknown. But even assuming that a model's parameterizations accurately reflect the effect of subgrid processes on the model's resolved grid, the low resolution and use of simplified thermodynamic formulations \citep{FraedrichLunkeit2008,Pascaleetal2011} within global climate models imply that their entropy budgets differ from that of Earth's atmosphere (see section \ref{sec:models}), and \add{care must be taken to ensure that comparisons across studies and models consider entropy sinks and sources at similar spatial and temporal scales \citep{Lucarini2014}}.
Finally, while detailed analysis of individual models is possible \citep{Pascaleetal2011}, standard outputs from model intercomparison projects only allow for the calculation of approximate entropy budgets, and this can lead to difficulty closing the budget \citep{Lembo2019}.

Despite the above challenges, 
there are a number of broad features of the atmosphere's material entropy budget that are  known with some confidence. The total material entropy production of the atmosphere makes up the vast bulk of the material entropy production of the climate system \add{(and indeed that of the entire Earth system). In section \ref{sec:material}, the material entropy production of the climate system was estimated to be in the range 35-60 mW m$^{-2}$ K$^{-1}$ \citep{Lembo2019}. The lower end of this range is only slightly higher than the estimates of the entropy production of RCE given in table \ref{table:RCE_budget}, suggesting that the bulk of the irreversible entropy production in the atmosphere is produced by processes acting vertically within each column. Indeed, \citet{Lucarinietal2011} and \citet{Lucarini2014} found that roughly 90\% of the irreversible entropy production in the climate system could be associated with vertical heat fluxes, with the remaining 10\% associated with horizontal heat transport.}

\add{As} in RCE, the atmosphere's entropy budget is dominated by moist processes. \citet{Lembo2019} found that entropy production associated with the hydrological cycle (terms $\tavg{\dot{S}_i^\text{mem}}$ and $\tavg{\dot{S}_i^\text{sed}}$ in our formulation) accounted for 80-90\% of the estimated total irreversible entropy production in an ensemble of state-of-the-art global climate models, consistent with theoretical expectations given the dominance of latent heat transport in the energy exchange between the surface and atmosphere at a global level \citep{Pauluis2002}. Of the component associated with moist processes, the bulk is due to phase change and vapor diffusion $\tavg{\dot{S}_i^\text{mem}}$; \citet{Lembo2019} estimated that the entropy production associated with precipitation sedimentation accounted for only roughly 4-6 mW m$^{-2}$ K$^{-1}$, consistent with an observational estimate of the dissipation owing to precipitation sedimentation given by \citet{Pauluis2012}.

Model-based estimates of the entropy production associated with frictional dissipation of the winds $\tavg{\dot{S}_i^\text{fric}}$ vary from roughly 6 mW m$^{-2}$ K$^{-1}$ \citep{Lembo2019}, similar to values found for disaggregated RCE, up to twice this value \citep{Pascaleetal2011}, with correspondingly large ranges in estimates of the rate of work performed by the atmospheric heat engine and its mechanical efficiency. The reason for this wide range is likely due to difficulties in estimating the frictional dissipation rate in global climate models \citep{Lembo2019}; such models often have multiple parameterizations that dissipate kinetic energy, and they may or may not include frictional heating within their thermodynamic formulation \citep{Pascaleetal2011}. Additionally, kinetic energy that is both generated and dissipated at scales smaller than the model grid is not included in the model's mechanical energy or entropy budgets. Frictional dissipation estimated from global climate models should therefore be considered to be only the portion of the dissipation that is associated with the large-scale flow; it is unclear to what extent one should compare such estimates to those derived from higher-resolution models such as presented in section \ref{sec:conv}.

In summary, while quantitative estimates remain uncertain, qualitatively, the entropy budget of the global atmosphere shares a number of similarities with the simpler case of RCE discussed in section \ref{sec:conv}. In particular, the dominance of entropy production associated with moist processes limits the mechanical efficiency of the global atmospheric heat engine, and it limits the rate at which work is done by the pressure gradient force. As we shall see below, this fact plays an important role in understanding the atmospheric heat engine's response to global climate change.





\subsection{The global atmospheric heat engine}
\label{sec:global_heat_engine}

\subsubsection{A thermodynamic perspective of the global atmospheric circulation}

Despite the long history of research describing the atmosphere as a heat engine \citep[e.g.,][ section \ref{sec:heat_engine}]{Brunt1926}, relatively few studies have expressed the \add{global} atmospheric circulation in traditional thermodynamic coordinates \citep[e.g., temperature-entropy ($T$-$s$) space;][]{Laliberteetal2015}. 
Rather, the global atmospheric circulation is more commonly characterized in terms of the meridional mass overturning streamfunction. This streamfunction may be constructed based on an Eulerian average at constant height or pressure, in which case it quantifies the average mass flow in the latitude-height plane. Alternatively, an isentropic averaging technique similar to that described in section \ref{sec:TC_work} but applied to the vertical dimension rather than a horizontal dimension may be used to re-express
the streamfunction as a function of latitude and an entropy-based vertical coordinate (specifically, potential temperature). This isentropic streamfunction provides a thermodynamic perspective on the global atmospheric circulation, and it may be used to quantify the global atmospheric heat engine.






\begin{figure}
\centering
\includegraphics[width=8.6cm]{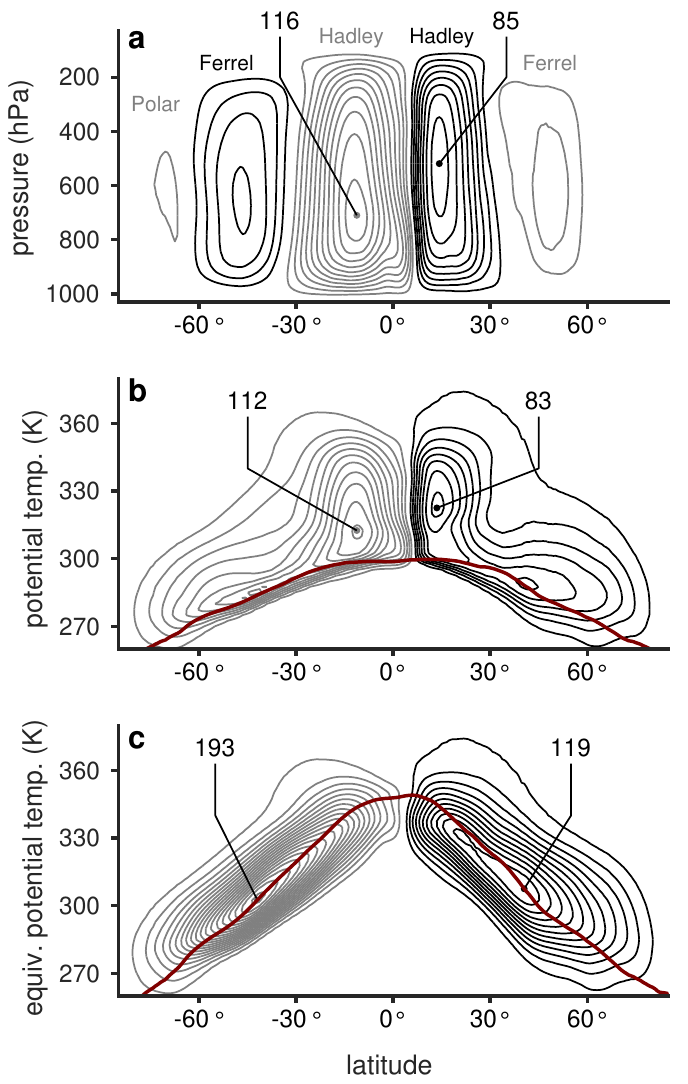}
\caption{Mean meridional mass overturning streamfunction estimated from the ERA-Interim reanalysis \citep{Dee2011} using 6-hourly snapshots for the years 1981-2000 and calculated using (a) pressure, (b) potential temperature, and (c) equivalent potential temperature as a vertical coordinate using the method described in \citet{Pauluis2010}. Black contours represent clockwise motion and gray contours represent anticlockwise motion. Contour interval is $10\times10^{9}$ kg s$^{-1}$ with zero contour omitted. Numbers give maximum streamfunction magnitude in each hemisphere in units of $10^9$ kg s$^{-1}$. Hadley, Ferrel, and Polar cells are labeled in (a). Thick maroon curves show mean (b) potential temperature and (c) equivalent potential temperature near the surface (at the pressure level $p = 0.9988p_s$, where $p_s$ is the surface pressure).}
		\label{fig:streamfunction}
\end{figure}

An estimate of the traditional Eulerian streamfunction reveals the three circulation cells known to characterize each hemisphere of Earth's annual-mean circulation: the Hadley, Ferrel, and Polar cells (Fig. \ref{fig:streamfunction}a; the Polar Cell in the Northern Hemisphere is too weak to be displayed at the contour interval shown). The Hadley and Polar cells are thermally direct; their mass fluxes imply warm air rising, cool air sinking, and \add{downgradient} energy transport. The Ferrel Cell, on the other hand, is thermally indirect; its implied energy transport is toward the equator. Such a cell is able to exist because eddies---motions in the atmosphere representing departures from a time or zonal mean---produce a poleward energy transport at midlatitudes that more than compensates for the equatorward energy transport implied by the mean circulation. In spite of the presence of the Ferrel Cell, the total energy transport by eddies plus the mean flow remains poleward \add{across the midlatitudes}.


The Ferrel cell does not appear when the streamfunction is calculated using isentropic averaging.
The lower panels of Fig. \ref{fig:streamfunction} show estimates
of the streamfunction based on mass fluxes averaged at fixed potential temperature $\theta$ (Fig. \ref{fig:streamfunction}b) and fixed equivalent potential temperature $\theta_e$ (Fig. \ref{fig:streamfunction}c) following \citet{Pauluis2008,Pauluis2010}. \add{As discussed in section \ref{sec:TC_work}, the potential temperature 
$\theta$ is approximately proportional to the logarithm of the entropy of dry air $s_d$. 
Isosurfaces of $\theta$ are approximately parallel to isosurfaces of dry entropy, and we therefore refer to such surfaces as dry isentropes. \addr{For an air parcel with a fixed total water content,} the equivalent potential temperature $\theta_e$ is proportional to the logarithm of the entropy $s$.
Since $s$ includes the entropy of both dry air and water within an air parcel, we refer to surfaces of constant $\theta_e$ as moist isentropes.}

Both the dry- and moist-isentropic mass streamfunctions are
characterized by a single, thermally direct overturning cell in each hemisphere. 
Physically, the Eulerian and isentropic mass streamfunctions differ because there is a strong tendency for poleward moving air to be warmer (and moister) than equatorward moving air at the same pressure level. A latitudinal exchange of air at a given pressure level contributes to the isentropic overturning but does not contribute to the Eulerian overturning. The dry- and moist-isentropic streamfunctions therefore provide a view of the global circulation of the atmosphere that is more directly connected to its meridional energy transport than the view provided by the Eulerian streamfunction. 

The moist-isentropic streamfunction captures energy transport associated with the latent heat content of moist air as well as its sensible heat content. As a result, the meridional mass transport by the atmosphere is $\sim 1.5$ times larger when viewed on moist isentropes compared to either the dry-isentropic view or the Eulerian view \citep{Pauluis2010}.  
The moist-isentropic streamfunction also differs from its dry-isentropic counterpart in that it has a roughly consistent ``depth'' in isentropic coordinates from the equator to the pole, with no obvious demarcation between the tropics and extratropics. This is consistent with the meridional energy transport of the atmosphere, which is also ``seamless'' between the tropics and extratropics \citep{Trenberth2003}.

On the basis of the moist-isentropic streamfunction, \citet{Pauluis2010} argue for a revised view of the global atmospheric circulation that includes a ``moist branch'' that involves \addr{warm and moist} air that is advected poleward in the subtropics and rises through the warm sectors of extratropical cyclones. Recent studies have highlighted the utility of the moist-isentropic view for understanding the thermodynamics of atmospheric circulations such as the monsoon \citep{Chen2018} and tropical cyclones (see section \ref{sec:TC_work}). Such analyses may be useful for quantifying the contribution of circulations of different scales to the global atmospheric heat engine \citep{Chen2020}.

\subsubsection{Theories for the global atmospheric heat engine}

A key goal of climate dynamics research is 
the development of a theory for the meridional heat transport in the atmosphere.
One approach toward achieving this goal is to relate the time-mean \add{meridional} heat transport $\tavg{F_H}$ to the meridional temperature gradient through an ``eddy diffusivity'' $K$, such that,
\begin{linenomath}\begin{equation*}
    \tavg{F_H} \propto K \frac{\Delta T}{\Delta y},
\end{equation*}\end{linenomath}
where $\Delta T/\Delta y$ gives a measure of the gradient of temperature $T$ in the meridional direction $y$ averaged over a suitable latitude band. The theoretical challenge is to understand the dependence of $K$ on the mean thermodynamic state of the atmosphere.

Theories for the eddy diffusivity $K$
go back at least half a century \citep[e.g.,][]{Green1970,Stone1972}.  
Of particular note for the present review is the study of \citet{Barryetal2002}, in which 
a scaling for $K$ was developed
by treating the atmospheric circulation as a heat engine. 

\citet{Barryetal2002} assumed that the net atmospheric energy flux out of the tropics $\tavg{F_H}$ could be related to the frictional dissipation rate associated with large-scale atmospheric circulations $\tavg{\dot{D}_{LS}}$ through an expression of the form,
\begin{equation}
    \tavg{\dot{D}_{LS}} \propto \frac{\Delta T}{T_0}\tavg{F_H}.
    \label{eq:Barry}
\end{equation}
\add{Here, large-scale circulations are those that are significantly affected by the Earth's rotation and are well resolved by global climate models.} As in the previous equation, $\Delta T$ represents a characteristic temperature difference across the midlatitude zone while $T_0$ is a characteristic temperature, giving $\Delta T/T_0$ the form of a heat engine efficiency. The theory is closed using a mixing length argument,
which expresses the diffusivity $K$ in terms of the dissipation rate and a characteristic length scale over which fluid parcels are displaced by eddies.
\citet{Barryetal2002} showed that, through the appropriate choice of mixing length, their expression for the diffusivity was able to account for changes in atmospheric heat transport in simulations with a comprehensive global climate model across a wide range of parameters.


A number of other diffusive theories for the atmospheric heat transport have been proposed [e.g., \citet{Green1970,Stone1972,Lapeyre2003,Held1996,Gallet2020}. See also \citet{Held2019} for a recent review]. Generally, such theories are developed on the basis of the budget of available potential energy rather than entropy (see section \ref{sec:APE}), but recent work by \citet{Changdissertation} casts both \citet{Barryetal2002} and \citet{Held1996} in a common entropy budget-focused framework, showing that they are both limiting cases of a more general theory for eddy diffusivity.

%

A common feature of many theories for the atmospheric eddy diffusivity, including that of \citet{Barryetal2002}, is that they do not explicitly consider the effect of moist processes. Indeed, the relation \eqref{eq:Barry} may be compared to similar relations used to develop theories of atmospheric convection discussed in section \ref{sec:moistRCE}, in which the rate of work performed by atmospheric convection is related to a forcing parameter through a thermodynamic efficiency. \add{As we have seen, this approach fails for moist convection because it neglects irreversible processes associated with water in all its phases, which account for the bulk of the material entropy production in the atmosphere. But such an approach, suitably adapted, may nevertheless have relevance to the larger-scale circulations that contribute to $\tavg{\dot{D}_{LS}}$. Such circulations are primarily driven by horizontal heating and temperature gradients, while, as noted above, the bulk of the entropy production in the atmosphere is associated with vertical heating and temperature gradients \citep{Lucarini2014}.}




\add{Indeed, a number of authors have sought to} adapt theories of the atmospheric eddy diffusivity to include moist thermodynamics \citep[see e.g.,][]{Lapeyre2004,O'Gorman2011}. \add{Recently, the use of diffusive closures based on the energy content of air (including latent energy) rather than its temperature have been identified as a promising direction \citep[e.g.,][]{Flannery1984,Hwang2010,Armouretal2019,Mooring2020}. However, theoretical justification for this approach remains incomplete,} and understanding atmospheric heat transport in a moist atmosphere remains an area of active research. 

Accounting for moist processes is particularly important in the context of global climate change: as the world warms, the concentration of water vapor in the atmosphere is expected to increase by roughly 7\% for each kelvin increase in temperature, following the Clausius-Clapeyron equation. This rapid increase in atmospheric humidity clearly must be taken into account in any theory for the atmospheric heat engine in a warming climate.

\subsubsection{The atmospheric heat engine under climate change}
\label{sec:future}

In the last few decades, the climate science community has collectively developed a large archive of simulation data containing projections of global climate change that is freely available to researchers \citep{Eyring2016}. This archive provides an opportunity to study how the atmospheric heat engine is affected by climate change, at least in the context of global climate models. 
While evaluating the entropy budget is challenging based only on the available outputs \citep{Lembo2019}, \citet{Laliberteetal2015} recently developed a technique for diagnosing the strength of the atmospheric heat engine using only standard model outputs.

Consider the fundamental thermodynamic relation \eqref{eq:fund_therm_rel_moist}, \add{applied to a parcel of air} and written in terms of enthalpy as,
\begin{equation}
    T\frac{ds}{dt} = \frac{dh}{dt}   - \alpha\frac{dp}{dt} - \sum_x g_x  \frac{dq_x}{dt}.
    \label{eq:fund_therm_rel_h}
\end{equation}
\add{Recall that thermodynamic equilibrium between phase $x$ and $y$ is defined by $g_x=g_y$, so that the last term on the right-hand side sums to zero for phase changes in equilibrium \citep{Pauluis2011}. Under the condition that all phase changes occur in thermodynamic equilibrium, the last term is therefore only non-zero when water is either added or removed from the air, and \eqref{eq:fund_therm_rel_h} may be written,
\begin{equation}
     T\frac{ds}{dt} = \frac{dh}{dt}   - \alpha\frac{dp}{dt} + (g_d-g_v)  \frac{dq_T}{dt},
     \label{eq:fund_therm_rel_h_equil}
\end{equation}
where $q_T$ is the total mass fraction of water. This equation applies equally to the addition or removal of water vapor in unsaturated conditions and the addition or removal of liquid (or solid) water in saturated conditions, for which $g_v = g_l$ (or $g_v=g_s$).

At first glance, the assumption of phase equilibrium suggests that \eqref{eq:fund_therm_rel_h_equil} is not well suited for application to the atmosphere, which often experiences phase changes far from equilibrium. But evaporation or sublimation at subsaturation may be taken into account provided the parcel to which \eqref{eq:fund_therm_rel_h_equil} is applied includes only water substance that is in thermodynamic equilibrium with the surrounding air. For example, by excluding falling precipitation from the parcel definition, evaporation of precipitation at subsaturation may be recast as the addition of water vapor to unsaturated air, which is included in the last term on the right-hand side of \eqref{eq:fund_therm_rel_h_equil}. On the other hand, melting and freezing that occurs outside of phase equilibrium cannot be treated this way, and irreversibility associated with the melt/freeze cycle is neglected by this framework.
}

Integrating \eqref{eq:fund_therm_rel_h_equil} with mass weighting over the atmosphere $\Omega_A$ \add{and taking a time mean over a statistically steady state}, the time derivative of the enthalpy vanishes, and the above equation may be transformed into a budget for work done by the atmosphere,
\begin{equation}
     \tavg{\dot{W}_\text{max}} = \tavg{\dot{W}_K} + \tavg{\Delta\dot{G}}.
     \label{eq:laliberte}
\end{equation}
where 
\begin{equation}
     \tavg{\dot{W}_K} = -\int_{\Omega_A} \tavgb{ \frac{dp}{dt}} \, dV
     \label{eq:laliberte_int}
\end{equation}
is the rate at which the atmosphere performs work to generate the kinetic energy of the winds
and
\begin{linenomath}\begin{align*}
     \tavg{\dot{W}_\text{max}} &= \int_{\Omega_A} \tavgb{\rho T\frac{ds}{dt}} \,  dV, \\
     \tavg{\Delta \dot{G}} &= \int_{\Omega_A} \tavgb{\rho(g_d-g_v)\frac{dq_T}{dt}} \, dV.
\end{align*}\end{linenomath}
Here the term $\tavg{\dot{W}_\text{max}}$ represents a measure of the maximum rate of work that could be performed by the atmosphere \addr{in the absence of moist processes} (all else being equal). The term $\tavg{\Delta \dot{G}}$ represents a ``Gibbs penalty'' related to the effects of moisture, and it primarily represents the power required to maintain the hydrological cycle \citep{Pauluis2011}. This budget is similar to \eqref{eq:TCworkterms} in section \ref{sec:TC_work}, but, because we consider the fluid velocity to be the barycentric velocity of the mixture of air and condensed water rather than the velocity of air, the work does not include the work required to lift water (see section \ref{sec:precip_sed}). The work required to lift water is included in the Gibbs penalty term $\tavg{\Delta\dot{G}}$.



\citet{Laliberteetal2015} devised a method for applying \eqref{eq:laliberte_int} to climate model output in order to estimate the strength of the global atmospheric heat engine and its changes under global warming. While the method allows for accurate closure of the budget, it does not distinguish between physical processes and the numerical production of entropy (see section \ref{sec:models}). In principle, the latter entropy source may be climate dependent and affect the results.
\addr{Moreover, \eqref{eq:laliberte} suffers from the same limitation as \eqref{eq:TCworkterms}, in that the terms $\tavg{\dot{W}_\text{max}}$ and $\tavg{\Delta \dot{G}}$ are sensitive to the specification of the reference entropies for dry air and liquid water in the definition of $s$.}
Notwithstanding these caveats, the authors found that, while the maximum rate of work $\tavg{\dot{W}_\text{max}}$ increased under warming, this was offset by an even larger increase in the power required to maintain the hydrological cycle $\tavg{\Delta\dot{G}}$ such that the rate of work done by the atmospheric heat engine in generating winds $\tavg{\dot{W}_K}$ decreased with warming. \add{
These results are consistent with a recent analysis of changes in the Lorenz energy cycle and entropy budget in a suite of global climate model simulations of future warming conducted by \citet{Lembo2019}.}

\citet{Laliberteetal2015} argued that the rapid increase in the power required to maintain the hydrological cycle $\tavg{\Delta\dot{G}}$ under warming could be related to the rapid increase in the moisture content of the atmosphere following the Clausius-Clapeyron relation. On the other hand, the maximum rate of work $\tavg{\dot{W}_\text{max}}$ is governed by the radiative cooling rate of the atmosphere, and this is known to increase at a more modest rate under global warming \citep[e.g.,][]{Allen2002}. The work performed by the atmosphere to generate winds $\tavg{\dot{W}_K}$ must then decrease with warming in order to balance \eqref{eq:laliberte}. 

\add{A similar reduction in kinetic energy generation and an increased dominance of moist irreversible processes was found in idealized simulations of climate warming induced by increased greenhouse gas concentrations \citep{Lucarini2010}, increased solar irradiance \citep{Lucarini2010b}, and increased ocean heat transport \citep{Knietzsch2015}. In these studies, a framework based on a variant of the dry-entropy budget (see section \ref{sec:latent}) originally developed by \citet{Johnson2000} is applied to define an efficiency $\eta$ relating the work $\tavg{\dot{W}_K}$ to the total heating owing to the combination of frictional dissipation, radiation, and latent and sensible heat fluxes within the climate system. \citet{Lucarini2009} further defined an irreversibility parameter as the ratio of the irreversible entropy production owing to frictional dissipation to that associated with down-gradient heat transport. The heat transport may be defined to include sensible and latent heat fluxes \citep{Knietzsch2015}, or to additionally include radiative fluxes within the climate system \citep{Lucarini2009}. According to \citep{Lucarini2010}, the increased importance of latent heat fluxes in a warming climate is associated with a decrease in efficiency $\eta$ and an increase in the irreversibility of the climate system. }





\begin{figure}
\centering
\includegraphics[width=8cm]{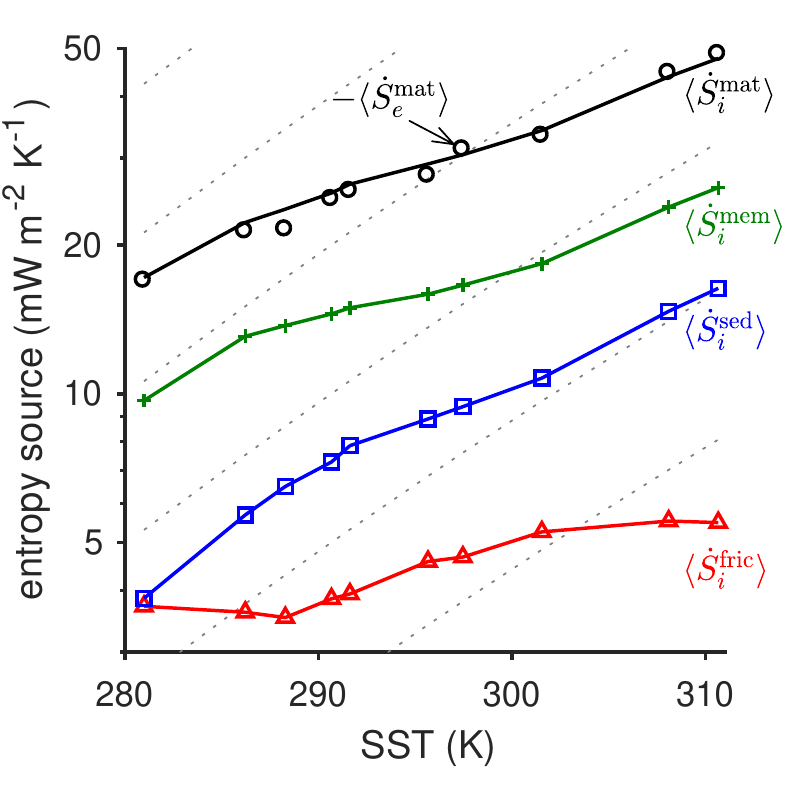}
\caption{Entropy budget as a function of sea-surface temperature (SST) in simulations of radiative-convective equilibrium taken from \citet{SinghOGorman2016} and plotted with log scale. Material entropy export $\tavg{\dot{S}_e^\text{mat}}$ (black circles), total irreversible entropy production $\tavg{\dot{S}_i^\text{mat}}$ (black line), and irreversible entropy production owing to frictional dissipation $\tavg{\dot{S}_i^\text{fric}}$ (red triangles), precipitation sedimentation $\tavg{\dot{S}_i^\text{sed}}$ (blue squares) and irreversible phase change \& mixing $\tavg{\dot{S}_i^\text{mem}}$ (green pluses). Dotted gray lines show Clausius-Clapeyron scaling, increasing in proportion to the saturation vapor pressure at the sea surface. Adapted from \citet{SinghOGorman2016}.}
		\label{fig:ent_bud_SST}
\end{figure}

The results of \citet{Laliberteetal2015} \add{and the other studies cited above} contrast with 
a recent estimate of trends in the generation and dissipation of kinetic energy in the atmosphere based on global climate models constrained by satellite and \textit{in-situ} observations, in which it was found that both the kinetic energy and its generation/dissipation rate increased over the period 1979-2013 \citep{Pan2017}. If accurate, this result suggests that recent warming has been associated with an intensification of the atmospheric heat engine. It should be noted, however, that the technique of estimating the atmospheric state using global climate models constrained by historical observations (known as reanalysis) is not well suited to evaluating climate trends \citep[e.g.,][]{Thorne2010}, and further work is needed to confirm the results of \citet{Pan2017} using other methods.

The decrease in $\tavg{\dot{W}_K}$ with warming seen in global climate simulations also contrasts with cloud-permitting simulations of RCE, in which the rate of work done by the atmospheric heat engine has been found to increase with surface temperature \citep{Romps2008,SinghOGorman2016}. This is despite the fact that the moisture content and radiative cooling rates vary similarly with temperature in both RCE and global climate simulations \citep{Jeevanjee2018}. 

\citet{SinghOGorman2016} examined the entropy budget in a series of simulations of RCE over a wide range of surface temperatures. Their results, reproduced in  Fig. \ref{fig:ent_bud_SST}, show that the magnitude of the irreversible entropy production terms associated with moist processes roughly scale with the total radiative entropy sink $\tavg{\dot{S}_e^\text{mat}}$, rather than with the Clausius Clapeyron equation. This allows the entropy production associated with frictional dissipation $\tavg{\dot{S}_i^\text{fric}}$ to also increase with warming.

The contrasting response of the atmospheric heat engine to warming in global climate models compared to high-resolution models run in RCE is puzzling and points to fundamental gaps in our understanding of the atmosphere's entropy budget.
A major difference between the two types of studies is one of spatial scale. Global climate models simulate the entire atmosphere, including the large-scale circulations that act to transport heat from the tropics to the polar regions.  But such models are generally run with horizontal resolutions too coarse to  explicitly represent cloud-scale circulations, and the work $\tavg{\dot{W}_K}$ includes only the work required to generate circulations that are resolved by the model grid (see section \ref{sec:models}). Simulations of RCE, on the other hand, are typically run on domains too small to contain large-scale circulations, and the work performed is entirely used to generate moist convection.
Future studies applying the technique of \citet{Laliberteetal2015} to a wider range of model types, including both realistic and idealized configurations, is clearly needed to better understand how circulations at different scales are affected by climate warming, and how this is manifest in changes to the atmospheric heat engine. A further outstanding question is whether the differences in mechanical efficiency associated with convective organization highlighted in section \ref{sec:org} may contribute to changes in the strength of the atmospheric heat engine in its response to climate change.



\subsection{Heat engines on other planets}
\label{sec:other_planets}






The rapidly growing zoo of detected exoplanets, in addition to our quirky companions in the Solar System, continues to astound the imagination. The assumptions that lead us to simple models in Earth's troposphere can be completely invalid or irrelevant on other planets. Nevertheless, heat engine concepts that consider entropy budgets, appropriately tailored, have been applied to rocky planets in our Solar System \cite{Lorenzetal2001,LorenzRenno2002,Goody2007,Titovetal2007,SchubertMitchell2013,Bannon:2017aa}, Jupiter \cite{LorenzRenno2002,Wichtetal2019}, tidally locked rocky exoplanets \cite{KollAbbot2016} and hot Jupiters \cite{Readetal2016,KollKomacek2018}.

An important complication in describing the characteristics of planetary heat engines is that, unlike for a traditional heat engine operating between two thermal reservoirs, the Carnot efficiency of a planetary heat engine is not an external parameter. The input temperature $T_\text{in}$ and output temperature $T_\text{out}$ are both functions of the climate (see section \ref{sec:carnot}). Even the effective emission temperature $T_e^*$ of a planet, defined as the temperature of a blackbody if it were to emit the same amount of radiant energy as the planet, is a function of its planetary albedo, which is climate dependent.

As a planet's climate changes, both its mechanical efficiency and Carnot efficiency may simultaneously change, perhaps substantially.
The history of Venus presents a possible example of such behavior; in its early history, Venus has been hypothesized to be water-rich before a runaway greenhouse effect occurred \citep{Walker1975}.
If there had been an active hydrological cycle, we can hypothesize that the mechanical efficiency of the Venusian heat engine (potentially near the Carnot efficiency at present) would be much lower in the past to account for the attendant irreversible entropy production associated with moist processes. However, the Carnot efficiency itself may have been lower and more Earth-like if the presence of water clouds and precipitation allowed for clear-sky patches to cool off the lower atmosphere to space. Thus mechanical efficiency as a fraction of the Carnot efficiency is a moving target that may obscure dramatic changes in climate.

\subsubsection{Rocky planets}

\citet{SchubertMitchell2013} and \citet{Bannon:2017aa} estimated the Carnot efficiency of the rocky planets with substantial atmospheres (Venus, Earth, Mars, Saturn's moon Titan). They found that Venus has a much higher Carnot efficiency than the other bodies considered, at around 24\%. Like the better-observed Earth, Mars and Titan are likely to have sedimentation-related material entropy production sources that significantly reduce the mechanical efficiency of the climate. 


\citet{Bannon:2017aa} used a creative approach to establish an upper bound on the Carnot efficiency and to study the importance of spatial variations of absorption and emission temperatures for the planetary heat engine. Using a variant of the transfer climate system definition (see section \ref{sec:fluxes}), they defined the entropic absorption temperature $T_\text{abs}$ by,
\begin{linenomath}\begin{equation*}
\frac{\tavg{\dot{Q}_\text{abs}}}{T_\text{abs}}  = \frac{1}{A}\int_\Omega \tavgb{\frac{\rho\dot{q}_\text{abs}}{T}}dV,
\end{equation*}\end{linenomath}
where $\dot{q}_\text{abs}$ is the heating of the climate system by shortwave radiation, with 
\begin{equation}
\tavg{\dot{Q}_\text{abs}}  = \frac{1}{A}\int_\Omega \tavg{\rho\dot{q}_\text{abs}} \, dV.\label{eq:heating_material2}
\end{equation}
The entropic emission temperature $T_e$ was defined analogously as a function of $\tavg{\dot{Q}_e}$, based on the temperature at which longwave radiation is emitted directly to space. Unlike \eqref{eq:heating_material}, \eqref{eq:heating_material2} does not include net heating by longwave radiation due to radiative exchanges within the climate system; it is purely defined based on \add{absorption} of photons \add{emitted by the sun}. \citet{Bannon:2017aa} showed that $T_e$ must be equal to or greater than the corresponding ``effective'' emission temperature $T_e^*$ necessary for a corresponding blackbody planet of equivalent albedo to be in radiative equilibrium with its star. Given a fixed stellar flux, orbital radius and planetary size, $T_e^*$ is only a function of the planetary albedo. Employing a thought experiment of the entropy balance of two idealized planets, \citet{Bannon:2017aa} demonstrated that the maximum entropy production of a planet occurs when $T_e = T_e^*$, which can be understood as the lowest temperature possible for the heat sink to space. When $T_e$ instead varies across the planet, entropy production decreases. In contrast, a \add{more variable} absorption temperature $T_\text{abs}$, or \add{more variable} albedo, \add{increases} entropy production. For a fixed $T_\text{abs}$, maximum entropy production occurs when $T_e = (3/4) T_\text{abs}$. \citet{Bannon:2017aa} further scaled these temperatures by a truly external temperature $T_\text{bb}$, which is the temperature the planet would have if it were a blackbody with zero albedo. The corresponding material entropy production can be scaled similarly. This permits a nondimensional comparison of entropy production between planets in a two-dimensional phase space of $T_e/T_\text{bb}$ and $T_\text{abs}/T_\text{bb}$, and it neatly demonstrates that Venus is indeed unique for its high Carnot efficiency, which is close to the upper bound.

Venus has a high albedo (around 76\%) and thick sulfuric acid cloud cover, such that little shortwave radiation reaches the surface. Venus' thin thermally direct overturning circulation likely does not penetrate to the stable atmospheric layer closest to the surface---it largely occurs aloft in the cloud region, 50-55 km, away from a solid frictional surface. Evidence suggests a possibility of convective overshooting in Venus' atmosphere, which would indicate a Venusian counterpart to Earth's hydrologic cycle \citep[e.g.,][]{Bakeretal1998,McGouldrickToon2008} and would suggest that the mechanical efficiency of the Venusian atmosphere is substantially lower than the Carnot efficiency. But dissipation due to friction around hydrometeors was estimated to be unimportant by \citet{LorenzRenno2002}.
The dominant dissipation mechanism that removes mechanical energy is hypothesized to be breaking internal gravity waves \cite{Izakov2010}. Taken together, the studies of \citet{LorenzRenno2002,Schubertetal1999,Bannon:2017aa} suggest that Venus has a very high absolute mechanical efficiency, but the role of moist convection is too poorly understood to have high confidence.


Mars, uniquely in the Solar System, has sporadic global dust storms that nearly shield the surface from the sun. This dust absorbs both shortwave and longwave radiation, adding to the longwave absorption from the CO$_2$ atmosphere. As the dust sediments back to the surface, it is also likely to be an important source of dissipation analogous to the dissipation of falling hydrometeors discussed in section \ref{sec:precip_sed}. Given the small difference between Mars' estimated average absorption temperature and emission temperature in the vertical, more work may potentially be produced during horizontal energy transport than vertical transport, which the existence of global dust storms and small dust devils seems to support \citep{SchubertMitchell2013,Jacksonetal2020}. The dominance of Earth's vertical production of irreversible entropy cannot be assumed of other planets. In particular, planets with shallow atmospheres and small surface-emission layer temperature differences may still have a strong lateral temperature gradient. This can lead to increased entropy loss to space, coincident with increased work production and frictional dissipation in the atmosphere. \citet{KollAbbot2016} showed that the temperature difference relevant to the heat engine model of a tidally locked rocky exoplanet is the permanent horizontal day-night temperature gradient and not the local vertical lapse rate.

The most Earth-like planet in terms of entropy sources and sinks is likely Saturn's moon Titan. It has an active methane cycle with some resemblance to Earth's hydrological cycle, which is a source of irreversible entropy generation due to drag around `rain'-drops. Similar to Mars however, there is a small difference between the average absorption and emission temperatures, rendering a rather low estimated Carnot efficiency of 4.1\% \citep{SchubertMitchell2013}. It is possible that the general circulation is able to produce more work than this Carnot efficiency would suggest if Titan has substantial seasonal cross-hemispheric heat transport, which was argued using energy balance and numerical modeling by \citet{Mitchell2012}.


\subsubsection{Giant planets}

Rocky planets can be generally assumed to be in energy balance with their star. They are much smaller on average than the fluid planets and are either geologically `dead' with a cold core, or the climate system is shielded from an active core by an insulating rocky mantle, as in the case of Earth. Either way, the geothermal heat flux on the rocky planets is typically an insignificant fraction of the energy received from the sun. Their well-defined solid surfaces bound the climate system on timescales shorter than the evolution of the lithosphere. These traits make single- and multi-layer energy balance models tractable and useful. Giant planets are much trickier systems. Gas and ice giants, like Jupiter, Saturn, Uranus and Neptune, are likely to lack any kind of solid lower boundary that could provide a large frictional drag on winds, and modeling efforts do not always carefully consider the need for a physically motivated dissipation mechanism \cite{Goodman2009}. In the absence of solid frictional surfaces, leading dissipation mechanisms to balance mechanical energy generation by thermally direct flows include turbulence and fluid instabilities, shocks \citep{DobbsDixonLin2008,LiGoodman2010}, Ohmic dissipation \citep{BatyginStevenson2010} and magnetic drag \citep{Pernaetal2010}.

The idealization of the atmospheric circulation as a thermally direct circulation similar to a Carnot engine is a reasonably good model for atmospheric layers with approximately adiabatic lapse rates, such that convective motions can move air quasi-adiabatically in the vertical as they advect heat downgradient. This is certainly a good model for the Earth's troposphere, but the Earth's stratosphere---\add{more generally any atmosphere in approximate radiative equilibrium}---does not exhibit thermally direct quasi-adiabatic motions \cite{KollKomacek2018}. The Brewer-Dobson overturning circulation in the stably stratified stratosphere, for example, is mechanically forced \cite{Haynes2005}, and it is thought that regions within the giant planets, as well as brown dwarf planets, should exhibit thermally indirect, wave-driven overturning circulations \cite{ShowmanKaspi2013}, where the waves may be excited by convective (thermally direct overturning) activity in an adjacent atmospheric layer. 

Even in regions of the atmosphere that are characterized by thermally direct circulations, the Carnot cycle is not necessarily a good model, particularly for heavily irradiated giant exoplanets. These so-called `hot Jupiters', believed to be in abundance throughout the universe, have extremely short and rapid orbits around their stars. The thermally direct overturning circulation occurs approximately within an isothermal layer column-wise, with a strong day-night gradient (where the day side is permanently irradiated because it is tidally locked to always face the star). \citet{KollKomacek2018} modeled the heat engine of hot Jupiters instead as an Ericsson cycle. Like the ideal Carnot cycle, an Ericsson cycle is heated and cooled during isothermal processes, but the other two legs are isobaric instead of isentropic. An ideal Ericsson cycle has the same efficiency $\eta_C$, but the mechanical efficiency of hot Jupiters could be larger or smaller depending on the unknown potential role of precipitation during the cycle (see section \ref{sec:mech}). Hot Jupiters are likely to host multiple layers of cloud decks (including silicate and titanium dioxide clouds) as well as hydrocarbon hazes \citep{Gaoetal2020}. 

Lastly, planets need not even be close to energy balance. This is observed for Jupiter, Saturn and Neptune, which emit around 80\%-160\% more energy than they receive from the sun \citep{Conrathetal1989,Ingersoll1990,Lietal2018}. These planets are still cooling, shrinking and stratifying from their formation \citep{Hubbard1968}. Given $\tavg{Q_\text{out}}>\tavg{Q_\text{in}}$, and assuming that average $T_\text{in}>T_\text{out}$ as on Earth, we could assume that these planets are in entropy balance and determine that they must be producing additional entropy irreversibly to balance the enhanced entropy export $\tavg{Q_\text{out}}/T_\text{out}$ to space. But why should we assume entropy balance under these conditions? More likely, these planets are secularly `ordering' (stratifying by density) as well as cooling, and accordingly losing net entropy to space. If the planets are indeed out of entropy balance, how could we measure that remotely? On a fluid planet, can we distinguish between a plausibly fast atmospheric adjustment to entropy balance and a core region that loses entropy over time? The fluid behaviors (including magnetohydrodynamics at depth) of the giant planets are still extremely unconstrained, so this is an area ripe for further observational missions.

\citet{Lucarini2009} and \citet{Bannon:2017aa} call for future climate models to routinely calculate and provide emission and absorption temperatures, in order to make possible quantitative heat engine analyses between simulations. As global climate model capabilities improve in realism as well as flexibility (i.e., \add{models capable of simulating} Mars, Jupiter, hot Jupiters, etc.), there is now potential for comparative planetary climatology to tackle the evolving Carnot and mechanical efficiencies of ancient and alien worlds. This approach may help bound the possible range in climates of exoplanets, which for the foreseeable future can be studied \add{observationally} only as point sources of light.


\section{Modeling the second law of thermodynamics}
\label{sec:models}

Given the impracticality of conducting controlled experiments on the climate system and the sparseness of our networks for 
observing the atmosphere, ocean, and land surface,
numerical models represent an essential component of the climate researcher's toolkit.
Climate models are used as numerical laboratories to test hypotheses about how the climate system operates, as state estimation tools to study aspects of the climate system that go beyond those accessible to observations, and as tools for projecting future climate change\footnote{Numerical weather prediction models, closely related to global climate models, are also used for weather forecasting.}.
Such models implement numerical approximations to physical laws including conservation of energy, mass, and momentum in order to solve for the evolution of the system. 
As discussed in section \ref{sec:irr_procs}, these conservation laws, combined with a suitable definition of entropy, are sufficient to specify the entropy budget. But developing models of the climate that produce a realistic entropy budget that satisfies the second law of thermodynamics remains challenging. In this section, we discuss some of the issues raised when attempting to model the second law in the context of two types of climate models: cloud-permitting models (section \ref{sec:CRMs}) and global climate models (section \ref{sec:GCMs}).



\subsection{Cloud-permitting models}
\label{sec:CRMs}

Cloud-permitting models are numerical models of the atmosphere with horizontal grid spacing $\lesssim 2$ km, giving them sufficiently high resolution to represent at least the largest convective cloud systems explicitly. This contrasts with global climate models, discussed in the next subsection, for which moist convection cannot be resolved on the model grid.
Because of their large computational cost, cloud-permitting models are usually run on regional domains or idealized domains that do not encompass the whole Earth\footnote{Recent advances in computing technology are beginning to allow for global-scale weather and climate models that approach cloud-permitting resolutions \citep[see e.g.,][]{Stevens2019,Wedi2020}.}. For example, section \ref{sec:conv} presents the results of idealized simulations of RCE using a cloud-permitting model on a domain roughly $200\times 200$ km$^2$ in size.

Although their horizontal grid spacing is much smaller than grid spacings typical for global climate models, \add{cloud-permitting models remain too coarse to properly resolve individual clouds and their associated turbulence. Explicit resolution of atmospheric turbulence down to the inertial subrange is estimated to require grid spacings of 100 meters or less \citep{Bryan2003b}, which is still many orders larger than} that required for direct numerical simulation (DNS) of the atmosphere\footnote{DNS has been applied to understand the detailed dynamics of cloud entrainment \citep[e.g.,][]{Mellado2018} and microphysics \citep[e.g.,][]{Vaillancourt2002}, but computational constraints currently limit the accessible Reynolds numbers far below the requirements for even a single cloud life-cycle, let alone the global atmosphere.}. As a result, cloud-permitting models do not explicitly resolve the diffusive molecular fluxes of heat, water, and momentum that are involved in irreversible entropy production in the atmosphere. Rather, the effects of these molecular fluxes must be approximated by the model's subgrid-scale turbulence parameterizations. Such parameterizations assume that the turbulent cascade of kinetic energy toward the molecular scale may be expressed in terms of the model's resolved-scale flow. If this assumption is satisfied, the frictional dissipation rate, and the associated irreversible entropy production, implied by the parameterized subgrid-scale momentum transports provide a good approximation to the dissipation rate and irreversible entropy source owing to viscosity in the atmosphere \citep{Romps2008}. In other words, we expect subgrid-scale momentum transport to produce entropy irreversibly, thereby satisfying the second law of thermodynamics \citep{Gassmann2019}.



In our simulation of RCE described in section \ref{sec:conv}, subgrid-scale fluxes of momentum are indeed associated with positive frictional dissipation and positive irreversible entropy production (see table \ref{table:RCE_budget}). Subgrid-scale fluxes of heat, however, are associated with a small but systematic sink of entropy. As has been noted by previous authors \citep{Goody2000,Romps2008,Gassmann2015}, parameterized turbulent heat transport does not necessarily produce an entropy source of the same sign as molecular diffusion of heat in the atmosphere. The reason for this is that the parameterization must account for heat transport by turbulent air motions on scales smaller than the grid in addition to diffusion of heat at the molecular scale. In particular, vertical turbulent heat transport is associated with the exchange of air parcels at different height levels. But in stably stratified conditions, air does not spontaneously move vertically; rather, work must be done against the stratification to produce vertical motion. In the atmosphere, and at resolved scales within numerical models, the energy required for this work is provided by the kinetic energy associated with turbulence. But for scales smaller than a model's grid, turbulence is not resolved, and the required work is supplied by the internal energy of the air itself. Such an energy conversion from internal energy to work corresponds to a sink of entropy, and if it were to occur spontaneously it would violate the second law of thermodynamics \citep{Gassmann2015}\footnote{Traditional parameterizations of subgrid-scale water vapor transport can also lead to local entropy sinks because of the work required to diffuse water vapor vertically \citep{Gassmann2015}.}.

In numerical models, parameterized turbulent heat transport does not occur in isolation; rather, it is associated with turbulent transports of momentum and mass. Previous authors have argued that the negative entropy production owing to parameterized turbulent heat transport may be reconciled with the second law by recognizing that turbulent heat transport and turbulent dissipation of kinetic energy are different aspects of the same turbulent cascade \citep{Akmaev2008,Priestley1947}. According to this view, the second law is satisfied provided that the total entropy production associated with all turbulent transports is positive. In shear-driven turbulence layers, it may be shown that this condition is guaranteed if the Richardson number (a nondimensional ratio representing the relative importance of buoyancy compared to shear) is below a critical value. In many turbulence parameterizations, the critical Richardson number is taken to represent the onset of shear-driven turbulence \citep[e.g.,][]{Lilly1962}, ensuring that subgrid-scale turbulent heat and momentum transports only occur when they would result in a net positive irreversible entropy production as required by the second law. 

\citet{Gassmann2015} and \citet{Gassmann2018} have recently argued against the above view, suggesting instead that the second law requires positive entropy production for both parameterized turbulent heat transport and parameterized turbulent momentum transport individually. The authors develop a formulation of turbulent heat transport that satisfies this constraint by including a subgrid work term in the mechanical energy equation. Effectively, this formulation shifts the energy source of work done by turbulence against the stratification from the internal energy of the fluid to the resolved-scale motion. \citet{Gassmann2018} provides evidence that this formulation allows for a more realistic simulation of a dry gravity wave, but it is at present not widely adopted within the field. Clearly, continued research is needed to further clarify the requirements placed by the second law of thermodynamics on the formulation of turbulence parameterizations used in cloud-permitting models \citep[see e.g.,][]{Gassmann2019}.

\subsection{Global climate models}
\label{sec:GCMs}

Global climate models, also known as general circulation models, are numerical models of the atmosphere, land, and ocean, that are used for both weather prediction and climate projection. Because they must cover the entire planet, global climate models are typically run at lower resolution than cloud-permitting models discussed above, and their horizontal grid spacing within the atmosphere ($\gtrsim 20$ km) is too coarse to resolve convective clouds. Since clouds and their associated circulations are responsible for a large portion of the irreversible entropy production in the atmosphere, evaluating and interpreting the entropy budget of a global climate model presents a particular challenge.


\citet{Johnson1997} was one of the first to explicitly link the entropy budget of global climate models to biases in their simulation of the atmosphere. 
The author argued that numerical dissipation in climate models leads to an artificial source of entropy that spuriously increases the material entropy production of the simulated atmosphere. In order to maintain a steady state, an opposing error in the simulated entropy import must also be present. \citet{Johnson1997} suggested that this could occur via a cold bias in the model's temperature field, providing an explanation for ``the general coldness of climate models''. As pointed out by \citet{Lucarini2011b}, however, a bias in the entropy import to the atmosphere is just as likely to be associated with a bias in the radiation field as in the temperature field. Moreover, the cold bias referred to by \citet{Johnson1997} is much reduced in more recent generations of global climate models \citep{IPCCch92013}.



\citet{WoollingsThuburn2006} investigated numerical entropy generation in climate-model simulations of a dry atmosphere in the absence of radiative heat transport or heat exchange with the surface---effectively a thermodynamically isolated atmosphere.
The authors found both positive and negative numerical entropy sources, contradicting Johnson's assumption that numerical entropy generation acts solely to increase the total internal entropy production of a simulated atmosphere. 
Moreover, numerical entropy sinks within a fluid that is otherwise isolated correspond to a local violation of the second law of thermodynamics. 
To prevent such occurrences, \citet{Liu2005} suggested an ad-hoc procedure in which the diabatic heating rate within a model is altered to ensure consistency with the second law.


A more rigorous solution is to employ numerical formulations of the governing equations that maintain their \add{Poisson}-bracket structure upon discretization \citep{Gassmann2008}, thereby 
reproducing exact conservation of energy and entropy (in the absence of non-conservative terms), and avoiding the problem of artificial numerical sources of entropy \citep{Gassmann2019}. But while such numerical formulations are beginning to be used in cloud-permitting models \citep{Gassmann2013}, they are not typically used in global climate models, which still suffer from numerical errors in their energy and entropy budgets \citep{Lucarini2011b,Irving2021}. \add{Similar numerical errors cause imbalances in the water budgets of global climate models \citep{Liepert2012}, with follow on effects for the energy and entropy budgets due to the latent heat carried by water substance.} Many global climate models also employ simplified thermodynamic formulations that neglect processes such as the heating owing to frictional dissipation \citep{Pascaleetal2011}. Care must be taken to evaluate the entropy budget of such models so that it may be compared meaningfully to that of the Earth \citep{Pauluis2002}.

As in the case of cloud-permitting models, irreversible entropy production in global climate models is not modeled explicitly; it occurs within parameterizations that calculate the effect of processes that occur at subgrid scales. Because of their low resolution, global climate models require parameterizations for processes such as ocean mesoscale eddies and atmospheric convection that are not required by higher-resolution models. Developing accurate parameterizations for these processes remains an ongoing challenge. For example, errors in moist convection parameterizations have been argued to be responsible for long-standing biases in the tropical precipitation distribution simulated by global climate models \citep[e.g.,][]{Oueslati2013}.
Evaluating the ability of parameterizations of moist convection to accurately represent the second law of thermodynamics therefore provides a potential pathway toward their improvement.
However, because convection parameterizations represent both reversible processes (e.g., the generation of kinetic energy by cloud motions), and irreversible processes (e.g., vapor mixing and irreversible evaporation at the cloud edge) within the atmosphere, evaluating their compliance with the second law remains a nontrivial theoretical challenge \citep{Gassmann2015}. 






The limitations in the representation of the second law in global climate models, and in cloud-permitting models as discussed in the previous subsection, lead to errors in their simulation of the \add{atmospheric and oceanic flow}.
Such errors are likely to be quantitatively small, but if they are systematic, they may nevertheless be consequential for the mean climate and its statistics. For instance, \citet{SinghOGorman2016} reported that the irreversible entropy source due to vapor diffusion in simulations of RCE was strongly sensitive to vertical resolution, and this resulted in resolution dependence of the simulated mechanical efficiency of moist convection.
Given these potential sensitivities, it is our view that a firm theoretical foundation for the representation of the second law of thermodynamics in climate models should be a goal of model developers. Increased availability and use of thermodynamic diagnostics for the evaluation of climate models, as recently advocated by \citet{Laliberteetal2015} and \citet{Lembo2019} provides one possible step toward this goal. \add{Promisingly, such diagnostics are beginning to be included in standard analysis packages for the evaluation of global climate models \citep[e.g., the Earth System Model Evaluation Tool;][] {Eyring2020}.}
 
 
\section{Variational approaches for climate and geophysical flows}
\label{sec:Hamiltonian}

The second law of thermodynamics implies that an isolated system evolves toward a state of maximum entropy. The eventual state of such a system may therefore be solved through an extremization procedure using variational methods. While Earth's climate is not isolated and exists far from equilibrium, similar variational approaches have nonetheless found a range of applications in the literature.

In particular, variational methods have been used to define measures of the amount of energy ``available'' to do work on the climate system. This literature involves an extremization of a particular energy reservoir under the constraint that the total energy of the (presumed isolated) climate system is constant. We discuss two examples of such measures in section \ref{sec:entropic_energies}.

Furthermore, certain long-lived coherent structures in planetary fluids exist in an ``inertial'' regime in which both forcing and dissipation are weak. Such flows are amenable to analysis through statistical mechanics techniques \add{where a quasi-steady solution is determined through the maximization of an entropy variable}, despite the forced-dissipative nature of the broader climate system. We discuss statistical mechanical approaches to geophysical fluid dynamics (GFD) in section \ref{sec:statmech}.

Lastly, a controversial hypothesis due to \citet{Paltridge1975} extends the idea of entropy maximization to forced-dissipative systems by arguing that such systems tend to maximize their entropy production rate: this is the Maximum Entropy Production (MEP) principle. We critically examine the MEP principle in section \ref{sec:MEP}, concluding that its physical basis remains unclear, and its application to the climate system remains speculative.

\subsection{Entropic energies}
\label{sec:entropic_energies}

The first law does not distinguish between heat and work, but the second law breaks that symmetry; according to the second law, work can be completely converted to heat but heat cannot be completely converted into work. Thus the second law of thermodynamics indicates that the universe is irreversibly and monotonically transforming energy from other forms into unusable internal energy.
In previous sections, this principle was expressed in terms of entropy, but it may also be expressed in energetic terms by defining a measure of the energy \textit{available} to drive motions in a fluid. In this section, we explore two parallel threads of research that seek to provide a definition of such an energy measure.

Section \ref{sec:APE} discusses a common approach in atmospheric science and physical oceanography, that of quantifying the Available Potential Energy (APE) of the climate system as a source of kinetic energy. Section \ref{sec:exergy} then describes exergy as an alternative and more formal measure of departure from thermodynamic equilibrium and briefly reviews some applications to the climate system. The reader is referred to \citet{Tailleux2013a} for a detailed review of APE, exergy and related concepts.

\subsubsection{Available potential energy}
\label{sec:APE}

\citet{Margules1905}\footnote{As translated by \citet{Abbe1910}.} and \citet{Lorenz1955} pioneered the quantification of an atmosphere's ability to drive motion \cite{McWilliams2019}. Lorenz defined the Available Potential Energy (APE) of the atmosphere $A$ as the component of the total potential energy $P$ that may be converted to kinetic energy of the general circulation\footnote{\add{Convective available potential energy (CAPE), introduced in section \ref{sec:convtheory}, has some similarities to the concept of APE, but it is defined as the energy available to an infinitesimal parcel on an ascent through the atmosphere rather than for the fluid as a whole.}}. Evaluating the APE for Earth's atmosphere, \citet{Lorenz1955} found it to be a very small fraction of the total potential energy $P$.

\begin{figure}
\centering
\includegraphics[width=6cm]{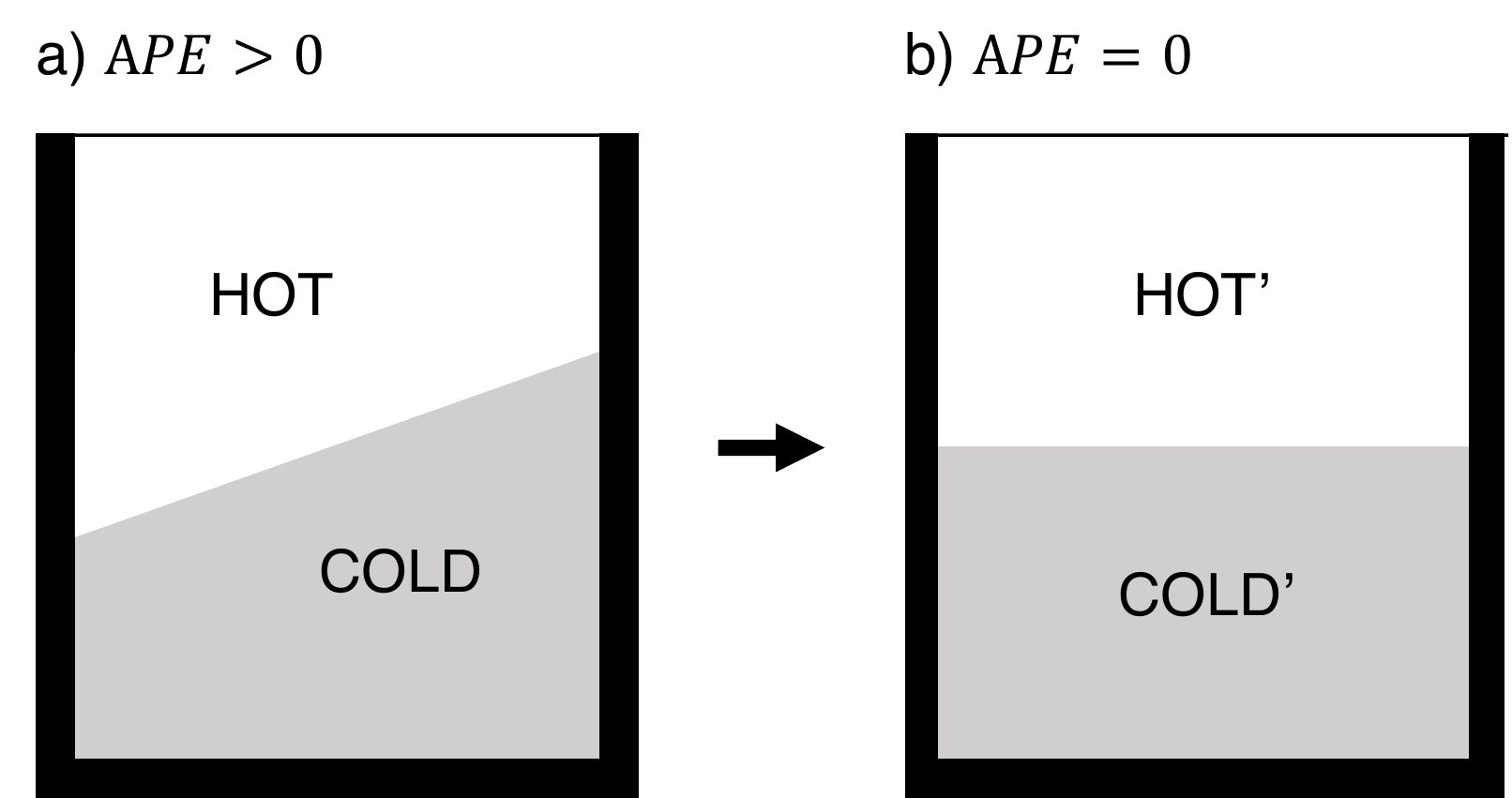}
\caption{Schematic of release of available potential energy. (a) Motionless initial water glass with a non-zero interface slope between stably stratified water masses. (b) Motionless final state after available potential energy $A$ was released, converted to kinetic energy $K$, and then converted to internal energy $IE$ \add{and non-available potential energy through frictional heating and the associated thermal expansion}.}
		\label{fig:waterglass}
\end{figure}

The concept of APE is easy to illustrate. Imagine a water glass containing hot water above cold water with a tilted interface between the two water masses (Fig. \ref{fig:waterglass}a). At an initial time $t_0$, the water is at rest and in hydrostatic balance. If the system is allowed to spontaneously evolve, one may intuit that there will be rapid turbulent motion as the water masses reduce the interface slope to zero. Thus, even though each column individually was in hydrostatic balance initially, the horizontal gradient of density provides a reservoir of potential energy that can be spontaneously converted to kinetic energy: it is ``available''. After sufficient time has passed, the turbulent motions cease and both the hot water and cold water masses become slightly warmer due to the frictional heating associated with the dissipation of the kinetic energy. But the potential energy is lower than that of the initial state\footnote{In the Boussinesq limit, there is no internal energy reservoir, and instead frictional dissipation of motion returns energy to the potential energy reservoir. Thus if the fluid considered were Boussinesq, the potential energy of the initial state and final state would be the same.}. Had the glass been stably stratified from the beginning, without horizontal gradients, it would have remained motionless, maintaining its initial potential energy. APE provides us with a mathematical tool to determine when and how much potential energy may be released by spontaneous fluid motion.

Lorenz's APE theory assumes an initially stable vertical profile, so energy cannot be converted to work from purely vertical rearrangement, and it requires the definition of a reference state that minimizes the total potential energy $P$ subject to some constraints. \citet{Lorenz1955,Lorenz1967} proposed that the appropriate reference state is one in which the atmosphere is reversibly and adiabatically re-arranged (holding mass constant) to a static state of minimum potential energy $P_{r}$, leaving the residual potential energy available to drive motion. Such a reference state is in mechanical equilibrium (motionless and in hydrostatic balance), but not in thermal equilibrium, given there remains a vertical temperature gradient associated with vertical stratification. The APE is then defined over the entire atmosphere \cite{PeixotoOort1992}:

\begin{equation}
A = \int_{\Omega_A} (P-P_{r}) \rho dV.\label{eq:APEdefinition}
\end{equation}
Letting $C(X,Y)$ indicate a rate of energy conversion from reservoir $X$ to reservoir $Y$, the Lorenz energy cycle is:

\begin{equation}
\tavg{\dot{A}} = \tavg{\dot{G}} - \tavg{C(A,K)}\\ 
\label{eq:APEbudget}
\end{equation}
\begin{equation}
\tavg{\dot{K}} = \tavg{C(A,K)} - \tavg{\dot{D}}.
\end{equation}
\add{In steady state (indicated by the time-averaging brackets)}, APE is generated at rate $\tavg{\dot{G}}$ and converted to kinetic energy \add{at rate} $\tavg{\dot{K}}$. Kinetic energy is ultimately removed by frictional dissipation at rate $\tavg{\dot{D}}$.

The original APE of \citet{Lorenz1955} was only defined for a global integral over small-amplitude perturbations from the resting state. The integrand in \eqref{eq:APEdefinition} may be locally positive or negative, but the APE is positive definite upon integration. 

\citet{VanMieghem1956} quickly pointed out limitations in the assumptions of the reference state, remarking: ``the hydrostatic hypothesis and the assumption of incompressibility which are commonly used in atmospheric dynamics are far more dangerous to introduce in energy studies''. \citet{Pearce1978} formulated a more complete APE that included the impact of energy available upon adiabatic rearrangement in the vertical direction and that is valid without assuming that the atmosphere is in hydrostatic balance.

The need to define a positive-definite local APE (an APE density) was addressed by \citet{HollidayMcIntyre1981} for a stratified, incompressible fluid. That same year, \citet{Andrews1981} developed a theory for local APE density valid for nonhydrostatic and compressible flows. He identified an additional energy reservoir in compressible atmospheres termed the available elastic energy. Lorenz's APE was shown to always be less than or equal to a volume integral of the APE density \cite{Andrews1981}. \citet{Shepherd1993} derived a quantity equivalent to the APE density using Hamiltonian methods and called it a ``pseudoenergy'' (we will briefly review Hamiltonian GFD in section \ref{sec:statmech}).

\add{The studies discussed above generally consider only the ``dry'' component of APE, neglecting the potential energy associated with the presence of water in the atmosphere. However,} when the effects of moist processes are included, the estimated APE production rate is far greater than the observed or estimated frictional dissipation rate \citep{Pauluis2007}, indicating that frictional dissipation $\dot{D}$ is not the only (nor even primary) sink of ``moist'' APE. \citet{Pauluis2007} formulated the first budget for APE that includes sources, sinks, and diffusion of water content in a moist atmosphere. He showed that heat and water diffusion, precipitation, and re-evaporation can each change APE. But despite the conceptual similarity between irreversible entropy generation and APE destruction, there is no direct mapping between the sign of the change in APE and the occurrence of irreversible moist processes; the sign of the APE change is also a function of the vertical position of the air parcels in which these processes occur relative to the vertical position of the same air parcels in the adiabatically-rearranged minimum-$P$ reference state. In a moist atmosphere, the vertical rearrangement of air parcels towards the reference state can be complex. For example, a moist parcel of air may be in a statically stable environment within the unsaturated boundary layer, but if it is lifted adiabatically until the water vapor begins to condense and release latent heat, the parcel may nevertheless acquire a higher altitude in the minimum-$P$ reference state, even if no horizontal gradients are present. 

Evaluating the APE involves applying a sorting algorithm to find the appropriate reference state, and this can be computationally intensive, particularly if moisture is considered \citep{Lorenz1979,RandallWang1992,HieronymusNycander2015,Stansiferetal2017,SuIngersoll2016}. Nevertheless, APE and the related Lorenz energy cycle are widely applied in the atmospheric (usually using a dry formulation) and oceanic literature \citep[see e.g.,][]{Lembo2019,Hughes2009,Lietal2007,Storchetal2012}. The zonal mean of APE has also been shown to scale with the eddy kinetic energy or `storminess' of the midlatitude storm tracks \citep{SchneiderWalker2006}, and this relationship has been used to help explain future changes in storminess projected by global climate models \citep{OGorman2010}.


\subsubsection{Exergetics}\label{sec:exergy}


Lorenz's and subsequent approaches that minimize potential energy to define a reference state may be characterized as a `mechanical' perspective \citep{HuangMcElroy2015}, in which the minimum potential energy reference state is not necessarily a state to which the atmosphere spontaneously tends. In contrast, the concept of exergy facilitates a second-law based thermodynamic perspective on the availability of energy to do work in a fluid which has been developed in parallel with APE theory \citep[e.g.][]{Keenan1951,DuttonJohnson1967,Dutton1973,LivezeyDutton1976,Karlsson1990,Marquet1991,Marquet1993,Kucharski1997,Fortak1998,Bannon2005,Bannon2012,Pengetal2015,HuangMcElroy2015,Marquetetal2020}. Some physics curricula and most engineering programs teach the concept of exergy \cite[coined by][]{Rant1956}, which is used widely in the energy industry \add{\citep{Hermann2006}}. In the climate literature it has also been called `static entropic energy' \citep{Dutton1973}, `static exergy' \citep{Karlsson1990}, `availability' \citep{Bannon2013}, `available energy' \citep{Bannon2005,Bannon2012}, or `available enthalpy' \citep{Marquet1991}, with small variations in formulation and assumptions.  The subfield of study generally may be called exergetics \citep{Karlsson1990}. The relationship between APE and exergy of the atmosphere has been discussed in \citet{Dutton1973,Marquet1991,Kucharski1997,Fortak1998,Tailleux2013a}.

Exergy $B$ is the amount of energy in an out-of-equilibrium system that can be converted to useful work upon moving to a reference state that is in thermodynamic equilibrium with its environment. The reference state has lower total energy than the original state. For an open system such as a power plant, we may imagine the system exporting energy reversibly until it reaches the reference state; the amount of energy exported is equal to the exergy. \add{\citet{Hermann2006} estimated the exergy available in various climate system components, disregarding practicality. For example, if we were theoretically able to extract the global gravitational and chemical exergy available in freshwater precipitation back to the salty ocean surface, that would yield 44 TW of power. Wind energy is currently extracted by surface-based wind turbines at a global rate of 743 GW \citep{GWEC2021}, with a theoretical (and undesirable) upper bound of \addr{at least 400 TW \citep{Marveletal2013}}.}

\add{For exergy calculations of the climate system as a whole, defining the ``environment'' becomes problematic. In this case, one may instead assume that the climate system is isolated in order to define an exergy that} measures the portion of energy within the system that can perform work on the system itself. As for the open-system case, the reference state has lower total energy than the initial state, and the difference is the exergy. But in the isolated-system case, the system cannot evolve to the reference state because it cannot export energy.

To define both APE and exergy, the reference state is reached assuming total mass is held constant. However instead of minimizing total potential energy, exergy is defined by minimizing the Gibbs free energy of the reference state at a reference temperature $T_r$. The reference state is then both static (mechanical equilibrium) and isothermal at $T_r$ (thermal equilibrium), with reference profiles in the vertical of pressure $p_r(z)$ and specific entropy $s_r(z)$. The difference in the Gibbs free energy between the reference state and the initial state is the exergy. The static exergy $B$ of a single fluid may therefore be defined \cite[e.g.,][]{Fortak1998,Bannon2005}:
\begin{eqnarray}
B = &&h(s,p)-h(s_r,p_r) - \alpha(p-p_r) - T_r (s-s_r)\nonumber\\
=&&[h(s,p)-h(s,p_r) - \alpha (p-p_r)] \label{eq:exergy}\\
&&+[h(s,p_r)-h(s_r,p_r) - T_r (s-s_r)]\nonumber
\end{eqnarray}
for enthalpy $h=u+p \alpha$. 

\begin{figure}
\centering
\includegraphics[width=8.6cm]{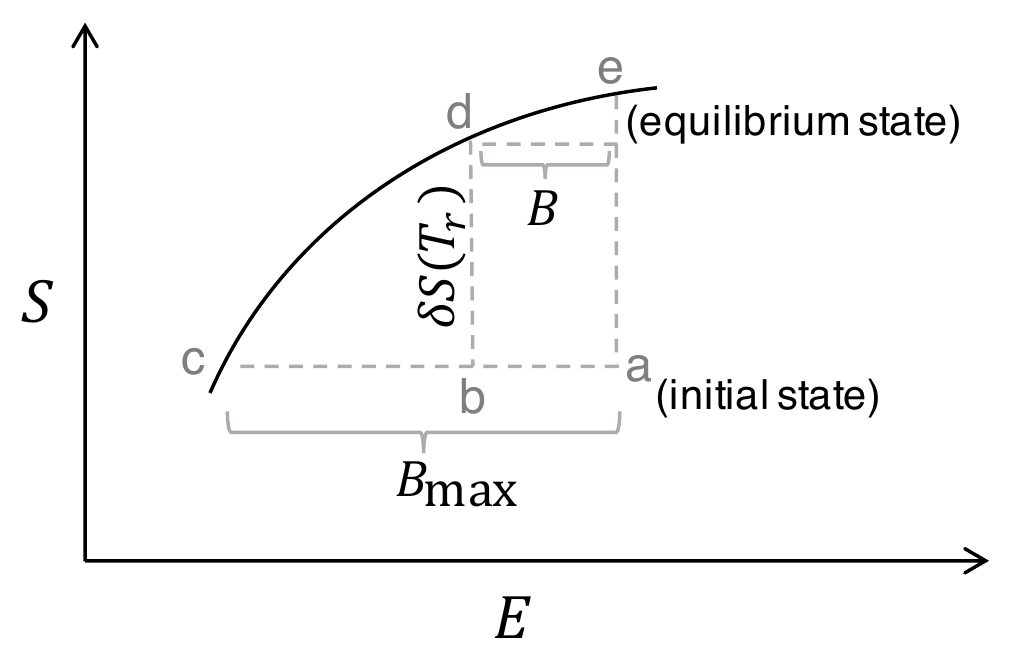}
\caption{Total energy $E$ vs. total entropy $S$. The solid black curve $S(E)$ is the entropy of an isolated system of total energy $E$ in thermodynamic equilibrium. For an isolated system, the initial state $a$ and final equilibrium state $e$ have the same total energy. The reference state $c$ maximizes exergy $B_\text{max}$ if it can be reached isentropically \citep{Karlsson1990}. If $T_r$ is chosen such that the reference state cannot be reached isentropically, exergy available for mechanical work $B<B_\text{max}$. Adapted from \citet{LandauLifshitz1978} Fig. 2 and \citet{HuangMcElroy2015} Fig. 5.}
		\label{fig:energyentropydiagram}
\end{figure}

The relationship between the energy of the reference state and the exergy is sketched in Fig. \ref{fig:energyentropydiagram}; the solid black curve gives the entropy $S$ as a function of total energy $E$ for an isolated system in thermodynamic equilibrium. The initial out-of-equilibrium $(S,E)$ state is at position $a$, and the system's total energy is fixed to lie along the line $a-e$. The lower-energy reference state can be identified as a position along the function $S(E)$ bounded by points $c$ and $e$. The function $S(E)$ can be interpreted as the partition between the reference state energy (to the left of the curved line) and the remaining exergy $B$ (to the right of the curved line), such that their sum equals the total energy. To be consistent with the definition of static exergy above, we note that we neglect the small kinetic energy portion of the initial state's total energy in the figure.

The reference temperature $T_r$ may be chosen such that the reference state can be reached reversibly, so that entropy production $\delta S=0$ \citep{Karlsson1990}, which is equivalent to selecting a $T_r$ to maximize exergy $B=B_\text{max}$ (position $c$ in Fig \ref{fig:energyentropydiagram}). A reference state at lower $T_r$ can only be reached by reducing the total entropy of the isolated system, which violates the second law; a reference state at higher $T_r$ cannot be reached reversibly (position $d$ in Fig. \ref{fig:energyentropydiagram}), and the subsequent entropy production $\delta S>0$ reduces the remaining exergy $B<B_\text{max}$. \add{Maximum exergy $B_\text{max}$} is the upper bound of work that can be extracted from a nonequilibrium system under purely reversible processes; irreversible processes, including moist irreversible processes that don't perform work, act to destroy exergy. The second law of thermodynamics may therefore be re-expressed in energy variables as $dB/dt\leq0$ for an isolated system: the universe is irreversibly destroying exergy as the `quality' of the fixed amount of energy decreases. This formulation makes clear the major difference between heat and work: all work can be converted to heat, but not all heat can be converted to work.

The reference state at temperature $T_r$ should not be confused with the final equilibrium state that the system would reach if isolated and allowed to evolve irreversibly (position $e$ in Fig. \ref{fig:energyentropydiagram}). This equilibrium state is motionless after all of the exergy has been dissipated to internal energy, maximizing entropy, but maintaining the total energy of the system at its initial value. 


Similar to the decomposition of APE \citep{Andrews1981}, exergy of an ideal gas can be split into two contributions \citep{Karlsson1990,Marquet1991,Bannon2005,Bannon2012}: an available elastic energy (first term in brackets of \eqref{eq:exergy}; zero in an incompressible atmosphere), and an available potential energy (second term in brackets of \eqref{eq:exergy}). \citet{Bannon2012} pointed out that the exergy could also be approximately partitioned into a component that could be released upon a stable resorting of each vertical column (the Available Convective Energy), and then upon completion of that process, the remaining horizontal gradients provide an Available Baroclinic Energy which drives the large scale horizontal flows. The difference between the traditional APE approach and an exergetic one can be reconciled if one considers the exergy within each layer of the atmosphere individually \cite{Kucharski2001,Pengetal2015}.

For moist atmospheres, one must also consider a chemical departure from a saturated reference state in order to define exergy. This requires an additional additive term \add{$g_x(q_x-q_{x0}$)} in \eqref{eq:exergy} for each \add{species $x$, to take into account the chemical potentials (\ref{sec:multi})}. Exergy correspondingly increases in a moist atmosphere due to an additional available chemical energy \cite[e.g.][]{Karlsson1990,Marquet1991,Bannon2005}. Because the climate system exists in a gravitational potential field, \add{the specific Gibbs free energy of a substance} at equilibrium is a function of the geopotential, such that \add{$g_x + g_{\earth} z =$}constant \add{for specific Gibbs free energy $g_x$ of} gas species $x$ (dry air and water vapor)\footnote{At equilibrium there are no condensed species aloft.}, gravitational acceleration $g_{\earth}$, and altitude $z$ \citep{HatsopoulosKeenan1965}. One must also decide whether water mass is conserved in the atmosphere upon rearrangement to the reference state \cite[e.g.,][]{LivezeyDutton1976,Marquet1993} or whether the atmosphere is, more accurately, an open system to water \cite{Bannon2005,Pauluis2011}.

There is little consensus over how to choose the reference temperature $T_r$ in exergy studies of the climate system \cite[e.g.,][]{Dutton1973,Karlsson1990,Bannon2012} and it may be arbitrary \citep{Kucharski1997}. Generally it is chosen to be around 250~K. \citet{Karlsson1990} chose the reference temperature as the one that minimized the entropy difference between the atmosphere and its equilibrium state, and \citet{Bannon2013} showed that this was equivalent to maximizing exergy. A difficulty arises because there is no external thermostat setting the temperature toward which our climate may evolve. In our previous power plant example, it is reasonable to assume that the plant is surrounded by a reservoir of roughly constant temperature; namely, the atmosphere. For the climate as a whole, the only technically correct reference temperature is the several-kelvin chill of deep space, and even then only after the death of the sun. Instead, one can imagine an Earth system that at some point suddenly becomes isolated, no longer receiving nor losing heat and entropy to space\footnote{This is a difficult thought experiment because radiation between components of the Earth system is so important (consider the atmospheric longwave warming of the surface). If the Earth system becomes thermodynamically isolated, can the components still exchange energy radiatively? If so, how does one deal with the radiative energy directed upward at the top of the atmosphere?}. Under this hypothetical condition, entropy would increase toward a maximum value while total energy would remain constant. What would the final state look like and how would an unforced atmosphere freely evolve toward it? It would certainly spin down due to friction; it would become saturated due to contact with a frozen water surface but cloudless because of hydrometeor fallout; and eventually, due to the very slow process of molecular heat transfer, the atmosphere would become isothermal (pressure and density would not homogenize because of the gravitational potential field). On even longer timescales, the various gases of the dry air mixture itself would begin to fractionate, as is hypothesized to be happening on the giant planets.


An exergetic budget provides an alternative way to formulate a mechanical efficiency of the climate system, which can be evaluated quantitatively using climate model output \citep{Karlsson1990,Bannon2012}. \citet{Karlsson1990} defines a climate efficiency as `the global net conversion to kinetic exergy divided by the global net inflow of static exergy under the assumption of local thermodynamic equilibrium', approximately $C(B,K)/\dot{B}$.
A dry exergetic analysis has been proposed by \citet{Lucarini2009} and recently applied to Earth's climate to study its seasonality \citep{HuangMcElroy2015}.

Both the APE and exergetic frameworks provide a perspective of irreversibility focused on energy rather than entropy. The literature that explores exergy in the climate system more closely appeals to fundamental thermodynamical concepts. 
However the study of the exergetics of the atmosphere is evidently hampered by the proposal of a unique terminology for nearly every paper published in the field (an early collection of which are tabulated in \citet{Marquet1991} Appendix B). It is likely that a consensus, yet forthcoming, may make exergetics more palatable in the classroom and assist in defining and quantifying useful measures of climate efficiency. Ideally this consensus will fall on terminology already widely used in physics and engineering, so as to promote more interdisciplinary collaboration and to avoid further conflation with the classical APE development of \citet{Lorenz1955}.



\subsection{Statistical mechanical approaches for steady flows}\label{sec:statmech}

So far the discussion of the atmospheric and oceanic circulations has emphasized their forced-dissipative nature and the consequences of the system being heated at a higher temperature than that at which it cools. This can be conceptualized as the climate system being in contact with two thermal reservoirs of different temperatures, which makes thermodynamic equilibrium impossible. Energy transfer between the reservoirs occurs via overturning circulations of various scales, from thunderstorms to the global atmospheric or oceanic circulation. 

Now we consider aspects of the atmosphere and ocean that cannot be conceptualized as being in contact with two different thermal reservoirs. Such systems do not develop overturning circulations to move heat downgradient, but they instead develop quasi-horizontal flows with characteristic organization and steadiness  dominated by inertial forces relative to weak forcing and dissipation. Examples include eye-eyewall mixing dynamics in tropical cyclones \citep{Schubertetal1999}; mesoscale eddies in the ocean \citep{VenailleBouchet2011}; the stratospheric polar vortex \citep{PrietoSchubert2001}; Rossby wave propagation at midlatitudes under a vorticity gradient \citep{Young1987}; and Jupiter's Great Red Spot \citep{Milleretal1992}. Such coherent structures are ubiquitous in turbulent geophysical flows. Unlike the heat engine analogy employed for the climate system, these phenomena can be thought of as being thermodynamically isolated or in contact with a single thermal reservoir, not two. 

The study of such fluid equilibria is a branch of geophysical fluid dynamics (GFD) that has adapted equilibrium statistical mechanics to the fluid equations: Hamiltonian GFD. This framework applies Hamilton's principle of stationary action for conservative systems to continuum fluid mechanics using statistical techniques valid in the thermodynamic limit (infinite particles). Equilibrium statistical mechanics has been used successfully to describe a wide range of simple fluid behavior with applicability to observed large-scale flows, \add{despite the viscous, dissipative nature of real fluids.}
More detailed treatments may be found in reviews by \citet{Salmon1988,Shepherd1990,Morrison1998,Shepherd2003,MajdaWang2006,Chavanis2009,BouchetVenaille2012}. \add{Equilibrium and nonequilibrium statistical mechanics is a branch of physics with substantial potential to solve problems in climate science \citep{Marston2011}, and the second law of thermodynamics is central to its formulation.}





\subsubsection{Theoretical development}\label{sec:statmechdev}

 


Use of Hamilton's principle of stationary action provides an alternative way to derive the equations of motion using the calculus of variations. However, the typical differential approach using Newton's second law easily incorporates terms for friction, viscosity and other nonconservative forces, and this is a limitation on the applicability of Hamiltonian formulations, which only exist for conservative systems. In spite of its idealism, Hamiltonian GFD is an important and illuminating part of the literature on quasi-steady flows, and variational methods used in Hamiltonian GFD have helped to explain the stability and longevity of some well-known enduring vortices. We'll briefly survey four useful equilibrium statistical mechanics approaches and then in section \ref{sec:statmechapplications} mention some interesting geophysical applications and successes.

It might be surprising that forced-dissipative geophysical flows are amenable to variational methods. Our ocean and atmosphere are very high Reynolds number, strongly stratified fluids on a rotating planet. Due to rotation and stratification in particular, such fluids at large scales exhibit an approximately two-dimensional (2D), non-divergent flow field, which manifests as jets, waves, and vortices \citep[e.g.,][]{Flierl1987}. 
\add{The constraints imposed by this 2D character promote heterogeneous structures in steady state geophysical flows at equilibrium.}

A non-divergent flow field is described by the inviscid, incompressible 2D Euler equation,
\begin{equation}
    \frac{d \mathbf{v}_h}{d t}=-\frac{1}{\rho_0}\nabla_h p \label{eq:barotropiceq}
\end{equation}
for horizontal flow field $\mathbf{v}_h$, where we continue to use $d/dt$ as the \add{Lagrangian} derivative. Energy conservation for this system is simply conservation of kinetic energy integrated over the domain. Mechanical work is limited to the inner product of a stress and a strain \add{between scales} \citep{FangOuellette2016}. Two key aspects of \eqref{eq:barotropiceq} make the application of Hamiltonian methods feasible. Firstly, in 2D turbulence, kinetic energy is, on average, transferred to larger scales, avoiding the build up of kinetic energy at the smallest scales that would occur in 3D inviscid flow. Secondly, taking the curl of \eqref{eq:barotropiceq} gives,

\begin{linenomath}\begin{equation*}
    \frac{d\omega}{dt}=0,
\end{equation*}\end{linenomath}
where $\omega = \nabla\times \mathbf{v}_h$ is the vorticity. The vorticity is materially conserved following fluid elements and conserved when integrated over the domain.
Depending on the geometry and boundary conditions of the system, the circulation (area integral of the vorticity) and in some cases zonal \add{and/or} angular momentum are also conserved\footnote{Common lateral boundary conditions in GFD are periodic boundaries in one or more dimensions. While these domains conserve linear momentum in the periodic dimensions, they do not conserve angular momentum. This is easy to confirm with a simple thought experiment. Imagine you have a tropical cyclone centered in the middle of your doubly-periodic domain and you are positioned far in one corner of it---this corner, of course, being joined by the three other corners of the domain because there are no walls. Measure your angular momentum as a function of your tangential speed and distance from the center of the domain. Maintaining that radius, rotate some angle around the cyclone's spin axis until you have crossed through the domain edge. You find yourself `re-entering' the domain from the opposite side. Though your angular momentum should not have changed, you are now closer to the storm center than you were before. Thus the domain lacks invariance to rotation and cannot conserve angular momentum.} following from Noether's theorem that every symmetry corresponds to a conservation law. 

Strongly stratified geophysical flows under the influence of rotation are well approximated by a system known as the quasigeostrophic (QG) equations, which share the properties of the 2D Euler equations discussed above. In the QG case, the relevant vorticity variable is known as the potential vorticity, and it includes a dependence on the stratification. The QG system is also dependent on a length scale known as the Rossby radius that does not appear in the 2D Euler equations but may nevertheless be included in the statistical mechanics frameworks we will discuss
\citep[e.g.,][]{Salmonetal1976,DiBattistaMajda2000,Weichman2006}.

A remarkable property of ideal 2D continuum fluids is that any domain integral that is only a function of vorticity (or potential vorticity) is conserved. The most commonly considered functionals are integrals of \add{$\omega^n$, the vorticity taken to some integer power $n$, giving an infinite set of invariants collectively referred to as the ``generalized enstrophy integrals'' \citep{Young1987}, or Casimirs.} The corresponding infinite set of symmetries is referred to as the particle-relabelling symmetry and is due to the Eulerian description of the fluid [this is because the Poisson bracket is singular; see \citet{Morrison1998} section IV or \citet{Salmon1998} section 7.11].
Canonical Hamiltonian methods fail here \add{because the system cannot be expressed in the canonical coordinates of a position-momentum pair}. However, a number of methods have been proposed to resolve this \add{difficulty} and apply Hamiltonian methods to 2D turbulent flows.  

One such approach relies on the ``selective decay hypothesis'' \citep{BrethertonHaidvogel1976,MatthaeusMontgomery1980}. This hypothesis is motivated by the observation that numerical experiments of 2D, unforced flow evolution from random initial conditions regularly yield coherent vortices, and with sufficient time they merge \citep[e.g.,][]{Onsager1949,McWilliams1984,Montgomeryetal1992}. \citet{Kraichnan1967} showed that 2D turbulence exhibits an energy cascade to smaller wavenumbers but an enstrophy cascade to larger wavenumbers. Authors pointed out that the direction of the cascade indicated the likely impact a small amount of viscosity may have on the various conserved quantities. Energy is a `rugged' quantity because it cascades to the largest scales. Total circulation and momentum are also rugged integrals. In contrast, enstrophy and higher orders of vorticity \add{largely cascade to smaller} scales. Thus a molecular (or numerical) viscosity that preferentially removes kinetic energy at small scales is likely to remove enstrophy much more rapidly than it removes total energy. 

Ultimately, any real fluid that freely evolves, unforced, will achieve a state of rest, as viscosity eventually converts all kinetic energy to internal energy. However, there is a long intermediate period where  enstrophy (and higher orders of the vorticity) has been appreciably removed \add{at small scales} while large-scale flows lose negligible energy \citep{McWilliams1984}. This is the period of interest to atmospheric scientists and oceanographers, who observe numerous long-lasting vortices in nature. These vortices have eddy turnover (inertial) timescales that are much shorter than the forcing and dissipation timescales, so they can be modeled as isolated systems. Thus researchers sought an equilibrium statistical mechanics that could predict the large-scale nature of such observed flows in the limit of weak forcing and weak dissipation.

The selective decay hypothesis posits that quantities dominated by distribution at the smallest scales preferentially decay while ``rugged'' quantities remain nearly conserved. This leads to a statistical equilibrium state for simple quasi-2D flows \citep[e.g.,][]{Leith1984,Young1987} that corresponds to a state of minimum enstrophy. This minimum enstrophy principle has proven to be really useful and often accurate, but it obscures the analogy to classical thermodynamic systems and doesn't explicitly employ the second law. Minimizing the enstrophy according to the selective decay principle or ``principle of minimum enstrophy'' allows one to avoid defining an entropy for the system entirely. It's a variational approach but not a statistical one.


A different approach to the application of Hamiltonian methods to GFD is to simply consider a finite number of vorticity levels. A number of authors have developed an equilibrium statistical mechanical theory for point vortices by maximizing an entropy measure defined based on the vorticity distribution \citep[e.g.,][]{Onsager1949,JoyceMontgomery1973,MontgomeryJoyce1974,Cagliotietal1995}. The vorticity levels in a point vortex system are discrete \add{and only a finite number of the infinite Casimirs are non-zero}, so the regular approaches apply. 



Other methods of truncating continuous fluid systems, such as spectral truncation of higher modes, also allow 2D fluids to be treated canonically \citep{Salmonetal1976,Kraichnan1975,Carnevale1982,MajdaHolen1997,DiBattistaMajda2000}. As for the point vortex system, the spectrally truncated framework maximizes entropy while conserving the quadratic integrals given by the energy and enstrophy \citep{Kraichnan1967,CarnevaleFrederiksen1987}. The Lagrange multipliers for energy and circulation can be considered an inverse temperature and a chemical potential, respectively, drawing close parallels to classic thermodynamic systems. However in statistical mechanics, such a ``temperature'' can be negative \citep{Onsager1949,JoyceMontgomery1973}, which (while stretching the helpfulness of the temperature analogy) in fact tends to correspond to the most interesting coherent structures.

The three approaches described above, the selective decay hypothesis, the point vortex method, and spectral truncation, all apply a canonical Hamiltonian approach by constraining the number of invariants to be finite. The challenge remained to derive an equilibrium statistical mechanical theory for a noncanonical representation of a fluid, namely a fluid in the continuum limit described with Eulerian coordinates that conserves all of the infinitely many dynamically relevant Casimirs. This was accomplished by \citet{Miller1990,Robert1991,RobertSommeria1991,Milleretal1992} and is referred to as \add{RSM (Robert-Sommeria-Miller) theory}, recently reviewed by \citet{BouchetVenaille2012}. It is a mean field theory that maximizes an entropy functional subject to infinite conserved quantities to identify an equilibrium probability distribution of the vorticity field. \add{A Liouville theorem can be satisfied if cast using vorticity in Fourier space. \citet{Milleretal1992} also demonstrated that a range of previous theories for specific problems were special cases of the more general RSM theory, including the \citet{LyndenBell1967} theory of star clusters and the \citet{Kraichnan1975} enstrophy-entropy theory}.

\add{To apply RSM theory to a given problem, we allow} the vorticity field $\omega$ to be composed of a range of vorticity levels $\sigma$, with a probability density $\rho(\mathbf{x},\sigma)$ such that at each point in the domain, $\int\rho(\mathbf{x},\sigma)d\sigma=1$. The area fraction of each vorticity level is conserved as the flow evolves. This results in a locally averaged vorticity field $\overline{\omega}(\mathbf{x}) = \int\rho(\mathbf{x},\sigma)\sigma d\sigma$.

The Boltzmann mixing entropy of RSM theory is:
\begin{equation}
    \mathcal{S} = -\int\int_\Omega \rho(\mathbf{x},\sigma) \log \rho(\mathbf{x},\sigma) dA d\sigma.
\end{equation}
Setting the first variation of this functional to zero subject to the constraints of energy, circulation and the Casimirs using Lagrange multipliers, one solves for an equilibrium probability distribution $\rho_{eq}(\mathbf{x},\sigma)$ that corresponds to the most-disordered macrostate---the macrostate corresponding to the largest number of microstates. To ensure that the stationary point of $\mathcal{S}$ is a maximum, one must further check to see that the second variation is negative. The equilibrium probability distribution of vorticity may be related to the large-scale flow field of the equilibrium state, and in principle it should recover the long-time average of the circulation derived from traditional time-dependent numerical models\footnote{All equilibrium statistical mechanics development requires an assumption of ergodicity, which remains unproven and doesn't always hold up well in experimentation \citep[e.g.,][]{Brandsetal1999}.}.





\add{While RSM theory provides a rigorous approach that conserves the infinite Casimir invariants, the truncated methods described previously remain useful in practical applications of statistical mechanics to real fluids.} Although there is no theoretical justification for the neglect of higher-order conserved Casimirs in an ideal fluid \citep{Milleretal1992}, \citet{MajdaHolen1997,DiBattistaMajda2000,Ellisetal2002,MajdaWang2006} argued that retaining higher order Casimirs is impractical and unnecessary when considering equilibria of forced-dissipative geophysical flows. \citet{Turkington1999} introduced an alternative approach to maximizing entropy \add{that doesn't} conserve the Casimirs. This theory conserves only energy, circulation, and the mean and extrema of vorticity-like variables, and it is based on an understanding of entropy from information theory \citep{Jaynes1957a,Jaynes1957b}.

The minimum enstrophy approach to Hamiltonian GFD embodied in the selective decay principle has been compared favorably to maximum entropy approaches of various forms by
\citet{HuangDriscoll1994,ChavanisSommeria1996,MajdaHolen1997,Schubertetal1999,Brandsetal1999,Nasoetal2010}. \citet{Nasoetal2010} showed an equivalence between the statistical equilibrium state of a truncated fluid system that only conserved energy, circulation and enstrophy for which the entropy is maximized, and the equilibrium state solved for when enstrophy is minimized. Minimum enstrophy theory continues to be used to make predictions of steady state flows \citep[e.g.,][]{Nasoetal2011,ContiBadin2020}. In fact there remains no clear consensus in the literature about a superior formulation of entropy-maximizing equilibrium methods \add{applied to geophysical flows}, in large part because successful applications are so domain- and scale-specific.




\subsubsection{Applications}\label{sec:statmechapplications}

Turning back to the observable flows that motivated much of the theoretical development, we briefly highlight a few applications to demonstrate the usefulness of statistical mechanics in understanding large-scale flows in the inertial limit. In particular, we will discuss four such examples: mesoscale [$\mathcal{O}$(100 km)] oceanic rings, the stratospheric polar vortex, tropical cyclone eye-eyewall dynamics, and Earth-sized vortices on Jupiter. Differential planetary rotation, stable stratification and column stretching \citep[including that due to topography, e.g.,][]{Salmonetal1976} are essential aspects of real quasi-steady coherent structures. \add{None of these examples are in thermodynamic equilibrium; however their quasi-steady state behavior allows them to be treated as inviscid fluids at equilibrium, as defined by a maximum Boltzmann entropy at a fixed total energy.}

\begin{figure*}
\centering
\includegraphics[width=12cm]{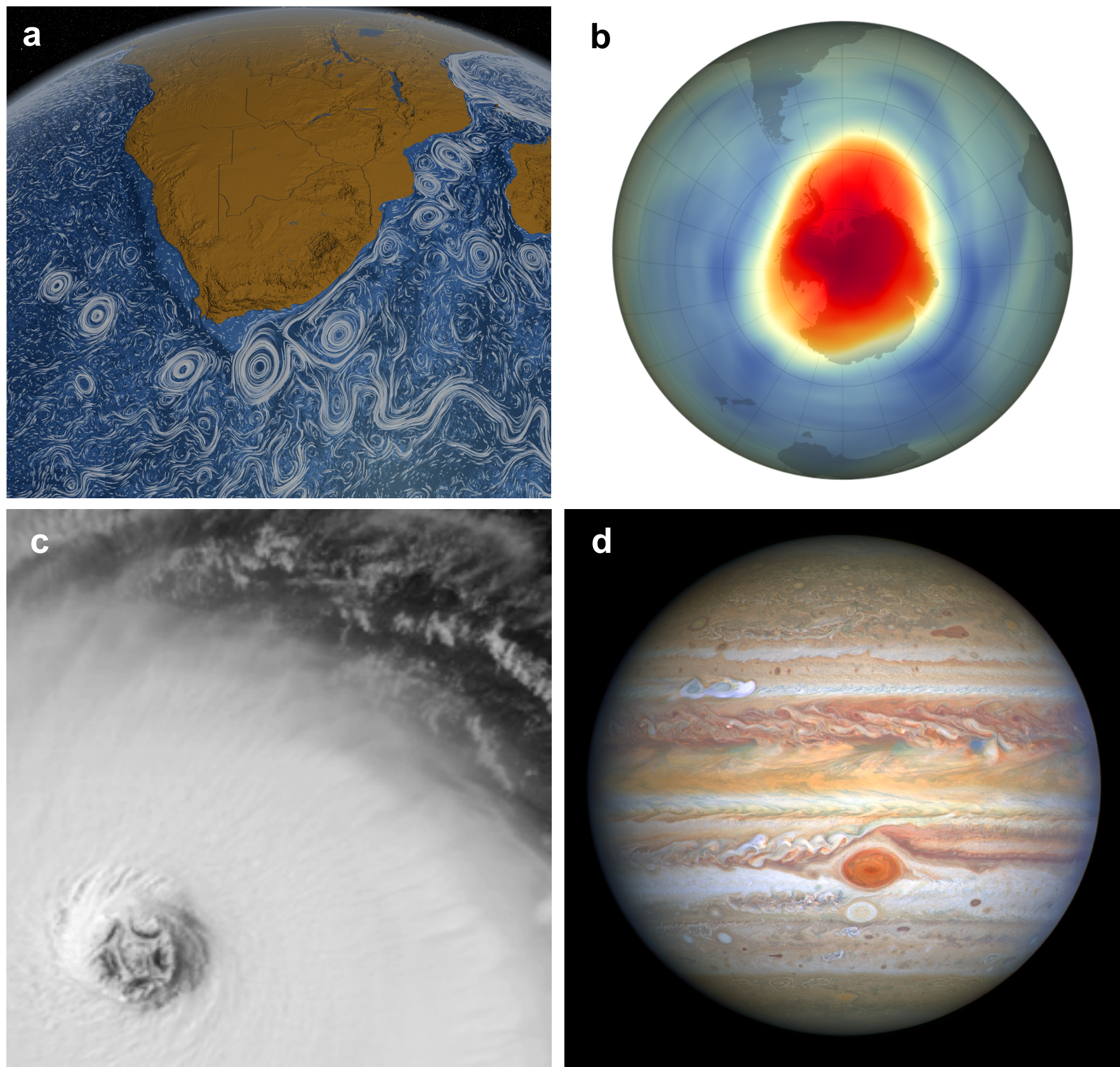}
\caption{Examples of geophysical flow phenomena that have been described using equilibrium statistical mechanics. (a) Streaklines of ocean surface flow exhibiting oceanic rings from a numerical model. Image credit: NASA/Goddard Space Flight Center Image Visualization Studio; (b) Southern Hemisphere ozone hole on October 12, 2018, indicating location of stratospheric polar vortex. Image credit: NASA Earth Observatory, NASA/NOAA Suomi NPP satellite; (c) Submesoscale vortices mixing the eyewall of Hurricane Isabel on September 12, 2003 at 1315 UTC as seen by GOES-12. Image credit: NASA/NOAA; (d) Jupiter and the Great Red Spot as seen by the Hubble Space Telescope on August 25, 2020. Image credit: NASA, ESA, STScI, A. Simon (Goddard Space Flight Center), M.H. Wong (University of California, Berkeley), and the OPAL team.} 
		\label{fig:examplevortices}
\end{figure*}

The ocean is full of coherent, relatively long-lived westward propagating vortices \citep[][and see an example in Fig. \ref{fig:examplevortices}a]{Cheltonetal2007} These vortices are surface-intensified, so they are not strongly damped by the frictional ocean floor. \citet{VenailleBouchet2011} used a 1.5 layer quasigeostrophic model of the ocean and used RSM theory to identify equilibrium flows in the inertial limit. They found that oceanic vortices (rings) can be considered as local equilibrium states, and their differential latitudinal drift (poleward for cyclones and equatorward for anticyclones) could be interpreted as an evolution toward potential vorticity homogenization. The RSM equilibrium states showed some similarity to observations, suggesting these flows are weakly forced and dissipated.

\citet{Davidetal2018} recently applied RSM theory to a forced-dissipative numerical ocean simulation and found that the maximum entropy principle, using only a truncated conservation of Casimirs, was able to account for some behavior of the nonequilibrium system. They suggest that statistical mechanics be reconsidered as a viable method of improving model parameterizations.


In the winter hemisphere, a stratospheric polar vortex (the ``polar night jet'') sets up and generally acts as a barrier to the mixing of chemical constituents between the polar cap and the midlatitudes. This is particularly consequential in the Southern Hemisphere because it maintains a region of extremely low ozone, contributing to the existence of the ozone hole in Austral spring (Fig. \ref{fig:examplevortices}b, a low-mixing regime). In the Northern Hemisphere, there are regular polar vortex breakdowns (``stratospheric sudden warmings'') during which the cold polar cap, previously maintained right on the pole and prevented from mixing by the jet-like edge of the polar vortex, suddenly experiences barotropic instability and mixes rapidly with warmer equatorward air via nonlinear wavebreaking (a high-mixing regime). It is of interest to see whether these two different regimes of polar vortex behavior can be captured by equilibrium statistical mechanics. \citet{PrietoSchubert2001} tested both maximum entropy theory and minimum enstrophy theory to determine which approach could more accurately predict a zonally symmetric equilibrium state for each regime compared to direct numerical integration of idealized initial vortices with passive tracers. The minimum enstrophy prediction was superior for the case where mixing occurred only within the polar cap. The more violent scenario of domain-wide mixing, reminiscent of a stratospheric sudden warming, was better predicted by the maximum entropy solution. 

In general, statistical mechanical techniques are limited to flows that are weakly forced and damped, and they don't apply to transient tropospheric weather systems. One of the most speculative geophysical applications of equilibrium statistical mechanics has been attempted at a very small scale in both space and time. Tropical cyclones exhibit the potential for barotropic instabilities in the eye/eyewall region, which results in highly asymmetric vorticity mixing (Fig. \ref{fig:examplevortices}c). \citet{Schubertetal1999} numerically integrated the evolution of an idealized TC eyewall vorticity ring using the 2D Euler equation. Barotropically unstable initial conditions led to similar polygonal mixing patterns. They compared the numerical well-mixed end state to solutions reached using both the minimum enstrophy theory and maximum entropy theory. Overall the maximum entropy theory predicted the numerical final state better. Whether the equilibrium state has any relevance to the actual evolution of a tropical cyclone is less clear, because such storms are very dynamic and transient (and very expensive to observe \textit{in situ}).

The Solar System's most famous vortex does not happen to be on Earth. The Great Red Spot of Jupiter (Fig. \ref{fig:examplevortices}d) has long been a source of inspiration for equilibrium statistical mechanics \citep{Milleretal1992}. Numerical integration of an annulus to which an external potential is applied, representing Jupiter's rapid rotation and zonal shear flow, demonstrated that a single large-scale coherent vortex can uniquely survive for many eddy turnover times in such an environment \citep{Marcus1988} and experimentation concurred \citep{Sommeriaetal1988}. \citet{Milleretal1992} compared the equilibrium solutions of RSM theory to the numerical experiments of \citet{Marcus1988} and found good qualitative agreement. \citet{Turkingtonetal2001} were able to retrieve realistic equilibria of Jovian anticyclones using the constrained theory of \citet{Turkington1999} if the initialized vorticity distribution skewed anticyclonically. \citet{BouchetSommeria2002} extended RSM theory to include quasigeostrophic flows in the limit of small deformation radius, and found a Great Red Spot-like vortex as a maximum entropy equilibrium structure. Under only slightly different parameters the vortex was absent, \add{suggesting a high sensitivity to environmental parameters}, and a possible explanation for the lack of such a vortex observed in Jupiter's Northern Hemisphere (which has a different zonal jet structure than the Southern Hemisphere). 

\begin{figure*}
\centering
\includegraphics[width=10cm]{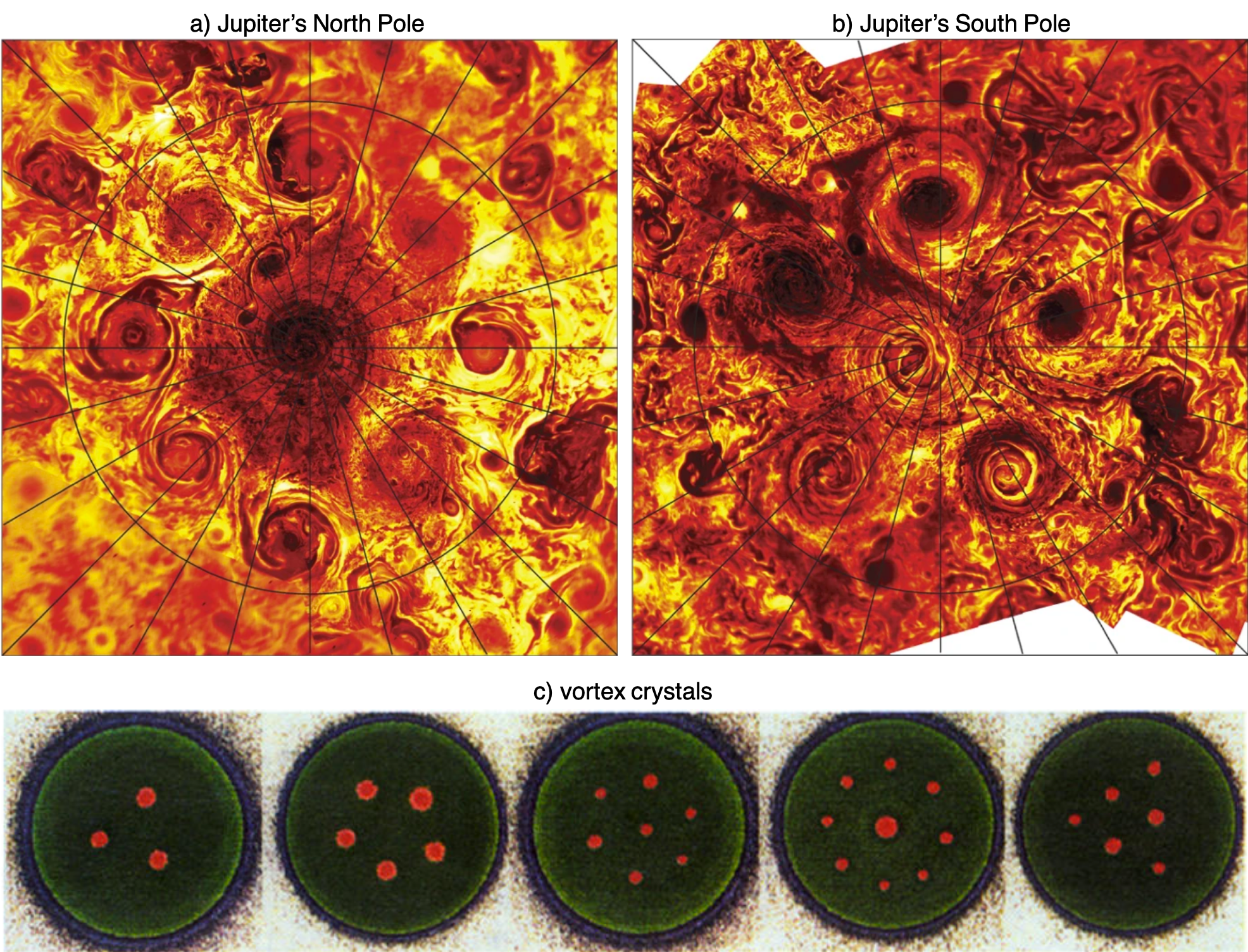}
\caption{Jupiter's (a) North and (b) South Polar Cyclones imaged by the NASA Juno mission in infrared \citep{Adrianietal2018}. (c) Vortex crystal experimental equilibria \citep{Fineetal1995}.} 
		\label{fig:Jupiter}
\end{figure*}

A new vortical puzzle has just been furnished by NASA's Juno mission to Jupiter, which captured images of the polar caps for the first time to reveal crystalline-like polygonal arrangements of cyclones centered on each pole \citep[Fig. \ref{fig:Jupiter}a,b;][]{Adrianietal2018}. The polygonal arrangements appear remarkably steady and may be the first geophysical observation of the strictly 2D vortex crystal phenomenon \cite{Adrianietal2018,Grassietal2018,TabatabaVakilietal2020}, \add{indicating a local equilibrium state (of a corresponding inviscid fluid)}, which was first identified by \citet{Fineetal1995} experimentally in a 2D electron plasma (Fig. \ref{fig:Jupiter}c). The potential of such remarkable plasma structures to have application to geophysical flows was noted by \citet{Schubertetal1999}.

Using numerical simulation and scaling theories, both shallow \citep{ONeilletal2015,ONeilletal2016} and deep models \citep{Garciaetal2020} suggest a range of statistically steady vortex-dominated behavior on giant planet caps. Such work is able to explain Saturn's isolated, steady polar cyclones. However, to date, no theory or modelling has been able to achieve a steady crystalline structure of polar vortices from random initial conditions as observed on Jupiter, though some recent progress has been made in achieving such states transiently \citep{Lietal2020}. This is a new area of research that is ripe for an equilibrium statistical mechanics study.





\citet{RobertSommeria1991} suggest that equilibrium statistical mechanical techniques could potentially be applied to a changing climate, given the separation of timescales between the rapid statistical adjustment of flows compared to the slowly varying Lagrange multipliers that represent the forced climate response. \citet{DiBattistaetal2001} showed that equilibrium statistical mechanics can predict `meta-stable' states of systems that are secularly driven, and \add{that} the most likely state varied with the forcing dynamically. Other work shows the viability of second-law based equilibrium tools for turbulent forced-damped flows \citep[e.g.,][]{Davidetal2018}. There is likely further opportunity to bridge the statistical and thermodynamical interpretations of entropy to better understand complex, out-of-equilibrium flows on Earth and other planets, given the domain-specific success of each approach.

\subsection{Maximum entropy production principle controversy}
\label{sec:MEP}



While statistical mechanics is well developed for systems in equilibrium, it is vastly less developed and accepted for systems out of equilibrium like the Earth system. Knowing the most probable end-state of a system does not necessarily yield information about the path that a system will take to get there other than that it must satisfy the second law. However, much theoretical and empirical work has sought a variational principle that could identify a preferential path that a system travels toward equilibrium in simple fluid systems, and for linear systems very close to, but not in, equilibrium \citep[e.g.,][]{Malkus1954b,MalkusVeronis1958,Busse1967,Palm1972,Prigogine1962,Jaynes1980}.

A more wide-ranging variational hypothesis was proposed by \citet{Paltridge1975,Paltridge1978,Paltridge1979} with application to the out-of-equilibrium steady state of climate. This ``maximum entropy production principle'' (MEP principle)\footnote{This should not be confused with the maximum entropy production principle established in statistical mechanics for the relaxation of simple systems toward equilibrium \citep[e.g.,][]{RobertSommeria1992,Robert2003}, nor the related `maximum caliber' hypothesis \citep{Presseetal2013,Ghoshetal2020}.} hypothesizes that the climate system adjusts itself to be in a state that maximizes entropy production \citep[see reviews by][]{Ozawa2003,KleidonLorenz2005,MartyushevSeleznev2006,OBrienStephens1995}. The MEP principle is a controversial hypothesis because it is not derivable from the equations of motion. None of the tools of the previous sections are clearly applicable here.


\citet{Paltridge1975} considered a highly idealized climate model where both the top of atmosphere and planetary surface are in energy balance. The hypothesis that the nonequilibrium climate system has sufficient degrees of freedom to be ``controlled by some minimum principle'' was proposed and tested. Paltridge constructed a ten-box climate model of the Earth spanning pole-to-pole with \add{four} free parameters: cloud areal fraction, \add{surface} temperature, sum of latent and sensible heat flux at the surface, \add{and meridional energy flux convergence}. Each box contained two energy balance equations; one for total energy balance (top-of-atmosphere), one for oceanic energy balance, \add{and a turbulent heat flux parameterization. For a given meridional energy flux distribution, the remaining parameters can be solved for.} Paltridge showed that minimization of entropy exchange with the environment [min$(dS_e/dt)$] between the ten boxes yielded realistic values of meridional heat flux. The only other constraint to the system was conservation of total energy.

\citet{Rodgers1976} argued that the simplicity of the model (lack of physics) suggested it should be broadly applicable to other planets but would clearly fail if applied to Mercury or Mars. Rodgers also pointed out that, in steady state, $dS/dt=0$, and minimizing $dS_e/dt$ is equivalent to maximizing internal entropy production. \citet{Paltridge1978} adopted this interpretation and extended the ten-box model to a two-dimensional global climate model with 380 cross-box energy flows. Upon minimizing entropy exchange (maximizing entropy production), the results were a surprisingly good fit to global distributions of temperature and cloud cover. His results suggested that the MEP principle was able to capture large-scale observed circulation patterns. However, \citet{Mobbs1982} noted that this is likely due to the model's albedo distribution being fixed to the observed albedo distribution. Even with this observational constraint, the two-dimensional ocean energy transport was quite different to that observed in some regions \citep{SohnSmith1994}.

Interest in the potential for MEP to provide closure for underconstrained climate problems prompted further development and application to energy balance models \citep[e.g.,][]{NicolisNicolis1980,Grassl1981,Mobbs1982,Gerardetal1990,SohnSmith1993,SohnSmith1994,Lorenzetal2001}. While such studies reported some successful applications, in general the results were mixed. Moreover, there are a number of difficulties with the MEP principle that, in our view, limit its applicability:


\paragraph{The MEP principle appears to fail when applied to vertical transport.}  Early applications of the MEP principle applied it to vertically-integrated representations of the climate system, for which only horizontal entropy exchanges are present \citep[e.g.][]{Lorenzetal2001}. However, the vast majority of the entropy production on Earth is associated with vertical entropy fluxes \citep{Lucarinietal2011}. When MEP is applied to vertical entropy exchange, it often gives unphysical predictions; for example, temperature distributions that are gravitationally unstable or atmospheric layers that destroy entropy \citep{OzawaOhmura1997,PujolFort2002,Herbertetal2013,Changdissertation}.

\paragraph{The MEP principle cannot account for previous climate states or future warming of Earth.} \citet{Grassl1981} demonstrated that Paltridge's procedure predicted negligible polar warming under a doubling of CO$_2$, because it lacked an ice-albedo feedback. \citet{Gerardetal1990} suggested that their MEP findings are consistent with relatively steady global temperature across deep time, and inconsistent with glaciation periods. \citet{Paltridgeetal2007} attempted to apply the MEP principle to a global climate model with a water vapor feedback and found an implausible reduction of cloud cover upon a doubling of CO$_2$ concentration \citep{Paltridge2009}. 

\paragraph{The specification of the MEP optimization problem is ambiguous.} The MEP principle has also been tested in global climate models \citep[e.g.,][]{Kleidonetal2003,Kleidonetal2006,ItoKleidon2005,FraedrichLunkeit2008,Pascaleetal2012a,Pascaleetal2013}. \citet{Kleidonetal2003} considered a simplified global climate model with no moisture or radiation in which the only irreversible entropy production occurs through frictional dissipation. The authors argued that the MEP principle could be used to determine the ``correct'' surface friction parameters in the model. However, using the irreversible entropy production in the atmosphere to set the surface friction parameters implicitly sets land/ocean areal fraction and roughness as primarily functions of the atmospheric dynamics, which is implausible \citep{Changdissertation}.
More generally, the MEP principle is unconstrained, meaning that aside from mass and energy conservation, there is no way to consider external factors of a planet like its size or rotation rate, which are among demonstrably important constraints on the general circulation \citep{Goody2007}. Adding constraints to the problem eventually obviates the need for a variational principle entirely \citep[e.g.,][]{Goody2007,Changdissertation}.


\paragraph{The MEP principle lacks a sound physical basis.} In spite of several efforts \citep[e.g.,][]{Paltridge1979,Paltridge2001}, no theoretical justification had been found for MEP, rendering its acceptance among climate scientists and physicists quite limited. \citet{Dewar2003,Dewar2005} attempted to provide a theoretical basis for the MEP principle using the Shannon entropy from information theory \citep{Jaynes1957a,Jaynes1957b}.
This development was severely challenged with the publication of technical rebuttals to the Dewar papers \citep{Bruers2007,GrinsteinLinsker2007}. \citet{GrinsteinLinsker2007} argued that key assumptions in the derivation of \citet{Dewar2005} required the system in question to be very close to equilibrium, which \citet{Dewar2009} conceded. \add{\citet{Volk2010} further argued that such a principle cannot be agnostic to the ways in which entropy is produced, because of the dominant role of moisture in atmospheric entropy production.}

Despite the difficulties described above, MEP research is ongoing, and proponents argue that it has applications as wide ranging as the evolution of river networks, economics, biotic activity and the Gaia hypothesis \citep{KleidonLorenz2005}. More recent work recasts MEP as an ``inference algorithm'' rather than a physical principle, and claims that it is effectively not falsifiable \citep{Dewar2009,DykeKleidon2010}. 
\add{\citet{Kleidonetal2014,Kleidon2016} recently suggested that one may interpret the MEP principle as a ``maximization of power'' instead.} 



The MEP principle is understandably appealing because it allows for a solution of the steady state of the Earth's climate without solving for its complicated dynamical evolution. Yet the principle lacks theoretical justification as well as consistent numerical and observational success. We conclude that the validity and usefulness of the MEP principle remain aspirational.

\section{Conclusions \& perspectives}

\label{sec:conclusions}
  
 We have reviewed the key scientific developments in  the application of the second law of thermodynamics to the climate system. The climate system may be defined in a number of ways, each of which differs in the extent to which radiation is treated as part of the system rather than as part of the surroundings \citep{Bannon2015,Gibbins2020}. By focusing exclusively on matter within the climate system, the second law may be seen to provide a direct constraint on the rate at which the climate system's heat engine performs work \citep{Goody2003,Pauluis2002}.
 
 The heat-engine perspective on atmospheric circulations was shown to allow for theoretical constraints on convective updraft velocities \citep{Emanuel1996}, tropical cyclone intensity \citep{Emanuel1986}, and the atmospheric meridional heat transport \citep{Barryetal2002}. While the theory of potential intensity of tropical cyclones has been quite successful \citep{Emanuel2018}, \add{the development of a first principles theory for convective updraft velocities based on the second law is made challenging by the dominance of moist irreversible processes in the atmosphere's entropy budget \citep{Pauluis2002,SinghOGorman2016}. Ongoing work confirms the important role of moist processes such as water-vapor mixing and cloud microphysics in constraining cloud updrafts  \citep{Parodi2009,Singh2013,Singh2015,Seeley2015}. The success of theories for the meridional heat transport by the atmosphere is also limited by the extent to which they account for the effects of moist processes on the atmospheric circulation. Such moist processes are} particularly influential in governing the response of the atmospheric circulation to global climate change
 \citep{Laliberteetal2015,SinghOGorman2016}.
 
 Interrogating the entropy budget and heat-engine characteristics of climate models and of other planets may provide a pathway toward better understanding of the Earth's climate system \citep{Lucarini2010,Lembo2019}. The heat engines of other planets differ in fundamental ways from that of the Earth, challenging our implicit assumptions about the thermodynamics of planetary circulations \citep[e.g.,][]{KollKomacek2018}. Global climate models are rarely developed with the second law in mind; entropy-based diagnostics therefore provide opportunities for model evaluation \citep{Lembo2019}. However, challenges remain in accurately modeling the second law in global climate models because their entropy production occurs in complex subgrid scale parameterizations which represent both reversible and irreversible processes \citep[e.g.,][]{Gassmann2019}. The rapid increase in computing power each decade continues to reduce the need for such subgrid scale parameterizations as smaller scales are explicitly resolved. But it will be many decades to centuries before direct numerical simulation of the atmosphere and ocean at Reynolds numbers characteristic of geophysical flows is possible.
 

We have also described the application of variational methods to the climate system. We discussed states of minimum free energy \citep{Bannon2005} and minimum potential energy \citep{Lorenz1955} as methods for determining the atmosphere's ability to perform work. A statistical-mechanics formulation for two-dimensional geophysical flows was also described in which concepts such as entropy are used in a completely different context \citep{BouchetVenaille2012}. \add{This approach considers only the hydrodynamic degrees of freedom of important climate subsystems like the polar vortices, which lack two distinct thermal reservoirs and a corresponding heat engine analogy.} Nevertheless, the idea of irreversibility is key to \add{the prediction of these out-of-equilibrium, steady-state flows}, and \add{the statistical GFD approach} may be described as an application of the second law to the climate system.

Finally, we discussed the maximum entropy production (MEP) principle proposed by \citet{Paltridge1975,Paltridge1978}. Although we argue that a sufficient basis for accepting the MEP principle has not been established, we note that it has motivated a great deal of research into Earth's entropy budget  \citep[e.g.,][]{Peixotoetal1991,StephensOBrien1993,Goody2003,Pascaleetal2011}, which has no doubt contributed to an improved understanding of irreversible processes within the climate system.
  
 This review has also highlighted a number of promising directions for future research. In particular, we have shown that global climate models and higher-resolution models run in RCE appear to disagree as to whether the work performed by the atmospheric heat engine will increase or decrease in a warmer climate \citep{SinghOGorman2016,Laliberteetal2015}. Further analysis of the atmospheric heat engine in both types of models could shed light on this important question \citep{Lucarini2010,Lembo2019}. Further, we presented new analysis of the effect of convective organization on the mechanical efficiency of moist convection. To our knowledge, such an analysis has not been presented previously, but it may have important implications for our understanding of convective organization, particularly in the context of global warming \citep{Wing2014}.
 
 The range of research in classical thermodynamics and statistical mechanics that can be brought to bear on climate questions is broad and in many cases well-developed. 
 Indeed, statistical mechanics approaches have already been applied to climate models in the form of applications of Hamiltonian fluid mechanics to the problem of parameterization of ocean eddies \citep{Davidetal2018} and the use of statistical mechanics principles for stochastic parameterization as recently reviewed by \citet{Ghil2020}.  
 Building on these examples requires increased collaboration between climate scientists and physicists which we hope this review will help to foster.

\begin{acknowledgments}
The authors thank Almut Gassmann, Goodwin Gibbins, Isaac Held, Valerio Lucarini, Pascal Marquet, Jonathan Mitchell, David Raymond, Laurel R\'egibeau-Rockett, Bjorn Stevens, and Peter Weichman for helpful discussions and feedback during the preparation of this manuscript. Detailed reviews from Kerry Emanuel and two anonymous reviewers are also gratefully acknowledged.
MSS acknowledges support from the Australian Research Council (grant nos. DE190100866 \& DP200102954) and computational resources and services from the National Computational Infrastructure (NCI), both supported by the Australian Government.
\end{acknowledgments}

\bibliographystyle{abbrvnat}
\bibliography{references}

\end{document}